\documentclass{aa}  

\usepackage{graphicx}
\usepackage{txfonts}

\usepackage{textgreek}
\usepackage{xcolor}
\usepackage[
    breaklinks=true,
    colorlinks=true,
    citecolor=blue,
    urlcolor=blue,
    linkcolor=blue
]{hyperref}
\usepackage{booktabs}
\usepackage{cleveref}
\usepackage{natbib}
\usepackage{subcaption}
\usepackage{orcidlink}
\usepackage[normalem]{ulem}
\usepackage{lscape}
\usepackage{placeins}
\usepackage{cuted}
\usepackage{capt-of}

\bibpunct{(}{)}{;}{a}{}{,}

\crefname{figure}{Figure}{Figures}

\newcommand{\arsec}{\ensuremath{{}^{\prime\prime}}}

\crefname{chapter}{Chapter}{Chapters}
\crefname{section}{Section}{Sections}
\crefname{subsection}{Subsection}{Subsections}
\newcommand{\Autoref}[1]{%
  \hyperref[#1]{\Cref{#1}}%
}

\begin{document}

    \title{High-precision abundance analysis of a unique CEMP-no star in Sculptor}

    \author{Francisco Cuadra\inst{1, 2}\orcidlink{0009-0005-6617-8121}
        \and
        \'Asa Skúladóttir\inst{1, 2}\orcidlink{0000-0001-9155-9018}
        \and
        Romain Lucchesi\inst{1}\orcidlink{0000-0002-9750-1922}
        \and
        Ioanna Koutsouridou\inst{1, 2}\orcidlink{0000-0002-3524-7172}
        \and
        Irene Vanni\inst{1, 2}\orcidlink{0000-0001-9647-0493}
        \and
        Stefania Salvadori\inst{1, 2}\orcidlink{0000-0001-7298-2478}
        \and 
        Eline Tolstoy\inst{3}\orcidlink{0000-0002-3047-4843}
        \and
        Raghubar Singh\inst{1}\orcidlink{0000-0001-8360-9281}
        }

    \institute{Dipartimento di Fisica e Astronomia, Università degli Studi di Firenze, Via G. Sansone 1, I-50019, Sesto Fiorentino, Italy\\
        \email{\href{mailto:francisco.cuadra@unifi.it}{francisco.cuadra@unifi.it}}
        \and
        INAF/Osservatorio Astrofisico di Arcetri, Largo E. Fermi 5, I-50125, Firenze, Italy
        \and
        Kapteyn Astronomical Institute, University of Groningen, PO Box 800, 9700AV Groningen, The Netherlands
        }

    \authorrunning{F. Cuadra et al.}

    \date{Received 28 May 2026; accepted 29 October 2026}

    \abstract{
        The Carbon-Enhanced Metal-Poor stars (CEMP-no, low barium) are commonly accepted as the direct descendants of Population~III (Pop~III) stars which either ended their lives as faint supernovae (SNe; $E_{\textsl{SN}} \lesssim 10^{51}\,\text{erg}$) or were fast-rotating stars (spinstars).
        While CEMP-no stars are common in the Milky Way (MW) and ultra-faint dwarf galaxies (UFDs), they are surprisingly rare in dwarf spheroidal galaxies (dSphs).
        This could hide an environmental dependence of the properties of Pop~III stars, highlighting the importance of high-resolution studies of CEMP-no stars in dSphs.
        Here we present the first detailed study of a CEMP-no star with $[\text{Fe}/\text{H}] < -3$ in the Sculptor dSph.
        The high-resolution VLT/UVES spectra allowed us to derive one of the most precise abundance patterns obtained for a star belonging to a dwarf galaxy, providing chemical abundances for 17 elements and 4 upper limits.
        This star, DR20080, is extremely metal-poor, \mbox{$[\text{Fe}/\text{H}] = -3.26$}, and clearly enhanced in carbon, $[\text{C}/\text{Fe}] = +1.60$ (after correcting for internal mixing).
        Furthermore, it has low $[\text{Ba}/\text{Fe}] = -1.28$.
        The oxygen abundance of DR20080 ($[\text{O}/\text{Fe}] = +1.53$) is significantly higher than the average found in CEMP-no stars in the MW, while the iron-peak elements are under-abundant ($[\text{Mn}/\text{Fe}] = -1.02$; $[\text{Ni}/\text{Fe}] = -0.45$).
        The abundance pattern of DR20080 was compared to theoretical yield predictions of Pop~III and Pop~II SNe to infer the properties of its progenitor.
        The best result points toward a double Pop~III progenitor scenario, with the major contributor to the abundance pattern being a Pop~III faint SN with mass $m_{\text{PopIII}} = 43 \text{M}_{\odot}$ and explosion energy $E_{\text{SN}} = 0.6 \cdot 10^{51} \,\text{erg}$, while the minor contributor is a Pop~III hypernova of $m_{\text{PopIII}} = 19.4 \, \text{M}_{\odot}$ and $E_{\text{SN}} = 10 \cdot 10^{51} \,\text{erg}$.
        The unique properties of DR20080 have never before been observed in Sculptor or in any other environment within the MW and its satellite system.
        }


    \maketitle

%-------------------------------------------------------------------

    \section{Introduction}
    \label{sec:Introduction}     
        
        According to cosmology and Big Bang nucleosynthesis, the first stars formed in a neutral environment composed only of hydrogen and helium.
        All the stars with such a metal-free composition are called Population~III (Pop~III) stars.
        Their impact on the evolution of the Universe was significant, as they were the first source of ionising photons, which started the reionisation of the Universe, as well as of metals, enriching the primordial surrounding environment.
        Due to their metal-free composition, Pop~III stars are predicted to be more massive than present-day stars \citep[e.g.,][]{mckee_formation_2008,hirano_one_2014}. 
        As a result, their initial mass function (IMF) is expected to be top-heavy, although different simulations yield varying conclusions \citep[e.g.,][]{klessen_first_2023}. 
        Recent observational evidence supports this picture: \cite{rusta_pristine_2026} shows that the observed properties of the high-redshift Pop~III galaxy Hebe \citep{maiolino_search_2026, ubler_ga-nifs_2026} are consistent with a top-heavy IMF extending beyond $150 \,\text{M}_{\odot}$. 
        Such an IMF implies short stellar lifetimes, which is consistent with the persisting non-detection of metal-free stars in the local Universe \citep[e.g.,][]{rossi_ultra-faint_2021}.

        According to simple models for early chemical enrichment \citep[e.g.,][]{salvadori_probing_2019,vanni_characterizing_2023} and numerical simulations resolving the first SN explosions \citep[e.g.,][]{bromm_formation_2002,ritter_confined_2012}, even one Pop~III SN could provide enough metals to allow for normal Pop~II star formation \citep[e.g.,][]{ritter_confined_2012,vanni_characterizing_2023}.
        This means that second-generation low-mass stars ($\leq 0.8 \,\text{M}_{\odot}$) with lifetimes longer than the age of the Universe were able to form and survive to the present day \citep[e.g., ][]{omukai_protostellar_2000,bromm_fragmentation_2001,maio_transition_2010,schneider_metals_2010,schneider_first_2012}.
        This is crucial because it provides an indirect method to study Pop~III stars.
        Indeed, the atmospheres of second-generation low-mass stars retain almost unaltered, preserving the chemical yields released by the SNe of their Pop~III progenitors, which primarily depend on the progenitor mass and explosion energy \citep[e.g., ][]{heger_nucleosynthetic_2002,vanni_characterizing_2023}.

        Currently, the Milky Way (MW) stellar halo and its satellite galaxies, including dwarf spheroidal galaxies (dSphs) and ultra-faint dwarf galaxies (UFDs), represent the most promising environments for searching for surviving second-generation stars \citep[e.g.,][]{tumlinson_chemical_2006,salvadori_cosmic_2007,salvadori_carbon-enhanced_2015}.
        Indeed, these are the only nearby systems where high-resolution spectroscopy on individual stars is feasible and whose stellar populations are predominantly ancient \citep[e.g.,][]{tolstoy_star-formation_2009,bonifacio_most_2025}.
        However, due to the relatively large distances of the MW satellites, high-resolution spectroscopy can only be carried out for the brightest stars, typically the red giant branch (RGB) stars.
        
        One of the strategies for identifying second-generation stars is to look for the most metal-poor stars.
        Typically, stars are classified as very metal-poor when their iron abundance is $[\text{Fe}/\text{H}] \leq -2$.
        Among them, the Carbon-Enhanced Metal-Poor stars (CEMP; $[\text{Fe}/\text{H}] < -2$, $[\text{C}/\text{Fe}] > +0.7$, \citealt{beers_discovery_2005,aoki_carbonenhanced_2007}) are a particularly significant category, considering that their fraction grows towards lower metallicities, making up over 80\% of the MW halo at $[\text{Fe}/\text{H}] < -4$ \cite[e.g.,][]{lucatello_frequency_2006,lee_carbon-enhanced_2013,placco_carbon-enhanced_2014,yoon_galactic_2018}.
        
        The CEMP stars are further divided into subgroups according to their neutron-capture element abundances.
        These subgroups are associated with different formation channels, which also lead to different origins for their enhanced carbon abundances.
        Those CEMP stars that also show an enrichment in heavy elements like barium, formed through the slow neutron-capture process (s-process), are classified as CEMP-s stars.
        The primary mechanism behind the C-enhancement of a CEMP-s star is the mass transfer from a binary companion that has entered the asymptotic giant branch (AGB) phase, which is the main production site of s-process elements \citep{aoki_carbonenhanced_2007,spite_carbon-enhanced_2013,starkenburg_binarity_2014,hansen_elemental_2015}.
        On the other hand, CEMP-r stars are enriched in the rapid neutron-capture (r-process) elements like europium.
        It is important to notice that the s-process and r-process enrichment are not mutually exclusive. 
        Stars enriched both in Eu and Ba are termed CEMP-r/s, and they could be linked to the intermediate neutron-capture process (i-process; \citealt{hampel_intermediate_2016}).
        However, if a CEMP star shows no significant enrichment in neutron-capture elements, despite its high carbon content, it is classified as a CEMP-no star \citep[e.g.,][]{beers_discovery_2005}.
        
        The CEMP-no stars are commonly accepted as the direct descendants of those Pop~III stars which either were fast-rotating stars (spinstars, e.g., \citealt{meynet_early_2006, meynet_are_2010}) or ended their lives as faint SNe \citep[fSNe; e.g.,][]{iwamoto_first_2005,vanni_characterizing_2023}.
        These lower-energy SNe are believed to pollute their surrounding environments mainly with C and other light elements ($Z < 20$), resulting in very high $[\text{C}/\text{Fe}]$ ratios.
        Depending on the explosion energy, other light elements, such as O, Na, Mg and Si, may also be enhanced \citep{norris_most_2013}.
        The study of CEMP-no stars is therefore of primary importance for attempting to characterise the properties of Pop~III stars, such as their IMF and the typical energy distribution of their SNe \citep[e.g., ][]{koutsouridou_energy_2023}.

        However, CEMP-no stars, which are quite common in the MW and in UFDs, are, surprisingly, extremely rare in dSphs \citep[e.g., ][see, in particular, Fig. 14 of \citealt{lucchesi_extremely_2024}]{starkenburg_extremely_2013,skuladottir_first_2015,skuladottir_tracing_2024,lucchesi_extremely_2024}.
        The reason behind this distinction remains uncertain \citep[e.g.,][]{salvadori_carbon-enhanced_2015}, which makes the study of CEMP-no stars discovered within dSph galaxies even more crucial.
        Such research could reveal whether these stars share common characteristics with those found in UFDs and in the MW halo or if they should be categorised into subpopulations based on their formation environments.
        This insight could help determine whether and how the chemical enrichment by Pop~III stars is influenced by the environment in which they formed.

        The Sculptor dSph, a satellite of the MW, was discovered by \cite{shapley_two_1938} and is located at a distance of $86 \pm 5 \,\text{kpc}$ \citep{pietrzynski_araucaria_2008}. 
        It has a total mass of $(3.4 \pm 0.7) \cdot 10^8 \,\text{M}_{\odot}$, with a mass-to-light ratio of $158 \pm 33 (M/L)_{\odot}$, which reveals its dark matter dominated nature \citep{battaglia_kinematic_2008}, and a mean systemic line-of-sight (los) velocity of $v_{\textsl{los}} = 111.2 \pm 0.3 \,\text{km s}^{-1}$ \citep[][using data from \citealt{tolstoy_3d_2023}]{arroyo-polonio_chemo-dynamics_2024}.

        \begin{figure}[t]
            \centering
            \includegraphics[width=\hsize]{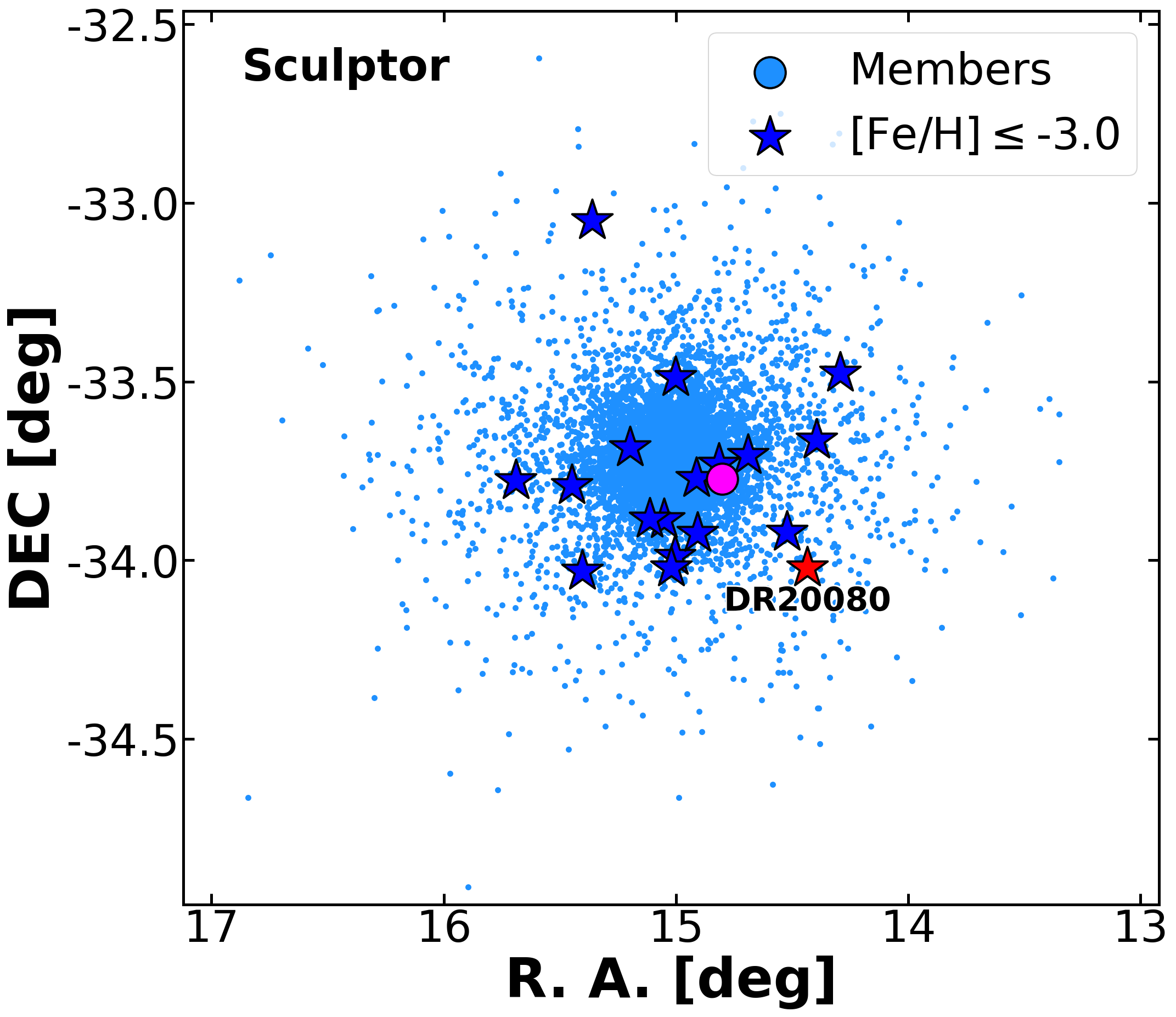}
            \caption{Member stars of the Sculptor dSph from \cite{battaglia_gaia_2022} based on positions from \textit{Gaia} eDR3 (\textit{blue dots}). 
            \textit{Star symbols} indicate known members with $[\text{Fe}/\text{H}] \leq -3.0$ (data from \citealt{tolstoy_3d_2023} and \citealt{skuladottir_tracing_2024}). 
            \textit{Red star}: DR20080, the CEMP-no star analysed in this work; \textit{magenta circle}: ET0097, the other CEMP-no star in the Sculptor dSph analysed in high-resolution \citep{skuladottir_first_2015}.
            }
            \label{fig:DR20080_position}
       \end{figure}
        
        Since its discovery, Sculptor has been the subject of numerous photometric and spectroscopic studies of its resolved stellar population.
        Particularly, an analysis of the colour-magnitude diagram reveals a star formation peak around $12\,\text{Gyr}$ ago, followed by a gradual decline in the star formation rate until it ceased entirely around $8 \,\text{Gyr}$ ago \citep{de_boer_star_2012,bettinelli_star_2019}.
        The stellar population of Sculptor is divided in two main populations, with the metallicity distribution functions showing peaks for $[\text{Fe}/\text{H}] = -1.41$ and $[\text{Fe}/\text{H}] = -2.00$, and a small sub-population at $[\text{Fe}/\text{H}] = -2.9$ which may be the residual of a past merger with a UFD \citep{arroyo-polonio_chemo-dynamics_2024}.

        During recent years, several studies in Sculptor focused on the carbon abundance of its stars \citep{frebel_linking_2010,tafelmeyer_extremely_2010,starkenburg_extremely_2013,jablonka_early_2015,kirby_carbon_2015,simon_chemical_2015,chiti_detection_2018,skuladottir_first_2015,skuladottir_tracing_2024}. 
        What was found is a general lack of CEMP stars, with the only exception of the low-resolution study from \cite{chiti_detection_2018}.
        However, it should be noted that, even in this case, no stars with $[\text{C}/\text{Fe}] > +0.90$ were found.
        So far, the only star in Sculptor that was confirmed to be a CEMP-no star through high-resolution analysis is ET0097, with $[\text{Fe}/\text{H}] = -2.03$ and $[\text{C}/\text{Fe}] = +0.90$ \citep{skuladottir_first_2015}.
        
        Here we present the second high-resolution follow-up of a CEMP-no star in the Sculptor dSph, DR20080.  
        This high-precision study aims to derive the chemical abundances of as many elements as possible, compare them with those of other metal-poor stars in Sculptor and across the MW and its satellite system, and fit them to theoretical yield models to constrain the properties of DR20080’s progenitors.

%--------------------------------------------------------------------
    \section{Data acquisition}
    \label{sec:Data_acquisition}
 
        In this work we present the results of the high-resolution follow-up of DR20080 (RA $00\text{h}\, 57\text{m}\, 44.16\text{s}$\,; DEC $-34\text{d}\, 01\text{m}\, 14.3\text{s}$), the only extremely metal-poor ($[\text{Fe}/\text{H}] < -3$) CEMP-no star discovered in the Sculptor dSph (see \autoref{fig:DR20080_position}).
        DR20080 is an RGB star, which was initially identified as a metal-poor star by a large survey that exploited the spectrograph FLAMES at the Very Large Telescope (VLT) to investigate the \ion{Ca}{ii} NIR triplet \citep{tolstoy_3d_2023}.
        It was followed up with the multi-wavelength medium-resolution spectrograph VLT/X-SHOOTER for the confirmation of the FLAMES measurement. 
        These medium-resolution spectra, analysed by \citet{skuladottir_tracing_2024}, allowed the authors to derive a metallicity of $[\text{Fe}/\text{H}] = -3.00 \pm 0.14$ (which confirmed the metal-poor nature of DR20080), but also to add an estimate of the carbon abundance $[\text{C}/\text{Fe}]_{\textsl{corr}} = 1.02 \pm 0.17$ from the CH $\sim 4300 \,\text{\AA}$ molecular band.
        These abundances, together with the upper limit $[\text{Ba}/\text{Fe}] < -0.5$, revealed the CEMP-no nature of DR20080, making it the second CEMP-no star confirmed in the Sculptor dSph and the most metal-poor of the two.
        However, further analysis outside of Fe, C and Ba was not feasible, due to both the low signal-to-noise ratio ($S/N$) and the modest spectral resolution of the X-SHOOTER spectrum ($S/N \sim 12\,\, \text{at}\, \, 4400 \, \text{\AA}$ and $R \sim 5400$, see \citealt{skuladottir_tracing_2024}).
        Moreover, in the recent work by \cite{arroyo-polonio_chemo-dynamics_2024}, the authors showed that within the Sculptor dSph there is a sub-population of stars, which includes DR20080, that may have originated outside Sculptor and was then accreted through a past merger.

        \begin{table}[t]
    		\caption{Log of the VLT/UVES observations.}
    		\label{tab:log_UVES_observations}
    		\centering
    		\begin{tabular}{cccccccc}
    			\toprule
    			Date & Exposure Time & Airmass & Seeing \\
       
                $[\text{dd}/\text{mm}/\text{yyyy}]$ & $[\text{s}]$ & (average) & $[\arsec]$ \\
    			\midrule
    			25/10/2022 & 3000 & 1.021 & 0.58 \\
    			29/10/2022 & 3000 & 1.062 & 1.01 \\
                18/11/2022 & 3000 & 1.020 & 1.17 \\
                18/11/2022 & 3000 & 1.067 & 1.18 \\
                20/11/2022 & 3000 & 1.022 & 0.78 \\
                21/11/2022 & 3000 & 1.035 & 0.76 \\
                21/11/2022 & 3000 & 1.098 & 0.75 \\
                22/11/2022 & 3000 & 1.020 & 0.68 \\
    			\bottomrule
    		\end{tabular}
    	\end{table}   

        \begin{table}
    		\caption{Wavelength ranges of all spectral bands and the corresponding $S/N$.}
    		\label{tab:Bands_and_SN}
    		\centering
    		\begin{tabular}{cccc}
    			\toprule
    			Spectral Band & Wavelength & $S/N$ & Line-free Regions \\
    			& $[\text{\AA}]$ & $[\text{pxl}^{-1}]$ & $[\text{\AA}]$ \\
    			\midrule
    			B  & $3700 - 4500$ & $16$ & $4437.0 - 4441.0$ \\
    			Rl & $4630 - 5585$ & $40$ & $5438.5 - 5444.0$ \\
                Ru & $5688 - 6625$ & $52$ & $5868.3 - 5874.3$ \\
    			\bottomrule
    		\end{tabular}
    	\end{table}  

        In the present work, we performed a high-resolution follow-up of DR20080 that was carried out through a series of eight observations, between October and November 2022, using the high-resolution VLT/UVES spectrograph.
        Each of these eight observations was conducted with an exposure time of 3000 seconds, as listed in \autoref{tab:log_UVES_observations}, and simultaneously covered three spectral bands:  Blue (B) band, Lower Red (Rl) band, and Upper Red (Ru) band. 
        The same table also reports the date of observations and the observational conditions.
        The mean airmass was calculated as the mean of the six airmass values of each exposure (one value for the start of the observation and another for the end, for all three spectral bands).        
        The mean seeing value per observing night was derived as the average between the values on the three spectral bands.
        
        The instrumental configuration consisted of a $1.2\arsec$ slit and 2$\times$2 pixel binning, with the DIC1 dichroic and the CD\#2/390 and CD\#3/564 settings for the blue and red arms, respectively.
        This configuration is a standard UVES setting centered on the wavelengths $3900 \, \text{\AA}$ in the blue arm, and $5640 \, \text{\AA}$ in the red arm.       
        The spectral resolution for all three bands was $R \sim 35000$.              
        The final wavelength ranges adopted for the subsequent analysis are listed in \autoref{tab:Bands_and_SN}.
        The edges of the full wavelength coverage were trimmed in order to exclude regions with lower data quality, particularly at the boundaries of the spectral ranges.        
        The 8 exposures per spectral band were then coadded to improve the $S/N$, obtaining the final values reported in \autoref{tab:Bands_and_SN}.
        During this process, we also performed a cosmic rays removal.
        The $S/N$ was determined in selected line-free regions of each spectral band, whose wavelength intervals are reported in \autoref{tab:Bands_and_SN}.

%-----------------------------------------------------------------
    \section{Data analysis}
    \label{sec:Data_analysis}

            \begin{table*}[t]            
                \caption{Apparent magnitudes of DR20080 in three spectral bands and corresponding reddening-corrected magnitudes.
                }
                \label{tab:photometry_GAIA_dustmaps}
                \centering
                \begin{tabular}{cccccccccc}
                    \toprule
                    \textit{Gaia} ID & RA & DEC & G & BP & RP & E(B$-$V) & $\text{G}_{\text{0}}$ & $\text{BP}_{\text{0}}$ & $\text{RP}_{\text{0}}$ \\
                     & $[\text{h}:\text{m}:\text{s}]$ & $[\text{d}:\text{m}:\text{s}]$ & $[\text{mag}]$ & $[\text{mag}]$ & $[\text{mag}]$ & $[\text{mag}]$ & $[\text{mag}]$ & $[\text{mag}]$ & $[\text{mag}]$ \\
                    \midrule
                     $5003180465076613376$ & $00 : 57 : 44.16$ & $-34 : 01 : 14.3$ & $17.84$ & $18.41$ & $17.17$ & $0.0178$ & $17.80$ & $18.36$ & $17.14$ \\
                    \bottomrule
                \end{tabular}
                \tablefoot{Apparent magnitudes are taken from \textit{Gaia} DR3, while the dust extinction E(B$-$V) is from dustmaps \citep{green_m_dustmaps_2018}.
                The last three columns are the apparent magnitudes corrected for reddening (see \Autoref{subsec:Stellar_parameters}).
                }
            \end{table*}
            \begin{table*}[t]            
                \caption{Values of the $T_{\textsl{eff}}$ values obtained with \textit{Gaia} photometry along with the mean $T_{\textsl{eff}}$ and its standard deviation.}
                \label{tab:Teff}
                \centering
                \begin{tabular}{cccccc}
                    \toprule
                    Star & $T_{\textsl{eff}}(\text{BP} - \text{RP})_{\text{0}}$ & $T_{\textsl{eff}}(\text{BP} - \text{G})_{\text{0}}$ & $T_{\textsl{eff}}(\text{G} - \text{RP})_{\text{0}}$ & $T_{\textsl{eff}}$ & $\Delta T_{\textsl{eff}}$ \\
                     & $[\text{K}]$ & $[\text{K}]$ & $[\text{K}]$ & $[\text{K}]$ & $[\text{K}]$ \\
                    \midrule
                    DR20080 & $4658$ & $4564$ & $4793$ & $4672$ & $\pm 94$ \\
                    \bottomrule
                \end{tabular}
            \end{table*}
    
        \subsection{Tools}
        \label{subsec:Tools}

            The abundance analysis carried out in this work was conducted with the aid of \textit{ASTRA} (Analysis of STellar Radiation and Atmospheres) from Romain Lucchesi, already used in \cite{lucchesi_extremely_2024}. 
            This tool wraps together the MARCS atmospherical models \citep{gustafsson_grid_2008} and for the spectral analysis \textit{DAOSPEC} \citep{stetson_daospec_2008}, to perform the equivalent width ($EW$) method, and \textit{Turbospectrum} \citep{plez_turbospectrum_2012}, to perform the spectral synthesis method.

            We adopted the MARCS models with a standard chemical composition, corresponding to $[\alpha/\text{Fe}] = +0.4$ at $[\text{Fe}/\text{H}] = -3.00$, assuming one-dimensional (1D) and Local Thermodynamic Equilibrium (LTE) approximations.
            The spectral analysis was performed through a combination of $EW$ and spectral synthesis methods.
            The line list given to \textit{DAOSPEC} is a compilation of those presented in \cite{tafelmeyer_extremely_2010,van_der_swaelmen_chemical_2013,jablonka_early_2015}, and is the same as that adopted by \cite{lucchesi_extremely_2024} for the analysis of metal-poor stars.
            The literature line list used for the chemical abundances derivation through spectral synthesis was a combination of the VALD database\footnote{\url{https://vald.astro.uu.se/vald.php}} \citep{piskunov_vald_1995,ryabchikova_vienna_1997,kupka_vald-2_1999,kupka_vald-2_2000} and, for molecular lines, we adopted the line lists from \cite{sneden_line_2014} and \cite{masseron_ch_2014}.

            After our initial LTE analysis, we applied state-of-the-art 1D Non-LTE (NLTE) corrections (see \Autoref{subsec:NLTE_corrections}) using the online tools from \textit{MPIA}\footnote{\url{https://nlte.mpia.de/gui-siuAC_secE.php}} \citep{kovalev_m_notitle_2018} and \textit{Inasan}\footnote{\url{https://spectrum.inasan.ru/nLTE/}} \citep{mashonkina_influence_2016,mashonkina_1d_2023}, which take the stellar parameters as input and output line-by-line corrections for a specific ion.

        \subsection{Line-of-sight velocity}
        \label{subsec:l-o-s_velocity}

            The membership of DR20080 in the Sculptor dSph had already been established through the study of proper motions conducted by \citet{battaglia_gaia_2022} using \textit{Gaia} eDR3.
            We performed los velocity (often called radial velocity) measurements (properly corrected for the heliocentric velocity) for all our 24 spectra across the three spectral bands (see \autoref{App:Rad_Vel} for more details). 
            The mean los velocity value of DR20080, $v_{\textsl{los}} = 120.15 \pm 0.27 \,\text{km s}^{-1}$, was measured as the arithmetic mean of all the velocity measurements, while the uncertainty was estimated as $\Delta v_{\textsl{los}} = \dfrac{\sigma}{\sqrt{n_{\textsl{obs}} - 1}}$, where $\sigma$ is the standard deviation between all the measurements and $n_{\textsl{obs}} = 24$ is the total number of exposures.
            This value falls within $1\sigma$ from the centre of the normalised distribution of los velocities of stars belonging to Sculptor \citep{tolstoy_3d_2023}, which further verifies that DR20080 indeed belongs to the Sculptor dSph.
            Moreover, this is fully consistent with the value $v_{\textsl{los}} = 120.63 \pm 0.48 \,\text{km s}^{-1}$ reported in the same paper and measured through the VLT/FLAMES spectra (obtained on 15/12/2018, program ID: 0102.B-0786), showing no substantial change of the los velocity with time. 
            Although binarity cannot be excluded at this stage, there is currently no indication that DR20080 belongs to a binary system.

            An arithmetic mean was also performed on the individual measurements for each band.
            The result for the B-band was $v_{\textsl{los}}^{\textsl{B}} = 120.65 \pm 0.51 \, \text{km s}^{-1}$, for the Rl-band $v_{\textsl{los}}^{\textsl{Rl}} = 120.32 \pm 0.27 \, \text{km s}^{-1}$, and for the Ru-band $v_{\textsl{los}}^{\textsl{Ru}} = 119.47 \pm 0.51 \, \text{km s}^{-1}$; the first two results are consistent, while the result for the Ru-band, which had a significantly lower number of identified lines, is slightly inconsistent. 
            However, the deviation was not considered significant enough to invalidate the overall mean result across the three bands.

        \subsection{Stellar parameters}
        \label{subsec:Stellar_parameters}
            
            The stellar parameters that characterise DR20080 were estimated through photometry based on \textit{Gaia} DR3 \citep{gaia_collaboration_gaia_2021}. 
            This choice was guided by the extremely low metallicity of this star.
            Indeed, as shown in \citet{mucciarelli_facing_2020}, spectroscopic parameters show an inconsistency at lower metallicities if compared with those derived by photometry, probably caused by metallicity-dependent 3D NLTE effects of \ion{Fe}{i}  \citep[e.g.,][see also Fig. 10 in \citealt{koutsouridou_large_2025}]{mashonkina_non-lte_2011,bergemann_non-lte_2012}.
            We used the three \textit{Gaia} filters G ($3300 - 10500\,\text{\AA}$), BP ($3300 - 6800\,\text{\AA}$) and RP ($6400 - 10500\,\text{\AA}$), whose measured magnitudes are reported in \autoref{tab:photometry_GAIA_dustmaps}, along with their values after reddening correction ($\text{G}_{\text{0}}$, $\text{BP}_{\text{0}}$ and $\text{RP}_{\text{0}}$), where the extinction value (E(B$-$V)) along the los was derived with \textit{dustmaps} \citep{green_m_dustmaps_2018} and the extinction coefficients were calculated as reported in \citet{gaia_collaboration_gaia_2018-1}.
            The uncertainties on the \textit{Gaia} apparent magnitudes, being reported as $\lesssim0.01\,\text{mag}$ for each filter \citep{sariya_gaia-based_2021}, are considered negligible compared to other sources of errors in this analysis.

            \subsubsection{Effective temperature}
            \label{subsec:Effective_temperature}
            
                The photometric effective temperature, $T_{\textsl{eff}}$, was obtained using the \textit{Gaia} calibration of \citet{mucciarelli_gaia_2020}:
                \begin{equation}
                \label{eq:Mucciarelli_Teff}
                    \dfrac{5040}{T_{\textsl{eff}}} = b_{\textsl{0}} + b_{\textsl{1}} \, C + b_{\textsl{2}} \, C^2 + b_{\textsl{3}} \, [\text{Fe}/\text{H}] + b_{\textsl{4}} \, [\text{Fe}/\text{H}]^2 + b_{\textsl{5}} \, [\text{Fe}/\text{H}] \, C\,\text{,}
                \end{equation}
                where $b_{\textsl{i}}$ are the coefficients for giant stars reported in \citet{mucciarelli_exploiting_2021}, while $C$ is the selected photometric colour. 
                Initially, the value $[\text{Fe}/\text{H}] = -3$ from \citet{skuladottir_tracing_2024} was assumed.
                Once the more accurate value of $[\text{Fe}/\text{H}] = -3.26$ was derived through abundance analysis, we performed a second iteration with the updated metallicity.
                The values reported below refer to this second iteration.
                
                Using the reddening corrected magnitudes reported in \autoref{tab:photometry_GAIA_dustmaps}, we applied \autoref{eq:Mucciarelli_Teff} using the three corrected colours $(\text{BP} - \text{RP})_0$, $(\text{BP} - \text{G})_0$ and $(\text{G} - \text{RP})_0$.
                This led to the three effective temperature values reported in \autoref{tab:Teff}.
                The final estimate of the effective temperature, $T_{\textsl{eff}} = 4672 \pm 94 \,\text{K}$, was derived as the average and standard deviation of the values for these three temperatures.

            \subsubsection{Surface gravity}
            \label{subsec:Surface_gravity}

                \begin{table}[t]
                    \caption{LTE abundances of DR20080. 
                    }
                    \label{tab:LTE_abundances}
                    \centering
                    \begin{tabular}{ccccc}
                        \toprule
                        Ion & $N_{\textsl{Lines}}$ & $\log{\epsilon}$ & $[\text{X}/\text{Fe}]$ & $\Delta [\text{X}/\text{Fe}]$ \\
                        \midrule
                        \ion{Fe}{i}                & $61$ & $+4.24$  & $-3.26$  & $\pm 0.12$ \\
                        \ion{Fe}{ii}               & $5$  & $+4.25$  & $-3.25$  & $\pm 0.13$ \\
                        C$_{\textsl{obs}}$ (CH)    & $1$  & $+6.24$  & $+1.07$  & $\pm 0.21$ \\
                        C$_{\textsl{corr}}$ (CH)   & $1$  & $+6.77$  & $+1.60$  & $\pm 0.21$ \\
                        \ion{N}{} (CN)             & $1$  & $+6.29$  & $+1.72$  & $\pm 0.26$ \\
                        \ion{O}{i}                 & $1$  & $+6.96$  & $+1.53$  & $\pm 0.25$ \\
                        \ion{Na}{i}                & $2$  & $+3.19$  & $+0.21$  & $\pm 0.15$ \\
                        \ion{Mg}{i}                & $3$  & $+4.81$  & $+0.47$  & $\pm 0.12$ \\
                        \ion{Al}{i}                & $1$  & $+3.02$  & $-0.17$  & $\pm 0.29$ \\
                        \ion{Si}{i}                & $1$  & $+3.95$  & $-0.30$  & $\pm 0.30$ \\
                        \ion{Ca}{i}                & $6$  & $+3.25$  & $+0.17$  & $\pm 0.07$ \\
                        \ion{Sc}{ii}               & $2$  & $-0.19$  & $-0.08$  & $\pm 0.21$ \\
                        \ion{Ti}{i}                & $4$  & $+1.67$  & $-0.02$  & $\pm 0.11$ \\
                        \ion{Ti}{ii}               & $15$ & $+1.87$  & $+0.18$  & $\pm 0.15$ \\
                        \ion{V}{i}                 & $2$  & $<+0.57$ & $<-0.10$ & $-$        \\
                        \ion{Cr}{i}                & $4$  & $+2.15$  & $-0.23$  & $\pm 0.16$ \\
                        \ion{Mn}{i}                & $2$  & $+1.15$  & $-1.02$  & $\pm 0.18$ \\
                        \ion{Co}{i}                & $4$  & $+1.75$  & $+0.02$  & $\pm 0.18$ \\
                        \ion{Ni}{i}                & $1$  & $+2.51$  & $-0.45$  & $\pm 0.14$ \\
                        \ion{Cu}{i}                & $3$  & $<+1.33$ & $<+0.40$ & $-$        \\
                        \ion{Zn}{i}                & $2$  & $<+1.80$ & $<+0.50$ & $-$        \\
                        \ion{Sr}{ii}               & $2$  & $-2.12$  & $-1.73$  & $\pm 0.22$ \\
                        \ion{Ba}{ii}               & $1$  & $-2.36$  & $-1.28$  & $\pm 0.17$ \\
                        \ion{Eu}{ii}               & $3$  & $<-2.24$ & $<+0.50$ & $-$        \\
                        \bottomrule
                    \end{tabular}
                    \tablefoot{The solar values are from \cite{asplund_chemical_2009}. 
                    For \ion{Fe}{i} and \ion{Fe}{i} abundances and errors, $[\text{Fe}/\text{H}]$ and $\Delta [\text{Fe}/\text{H}]$ values are listed instead of $[\text{X}/\text{Fe}]$ and $\Delta [\text{X}/\text{Fe}]$.   
                    C$_{\textsl{obs}}$ is the observed C abundance, \mbox{C$_{\textsl{corr}}$} when corrected for internal mixing \citep{placco_carbon-enhanced_2014}.
                    }
                \end{table}
        
                \begin{table}[t]
                    \caption{NLTE corrections and chemical abundances of DR20080.      
                    }
                    \label{tab:NLTE_abundances}
                    \centering
                    \begin{tabular}{cccccc}
                        \toprule
                        Ion & $N_{\textsl{Lines}}$ & $\delta [\text{X}/\text{H}]_{\textsl{NLTE}}$ & $[\text{X}/\text{Fe}]_{\textsl{NLTE}}$ & $\Delta [\text{X}/\text{Fe}]$ \\
                        \midrule
                        \ion{Fe}{i}               & $25/61$ & $+0.19$  & $-3.08$  & $\pm 0.12$ \\
                        \ion{Fe}{ii}              & $3/5$   & $+0.01$  & $-3.24$  & $\pm 0.13$ \\
                        C$_{\textsl{corr}}$ (CH)  & $-$     & $-$      & $+1.42$    & $\pm 0.21$ \\
                        N (CN)                    & $-$     & $-$      & $+1.54$  & $\pm 0.26$ \\
                        \ion{O}{i}                & $-$     & $-$      & $+1.35$  & $\pm 0.25$ \\
                        \ion{Na}{i}               & $2/2$   & $-0.26$  & $-0.23$  & $\pm 0.15$ \\
                        \ion{Mg}{i}               & $3/3$   & $+0.04$  & $+0.33$  & $\pm 0.12$ \\
                        \ion{Al}{i}               & $1/1$   & $+1.00$  & $+0.65$  & $\pm 0.29$ \\
                        \ion{Si}{i}               & $1/1$   & $-0.09$  & $-0.57$  & $\pm 0.30$ \\
                        \ion{Ca}{i}               & $6/6$   & $+0.23$  & $+0.22$  & $\pm 0.07$ \\
                        \ion{Sc}{ii}              & $-$     & $-$      & $-0.26$  & $\pm 0.21$ \\
                        \ion{Ti}{i}               & $4/4$   & $+0.56$  & $+0.36$  & $\pm 0.11$ \\
                        \ion{Ti}{ii}              & $15/15$ & $+0.08$  & $+0.08$  & $\pm 0.15$ \\
                        \ion{V}{i}                & $-$     & $-$      & $<-0.28$ & $-$        \\
                        \ion{Cr}{i}               & $4/4$   & $+0.55$  & $+0.14$  & $\pm 0.16$ \\
                        \ion{Mn}{i}               & $2/2$   & $+0.17$  & $-1.03$  & $\pm 0.18$ \\
                        \ion{Co}{i}               & $4/4$   & $+0.79$  & $+0.63$  & $\pm 0.18$ \\
                        \ion{Ni}{i}               & $-$     & $+0.19$  & $-0.44$  & $\pm 0.14$ \\
                        \ion{Cu}{i}               & $-$     & $-$      & $<+0.22$ & $-$        \\
                        \ion{Zn}{i}               & $-$     & $-$      & $<+0.32$ & $-$        \\
                        \ion{Sr}{ii}              & $2/2$   & $+0.31$  & $-1.60$  & $\pm 0.22$ \\
                        \ion{Ba}{ii}              & $1/1$   & $+0.21$  & $-1.25$  & $\pm 0.17$ \\
                        \ion{Eu}{ii}              & $-$     & $-$      & $<+0.32$ & $-$        \\
                        \bottomrule
                    \end{tabular}
                    \tablefoot{In the case of \ion{Fe}{i} and \ion{Fe}{ii}, $[\text{Fe}/\text{H}]$ and $\Delta [\text{Fe}/\text{H}]$ are shown.}
                \end{table} 
            
                The photometric surface gravity, $\log{g}$, was determined through the standard relation:
                \begin{equation*}
                \label{eq:logg}
                    \log{g} = \log{g_{\odot}} + \log{\left(\dfrac{M_{\star}}{M_{\odot}}\right)} + 4 \log{\left(\dfrac{T_{\textsl{eff}}}{T_{\textsl{eff},\odot}}\right)} + 0.4 \left(M_{\textsl{bol},\star} - M_{\textsl{bol},\odot}\right)\,\text{,}
                \end{equation*}
                based on the previously obtained $T_{\textsl{eff}}$, as well as the magnitude $\text{G}_{\text{0}}$ and the assumed stellar mass ($M_{\star} = 0.8 \pm 0.2 \, \text{M}_{\odot}$).
                The bolometric absolute magnitude ($M_{\textsl{bol},\star}$) of DR20080 was derived through the equation:
                \begin{equation*}
                    M_{\textsl{bol},\star} = \text{G}_{\text{0}} - (m - M)_0 + BC\,\text{,}
                \end{equation*}
                where we used the distance modulus of the Sculptor dSph galaxy $(m - M)_0 = 19.67 \pm 0.16 \, \text{mag}$; \citep{gaia_collaboration_gaia_2018} and the bolometric correction, $BC$, calculated as indicated in the \textit{Gaia} data release documentation\footnote{\url{https://gea.esac.esa.int/archive/documentation/GDR2/Data_analysis/chap_cu8par/sec_cu8par_process/ssec_cu8par_process_flame.html}}. 
                We adopted the solar parameters $\log{g_{\odot}} = 4.44$, $T_{\textsl{eff},\odot} = 5772 \,\text{K}$ and $M_{\textsl{bol},\odot} = 4.74 \,\text{mag}$.
                The final value of DR20080 surface gravity results to be $\log{g} = 1.27 \pm 0.27$.

            \subsubsection{Microturbulence velocity}
            \label{subsec:Microturbulence_velocity}
                For the photometric microturbulence velocity, $v_{\textsl{turb}}$, the empirical relation presented by \citet{anthony-twarog_lithium-rich_2013} was used:
                \begin{equation*}
                \label{eq:v_turb}
                    v_{\textsl{turb}} = 2 - 0.2\,\log{g}\,\text{.}
                \end{equation*}
                The scatter around this relationship is $\Delta v_{\textsl{turb,rms}} = 0.17 \, \text{km s}^{-1}$, as reported in \citet{carretta_iron_2004}, so the total uncertainty was estimated considering the two independent sources of error $\Delta v_{\textsl{turb,rms}}$ and $\Delta \log{g}$.
                The final value of the microturbulence velocity was found to be $v_{\textsl{turb}} = 1.75 \pm 0.18 \, \text{km s}^{-1}$.

%-----------------------------------------------------------------
    \section{Chemical abundances}
    \label{sec:Chemical_abundances}

        The analysis was carried out thanks to a combination of $EW$ and spectral synthesis methods, giving as a result the abundances in 1D LTE approximation.
        Finally, state-of-the-art NLTE corrections were applied to all elements for which such corrections were available in the literature (see \Autoref{subsec:Tools} and \Autoref{subsec:NLTE_corrections}).
        
        To select which of the detected lines to use for the abundance derivation with $EWs$, we impose two main selection criteria:
        1)~$20\,\text{m\AA} \leq EW \leq 100\,\text{m\AA}$;
        2)~all those lines that presented an error $\frac{\Delta EW}{EW} > 25\%$ were rejected, where $\Delta EW$ is the uncertainty on the individual $EW$ derivation estimated by \textit{DAOSPEC} (see \citealt{stetson_daospec_2008}).
        However, whenever a line presented a blend, or if a specific element presented only a few detectable lines, the analysis was carried out through spectral synthesis.
        For a complete list of the lines that were used and the method adopted to analyse each of them ($EWs$ or spectral synthesis) see \autoref{tab:Linelist} in \autoref{App:Lines}.

        The abundance values are presented in \autoref{tab:LTE_abundances}, together with their associated uncertainties and the number of lines used to derive the abundance of each ion.
        All the $[\text{X}/\text{Fe}]$ values are referred to the solar abundances from \cite{asplund_chemical_2009}.
        For four of the elements, only the upper limits are available. 
        This applied to all those elements whose absorption lines were too small to be directly detected but lay within the range covered by the three spectral bands.
        In this case, \autoref{tab:LTE_abundances} displays the number of lines used to estimate the upper limit. 
        In \autoref{tab:NLTE_abundances} are reported the mean NLTE corrections ($\delta [\text{X}/\text{H}]_{\textsl{NLTE}}$) applied to obtain the final $[\text{X}/\text{Fe}]_{\textsl{NLTE}}$ values.
        When the available NLTE corrections in the literature do not cover all detected lines of a given element, we adopted the mean value of the available corrections.
        All the abundances $[\text{X}/\text{Fe}]_{\textsl{NLTE}}$ assume the NLTE metallicity $[\text{Fe}/\text{H}]_{\textsl{NLTE}}$.

        The structure of the next Subsections is as follows.
        The LTE abundance analysis is discussed in \cref{subsec:Metallicity_[Fe/H],subsec:CNO,subsec:alpha-elements,subsec:Odd_elements,subsec:Iron-peak_elements,subsec:n-capture_elements}.
        The NLTE corrections are described in \cref{subsec:NLTE_corrections}.
        Finally, in \cref{subsec:Uncertainties} we present the method used to derive the uncertainties.

        \subsection{Metallicity \texorpdfstring{$[\text{Fe}/\text{H}]$}{[Fe/H]}}
        \label{subsec:Metallicity_[Fe/H]}

            In the DR20080 spectra, it was possible to identify 181 \ion{Fe}{i} and 8 \ion{Fe}{ii} lines. 
            In this case, we imposed stricter selection criteria to maximise the accuracy and the precision of the abundance derivation.
            First of all, all the lines with excitation potential $\chi_{\textsl{exc}} < 1.4 \, \text{eV}$ were rejected, due to the increase in the spread and the uncertainty of measurements at low $\chi_{\textsl{exc}}$ values.
            Moreover, a $\sigma$ clipping was imposed through an iterative process, with a rejection of all those lines that were situated at more than $2.5\,\sigma$ from the $[\text{Fe}/\text{H}]$ mean value.
            Finally, we only included Fe lines at $\lambda > 4000 \, \text{\AA}$ to avoid the lowest $S/N$ region in the spectra.

            Thanks to this selection process, of the original 181 detected \ion{Fe}{i} lines, 61 (all analysed with $EW$ method) were used to derive the final abundance $[\text{Fe}/\text{H}] = -3.26 \pm 0.12$, as also reported in \autoref{tab:LTE_abundances}. 

            In the \ion{Fe}{ii} case instead, the derived LTE metallicity resulted to be $[\text{Fe}/\text{H}] = -3.25 \pm 0.13$. 
            Even if \ion{Fe}{i} and \ion{Fe}{ii} values are fully consistent within the uncertainties, \ion{Fe}{i} was assumed as the reference Fe abundance due to its best statistics.
            Nevertheless, both iron abundances confirm the extremely metal-poor nature of the star DR20080.

            Finally, our derived abundances are consistent with the metallicity determinations previously reported in the literature for this star: $[\text{Fe}/\text{H}] = -3.33 \pm 0.09$, in \cite{tolstoy_3d_2023}, and $[\text{Fe}/\text{H}] = -3.00 \pm 0.14$, in \cite{skuladottir_tracing_2024}.

        \subsection{Carbon - nitrogen - oxygen}
        \label{subsec:CNO}

            \begin{figure*}[p]
                \centering
                \begin{subfigure}[t]{0.87\textwidth}
                    \centering
                    \includegraphics[width=\textwidth]{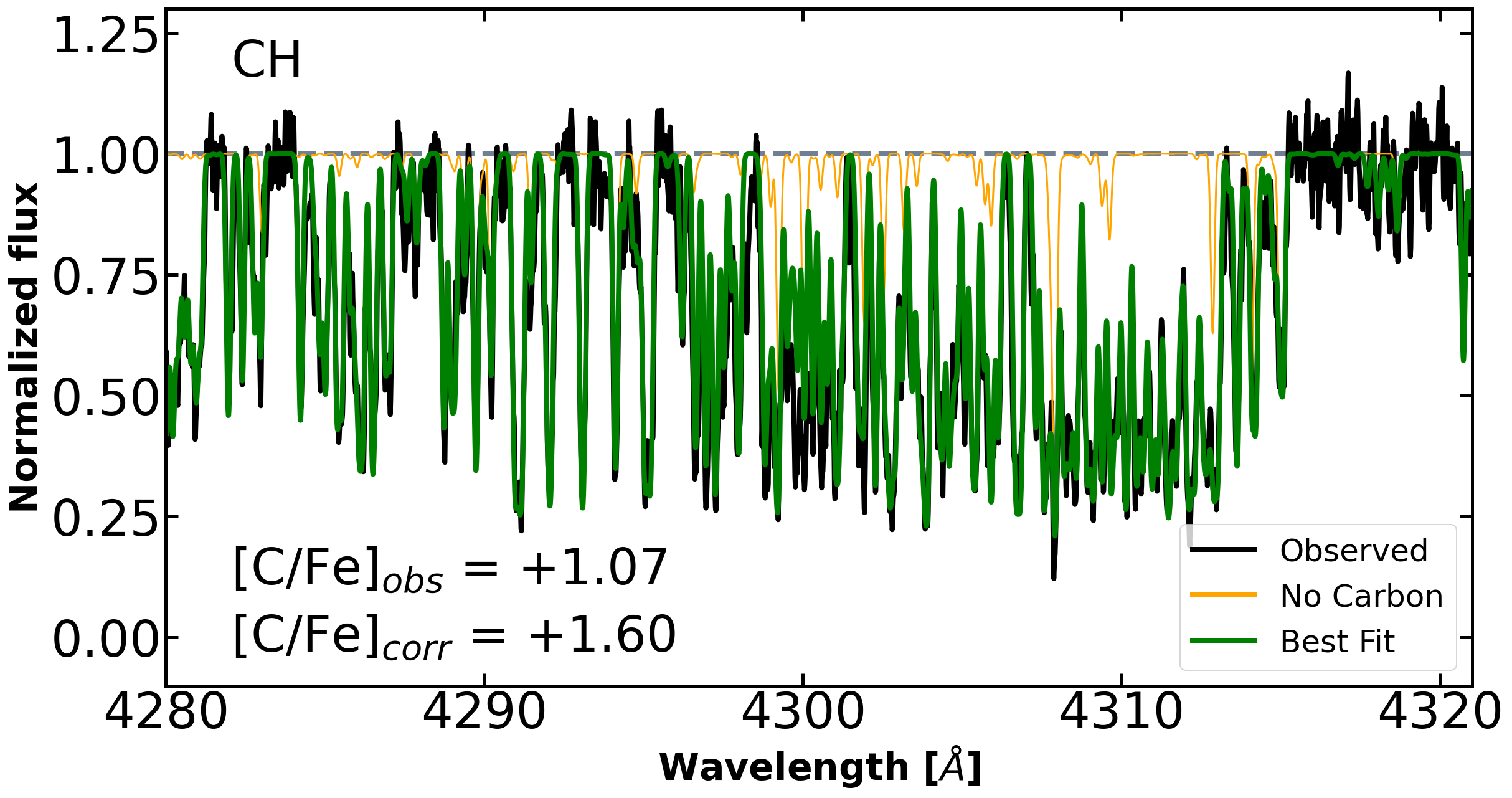}
                    \caption{The CH molecular band ($\sim 4300\,\text{\AA}$) $-$ B-band. Two abundances reported: the observed one and when corrected for mixing.}
                    \label{subfig:CH_4300}
                \end{subfigure}
                \\
                \vspace{0.5cm}
                \begin{subfigure}[t]{0.42\textwidth}
                    \centering
                    \includegraphics[width=\textwidth]{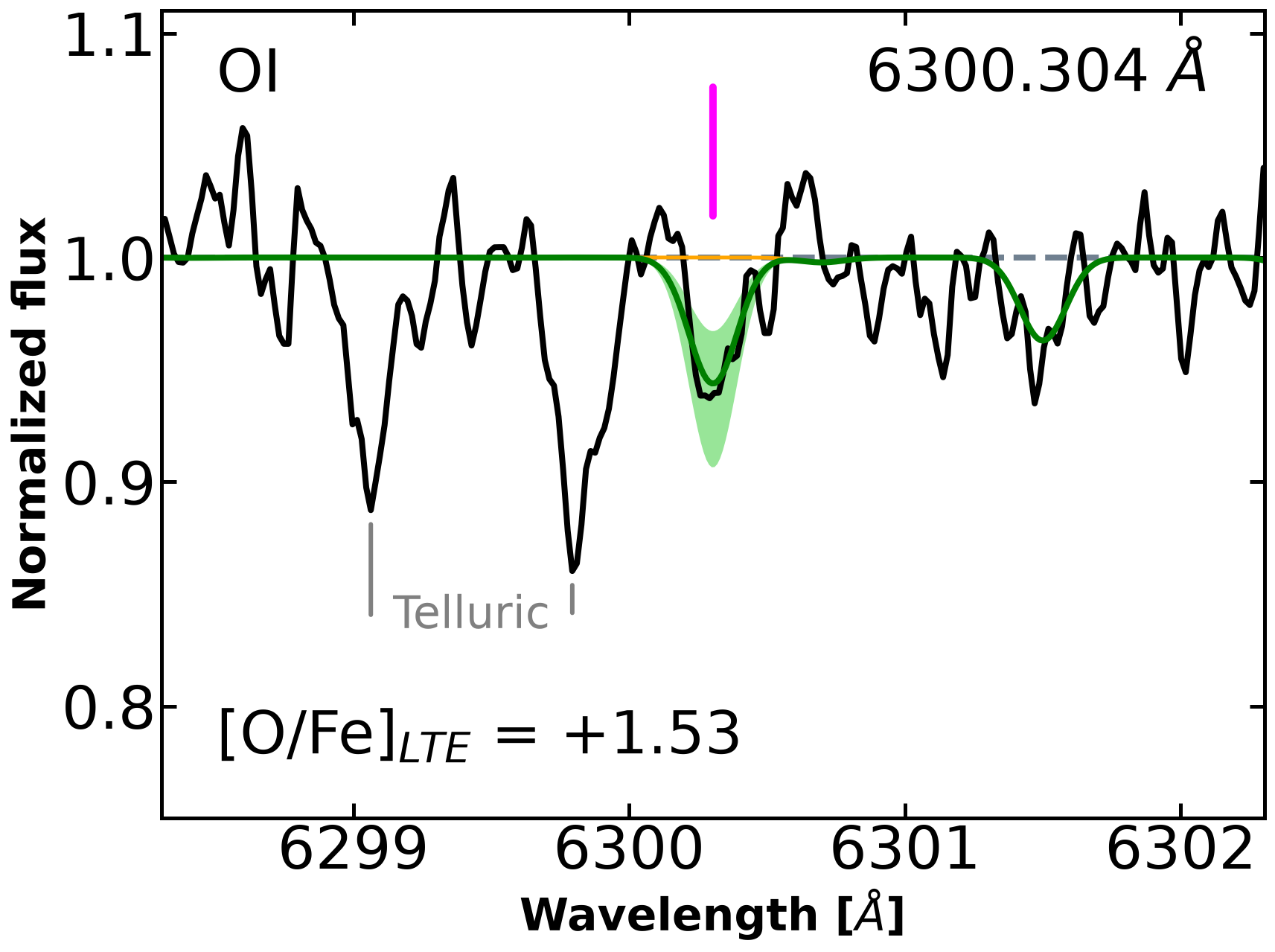}
                    \caption{\ion{O}{i} $(6300.304\,\text{\AA})$ $-$ Ru-band.}
                    \label{subfig:O_6300}
                \end{subfigure}
                \quad
                \begin{subfigure}[t]{0.42\textwidth}
                    \centering
                    \includegraphics[width=\textwidth]{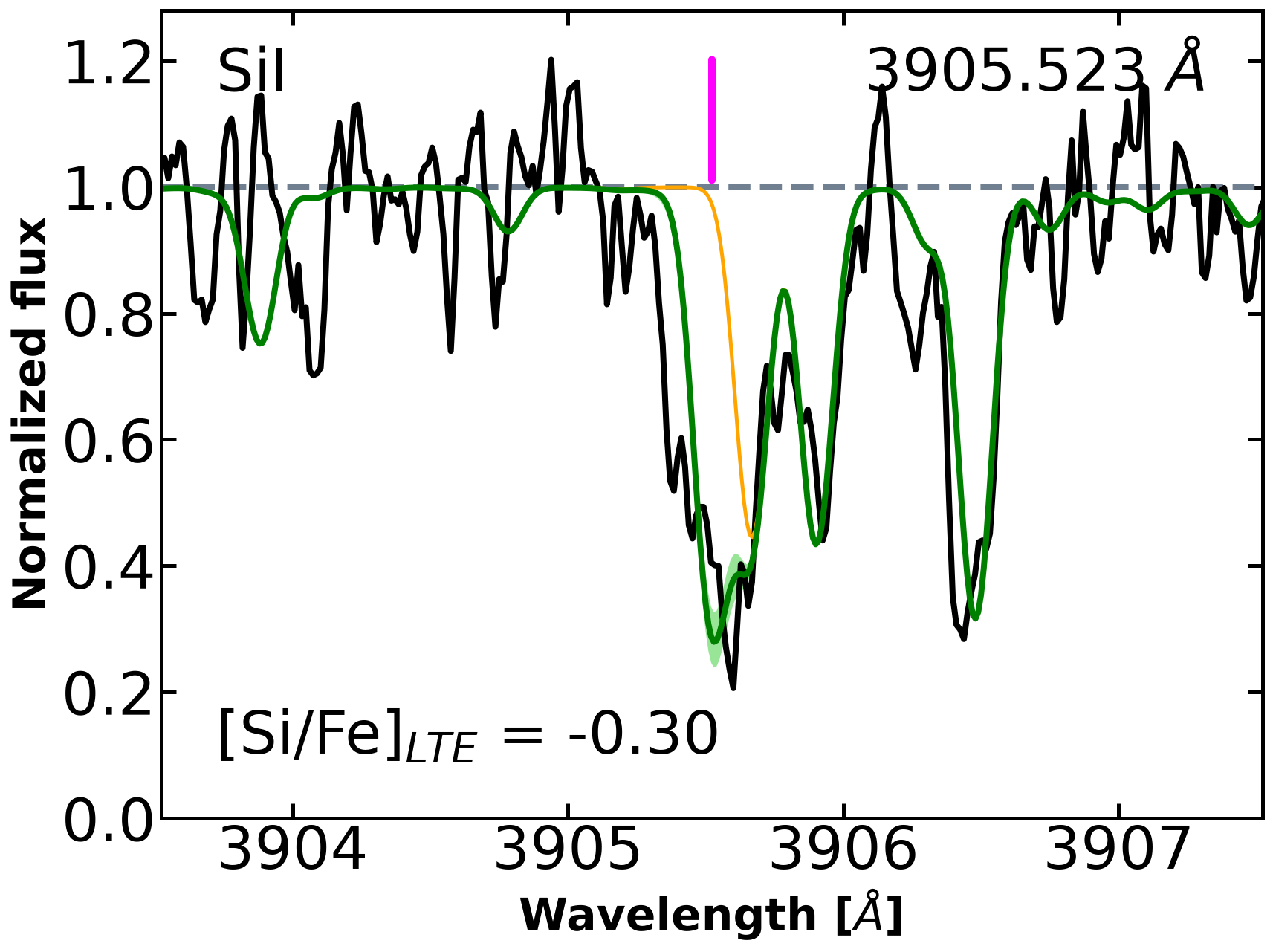}
                    \caption{\ion{Si}{i} $(3905.523\,\text{\AA})$ $-$ B-band.}
                    \label{subfig:Si_3905}
                \end{subfigure}
                \\
                \vspace{0.5cm}
                \begin{subfigure}[t]{0.42\textwidth}
                    \centering
                    \includegraphics[width=\textwidth]{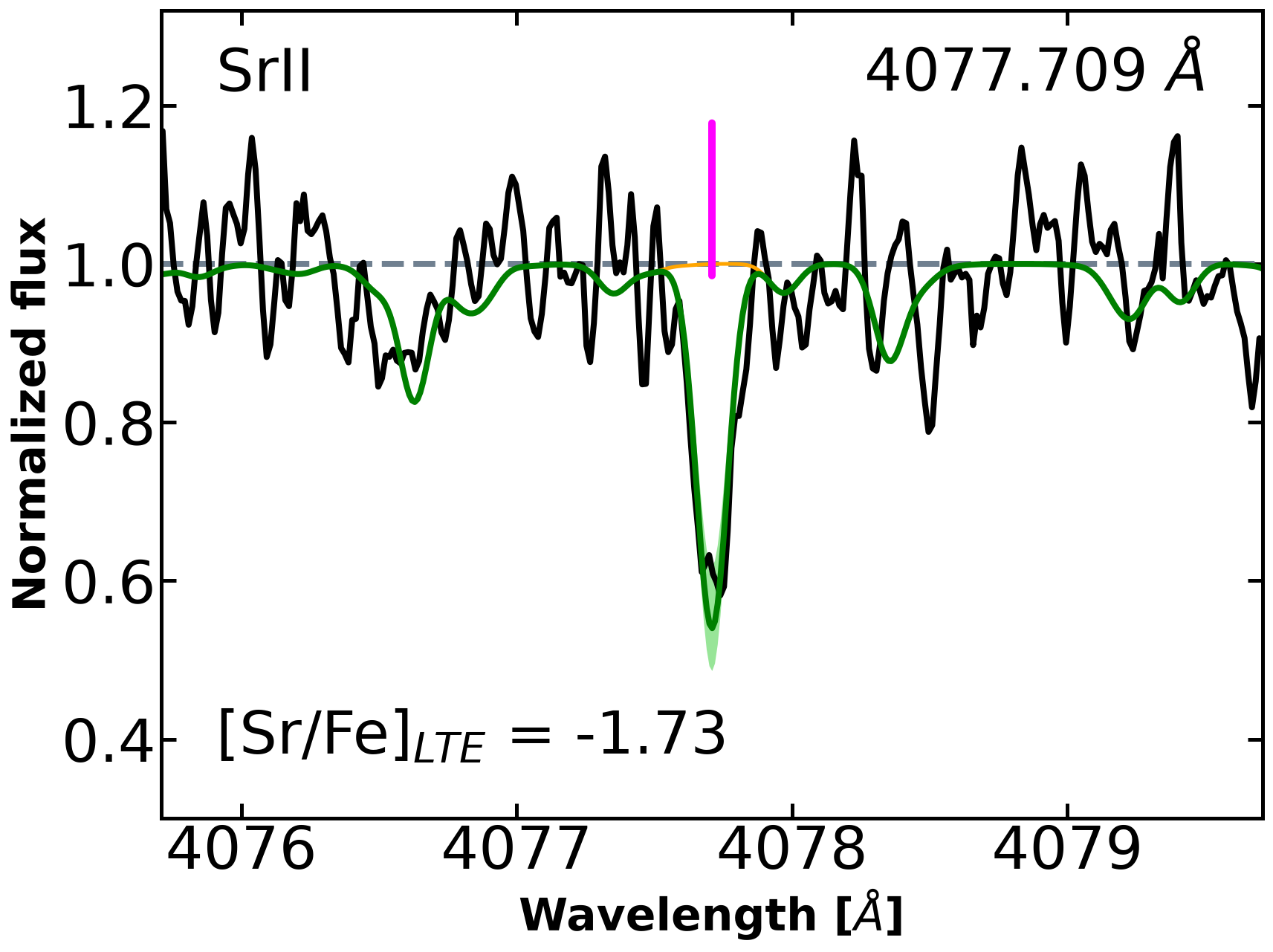}
                    \caption{\ion{Sr}{ii} $(4077.709\,\text{\AA})$ $-$ B-band.}
                    \label{subfig:Sr_4077}
                \end{subfigure}
                \quad
                \begin{subfigure}[t]{0.42\textwidth}
                    \centering
                    \includegraphics[width=\textwidth]{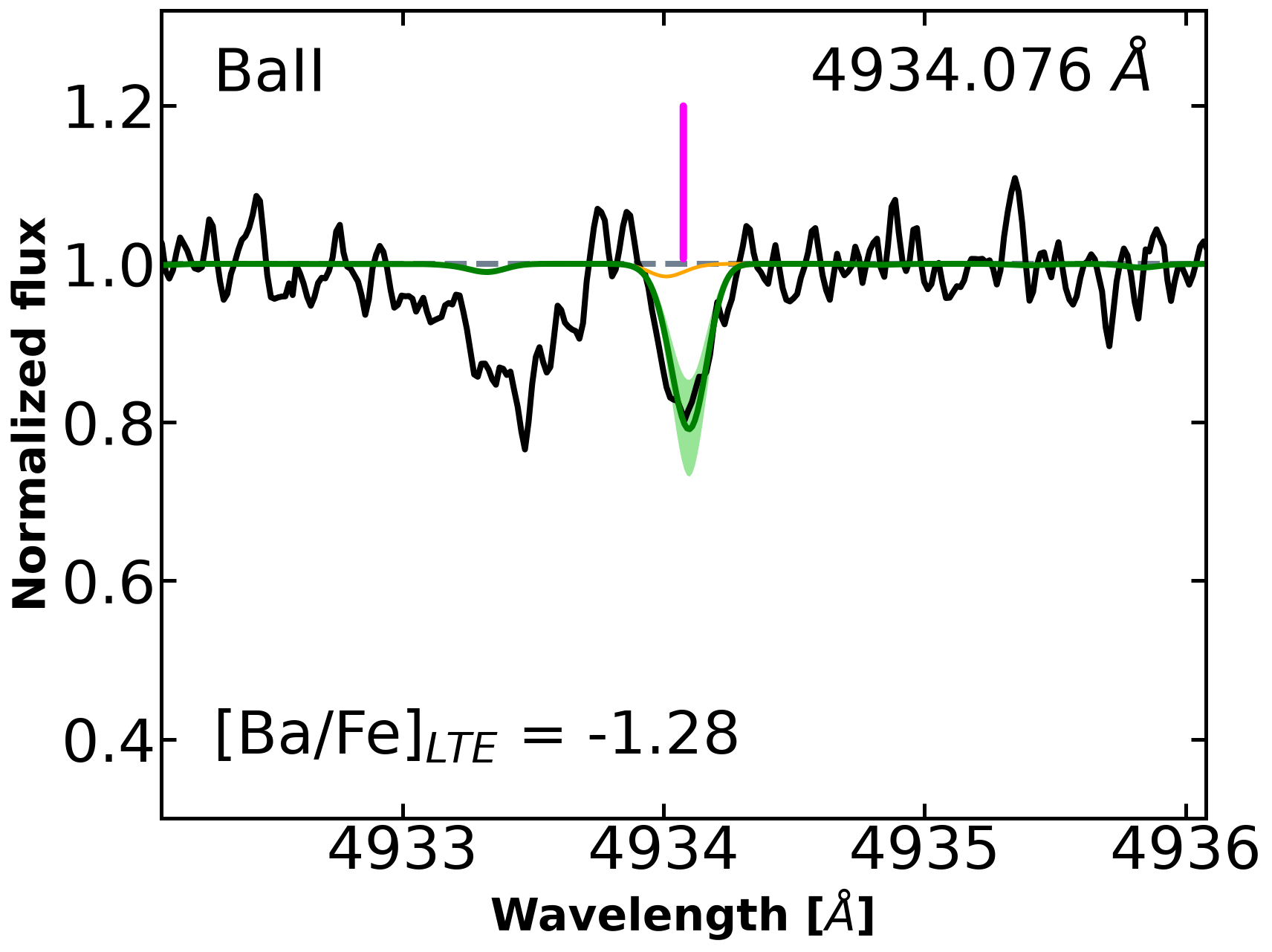}
                    \caption{\ion{Ba}{ii} $(4934.076\,\text{\AA})$ $-$ Rl-band.}
                    \label{subfig:Ba_4934}
                \end{subfigure}
                \\
                \caption{Examples of key spectral lines. 
                \textit{Black solid line} is the observed normalised spectrum, \textit{green} is the best fit, and \textit{orange} is a reference synthetic spectrum in the absence of element X.
                In (\subref{subfig:O_6300}), (\subref{subfig:Si_3905}), (\subref{subfig:Sr_4077}) and (\subref{subfig:Ba_4934}) the \textit{shaded green area} indicates the uncertainty on the abundance measurement, and the \textit{magenta segment} indicates the central wavelength of the transition.
                }
                \label{fig:spectral_lines}
            \end{figure*}

            The only detected \ion{O}{i} line is at $6300.304\,\text{\AA}$ (\autoref{subfig:O_6300}) and yields a clearly super-solar value of $[\text{O}/\text{Fe}] = +1.53 \pm 0.25$.
            This value is consistent with the upper limit $[\text{O}/\text{Fe}] < +2.10$, evaluated by using the $6363.776\,\text{\AA}$ line.
            To further test this result, we also analysed the available low-$S/N$ VLT/X-SHOOTER spectra \citep{skuladottir_tracing_2024} in the near-infrared (NIR) band ($9940 - 21010\,\text{\AA}$), searching for CO molecular lines.
            From these spectra, we derived an upper limit of $[\text{O}/\text{Fe}] < +3$ using CO lines around $16000\,\text{\AA}$ commonly employed in APOGEE \citep[e.g.,][]{smith_apogee_2021}.
            The two absorption features observed at shorter wavelength in \autoref{subfig:O_6300} appear consistent with telluric lines, known to be in that part of the spectrum; however, there is no evident sign of blending with the O~line.

            The carbon abundance was derived through the CH molecular band (\autoref{subfig:CH_4300}) around $4300\,\text{\AA}$. 
            The abundance derived from this band, however, shows a slight dependence on the oxygen abundance, due to the fact that carbon, in addition to being present in CH, can also be partially distributed into CO (and CN, though it has less of an impact on the abundance derived from CH). 
            For this reason, $[\text{C}/\text{Fe}]$ was derived upon completion of a second run of the abundance derivation, started only after oxygen abundance was derived and fixed at the value of $[\text{O}/\text{Fe}] = +1.53$. 
            The resulting value was $[\text{C}/\text{Fe}]_{\textsl{obs}} = +1.07 \pm 0.21$.
            However, this value represents the observed carbon abundance, which does not consider that the stellar atmosphere of an RGB star like DR20080 is affected by internal mixing.
            To take this process into account, we exploited the online tool\footnote{\url{https://vplacco.pythonanywhere.com/}} based on \cite{placco_carbon-enhanced_2014}, which generated the correction $\delta [\text{C}/\text{Fe}] = +0.55$, resulting in a final carbon LTE abundance of $[\text{C}/\text{Fe}]_{\textsl{corr}} = +1.60 \pm 0.21$. 
            This value is higher than what was found in \cite{skuladottir_tracing_2024} ($[\text{C}/\text{Fe}]_{\textsl{corr}} = +1.02 \pm 0.17$), confirming that DR20080 is a CEMP star.
            Thanks to the high spectral resolution of our observations, we could also estimate the isotopic ratio $\log{\left(\frac{^{12}\text{C}}{^{13}\text{C}}\right)} = 0.9 \pm 0.2$ (see also \autoref{fig:13CH_4213} in \autoref{App:Lines}).
            This result is consistent to what was found in \cite{gratton_mixing_2000} for stars with similar metallicity to that of DR20080. 
            However, it is important to consider that the surface $^{12}\text{C}/^{13}\text{C}$ ratio of DR20080 has been affected by internal mixing during its evolution and therefore does not directly reflect its initial isotopic composition.

            The nitrogen abundance was derived through the CN molecular band around $3883\,\text{\AA}$ (see \autoref{fig:CN_3883} in \autoref{App:Lines}). 
            Due to the distribution of carbon among the CO, CH, and CN molecules described before, CN band was strongly affected by carbon abundance, so it was necessary to fix it to the observed value $[\text{C}/\text{Fe}] = +1.07$ before running the abundance derivation on the CN band.
            The LTE value of the nitrogen chemical abundance was found to be $[\text{N}/\text{Fe}] = +1.72 \pm 0.26$.
            It has to be considered that, due to the before-mentioned internal mixing in RGB stars, the N surface abundance is expected to be enhanced with respect to its original value \citep{gratton_mixing_2000,spite_first_2005,placco_carbon-enhanced_2014}.
            However, in the case of nitrogen, there is no tool available to correct for this effect.            

            All the following abundance values were derived after imposing a fixed value of $[\text{O}/\text{Fe}] = +1.53$, $[\text{C}/\text{Fe}] = +1.07$ and $[\text{N}/\text{Fe}] = +1.72$.
            This was done to count for possible blends of molecular lines (CH, CN) in the spectra.

        \subsection{\texorpdfstring{\textalpha-elements}{alpha-elements}}
        \label{subsec:alpha-elements}

            From the 9 detected \ion{Mg}{i} lines, only 3 were kept to derive the abundance value of magnesium (one of them is plotted in \autoref{fig:Mg_5183} in \autoref{App:Lines}).
            The triplet in B-band ($3829.355\,\text{\AA}$, $3832.304\,\text{\AA}$ and $3838.294\,\text{\AA}$) was detected, however it was located in a bad $S/N$ region, while three other detected lines were heavily affected by blends, thus we decide to exclude them from the analysis.
            The resulting Mg abundance is $[\text{Mg}/\text{Fe}] = +0.47 \pm 0.12$.
    
            \ion{Si}{i} lines presented some challenges: the line at $3905.523\,\text{\AA}$ shows clear blends with CH lines, as can be seen in \autoref{subfig:Si_3905}, while the redder line at $4102.936\,\text{\AA}$ is strongly affected by the wings of H$\delta$.
            Considering that we were able to identify and fit the CH lines (see the yellow spectrum in \autoref{subfig:Si_3905}) that are blended with the first silicon line, we decided to assume the abundance value derived from this line, $[\text{Si}/\text{Fe}] = -0.30 \pm 0.30$, as the final LTE \ion{Si}{i} abundance of DR20080.
            An upper limit of $[\text{Si}/\text{Fe}] < +0.30$ was estimated from the second \ion{Si}{i} line, which still indicates a low Si value, since the MW mean value is $[\text{Si}/\text{Fe}] \simeq +0.40$ at this metallicity \citep{cayrel_first_2004}. 
            It is worth noticing that sub-solar $[\text{Si}/\text{Fe}]$ at these metallicities are extremely rare, especially in the case of CEMP-no stars \citep{cayrel_first_2004,norris_most_2013}.

            In total, 10 calcium lines were identified in the observed spectra. 
            Among these, 5 lines either met the selection criteria and/or were correctly synthesised showing no major blend, and were therefore used in the \ion{Ca}{i} abundance derivation.
            Thus, the final estimate for the Ca abundance is $[\text{Ca}/\text{Fe}] = +0.17 \pm 0.07$.

            Finally, titanium is the only element, alongside iron, for which lines from two different ionisation states have been identified.
            Specifically, 4 \ion{Ti}{i} and 15 \ion{Ti}{ii} lines contributed to the final abundances, $[\text{Ti}/\text{Fe}] = -0.02 \pm 0.11$ for \ion{Ti}{i} and $[\text{Ti}/\text{Fe}] = +0.18 \pm 0.15$ for \ion{Ti}{ii}.
            As for the case of \ion{Fe}{i} and \ion{Fe}{ii}, \ion{Ti}{i} and \ion{Ti}{ii} abundances are fully consistent within the uncertainties. 
            We chose \ion{Ti}{ii} as the reference indicator for titanium chemical abundance due to its better statistics and because, as highlighted by \cite{bergemann_ionization_2011}, \cite{sitnova_influence_2020} and \cite{mallinson_titanium_2022}, \ion{Ti}{ii} lines in metal-poor stars are less affected by NLTE effects and their corresponding corrections are more reliable.

        \subsection{Odd elements}
        \label{subsec:Odd_elements}

            Only 2 sodium lines were detected in the spectra of DR20080, both in the Ru-band.
            The resulting Na abundance is $[\text{Na}/\text{Fe}] = +0.21 \pm 0.15$.
            The only aluminium line that was considered in this analysis was the one at $3961.520\,\text{\AA}$.
            This yields an abundance of $[\text{Al}/\text{Fe}] = -0.17 \pm 0.29$.
            Even if also $3944.006\,\text{\AA}$ was detected, this line resulted to be heavily blended and, consequently, it was excluded.
            From the 3 scandium lines detected in the B-band spectrum of DR20080, 2 of them were kept in the final analysis.
            Thus, we derive a Sc abundance of $[\text{Sc}/\text{Fe}] = -0.08 \pm 0.21$.
            No vanadium lines were strong enough to be detected in the spectra of DR20080. 
            For this reason, we extracted the upper limit $[\text{V}/\text{Fe}] < -0.30$ from the \ion{V}{i} line $4379.230\,\text{\AA}$. 
        
        \subsection{Iron-peak elements}
        \label{subsec:Iron-peak_elements}

            From the 8 identified lines, 3 were considered to derive the final LTE \ion{Cr}{i} abundance, yielding an abundance of $[\text{Cr}/\text{Fe}] = - \pm 0.07$.
            Regarding manganese, the chemical abundance was derived using the two bluer lines among the $4030 - 4034 \,\text{\AA}$ triplet.
            The reddest line of the triplet was rejected due to heavy blending.
            Thus, the final estimate for the Mn abundance is $[\text{Mn}/\text{Fe}] = -1.02 \pm 0.18$.
            Among the 5 detected \ion{Co}{i} lines, all in the B-band, 4 were taken to derive the final abundance $[\text{Co}/\text{Fe}] = +0.02 \pm 0.18$, also reported in \autoref{tab:LTE_abundances}.

            Despite the identification of 6 \ion{Ni}{i} lines, only the line at $5476.904\,\text{\AA}$ was used to determine an abundance of $[\text{Ni}/\text{Fe}] = -0.45 \pm 0.14$, i.e., a sub-solar value. 
            Although the other lines were less reliable, either due to strong blends or to bad $S/N$, they were also consistent with a sub-solar abundance of nickel.
            This is a very unusual result, especially at these low metallicities where $[\text{Ni}/\text{Fe}] \sim 0$ \citep[e.g.,][]{cayrel_first_2004,skuladottir_tracing_2024}.

            Finally, neither copper nor zinc lines were detected in the spectra.
            For Cu, three undetected \ion{Cu}{i} lines were analysed to estimate the upper limit of its chemical abundance, $[\text{Cu}/\text{Fe}] < +0.40$. 
            Regarding Zn, we estimated the upper limit in the same way, but from two lines, obtaining $[\text{Zn}/\text{Fe}] < +0.50$ (see \autoref{tab:Linelist}).

        \subsection{Neutron-capture elements}
        \label{subsec:n-capture_elements}

            Both the detected strontium lines were considered when deriving the \ion{Sr}{i} LTE abundance (\autoref{subfig:Sr_4077}).
            The final abundance, $[\text{Sr}/\text{Fe}] = -1.73 \pm 0.22$, shows that strontium is not enhanced.

            Barium abundance, as the representative element of the s-process, is crucial for distinguishing between CEMP-s stars and CEMP-no stars.
            The only identified \ion{Ba}{ii} line, at $4934.076\,\text{\AA}$ (see \autoref{subfig:Ba_4934}), was analysed assuming solar isotopic fractions for Ba, yielding a very sub-solar abundance of $[\text{Ba}/\text{Fe}] = -1.28 \pm 0.17$.
            This result is also consistent with the upper limit $[\text{Ba}/\text{Fe}] < -0.5$ found in \cite{skuladottir_tracing_2024}.
            Thus, we can conclude that the star DR20080 is clearly not a CEMP-s star.

            No europium line was detected in the observed spectra of DR20080.
            Two \ion{Eu}{ii} lines were analysed to estimate the upper limit, at $3907.106\,\text{\AA}$ and $4129.708\,\text{\AA}$.
            The lowest of the two upper limits, $[\text{Eu}/\text{Fe}] < +0.50$, corresponds to the line at $4129.708\,\text{\AA}$.
            This result indicates that europium is not enhanced, ruling out the hypothesis of a CEMP-r star.
            Together with the sub-solar barium abundance, this confirms the CEMP-no classification of DR20080.

        \subsection{NLTE corrections}
        \label{subsec:NLTE_corrections}

            We mainly used two online tools to retrieve NLTE corrections for our abundance derivations, \textit{MPIA} and \textit{Inasan}.
            We used the \textit{MPIA} database for the corrections of: \ion{Fe}{i} and \ion{Fe}{ii} \citep{bergemann_non-lte_2012}, \ion{Mg}{i} \citep{bergemann_non-local_2017}, \ion{Si}{i} \citep{bergemann_red_2013}, \ion{Ca}{i} \citep{mashonkina_non-lte_2007}, \ion{Ti}{i} and \ion{Ti}{ii} \citep{bergemann_ionization_2011}, \ion{Cr}{i} \citep{bergemann_chromium_2010}, \ion{Mn}{i} \citep{bergemann_nlte_2008} and \ion{Co}{i} \citep{bergemann_nlte_2010}. 
            We also checked \ion{O}{i} corrections \citep{sitnova_influence_2013} using \textit{MPIA}, but it proved negligible for our stellar parameters.            
            The \textit{Inasan} online tool was instead used to derive the corrections for \ion{Sr}{ii} \citep{mashonkina_formation_2021} and \ion{Ba}{ii} \citep{mashonkina_even--odd_2019}.
            The \ion{Na}{i} correction \citep{lind_non-lte_2011} was retrieved using the tool \textit{Inspect-Stars}\footnote{\url{https://www.inspect-stars.com/}}.
            
            The NLTE corrections are provided by the respective tool for individual lines.
            If more than one line of the analysed element was present in the database, the final NLTE correction ($\delta [\text{X}/\text{H}]_{\textsl{NLTE}}$) was obtained by calculating the arithmetic mean of the corrections for the available individual lines. 

            In the case of Al, the NLTE correction came from \cite{nordlander_non-lte_2017}.
            No grids exist for Ni, however the corrections are expected to be similar as those for Fe, thus the correction was assumed to be the same as that of \ion{Fe}{i} ($\delta [\text{Ni}/\text{H}]_{\textsl{NLTE}} = \delta [\text{Fe}/\text{H}]_{\textsl{NLTE}} = +0.19$), as done, e.g., in \cite{skuladottir_zero-metallicity_2021}.
            For carbon, according to \cite{popa_non-local_2023}, the NLTE corrections are expected to be positive, in the range of $0.1-0.2 \,\text{dex}$.
            However, this result is very uncertain, as performing NLTE calculations for a molecular band, such as the CH band, is very challenging.
            Furthermore, the 3D corrections are expected to be negative \citep{amarsi_carbon_2019-1}.
            Given the lack of accurate data, we have chosen to take a conservative approach and apply no corrections to the carbon abundance.
            
            All the mean NLTE corrections applied, as well as the number of lines with respect to the total that presented such corrections, are reported in \autoref{tab:NLTE_abundances}, together with the final 1D NLTE abundance values $[\text{X}/\text{Fe}]_{\textsl{NLTE}}$, all referred to the NLTE metallicity $[\text{Fe}/\text{H}]_{\textsl{NLTE}} = -3.08$.

        \subsection{Uncertainties}
        \label{subsec:Uncertainties}

            The derivation of the uncertainties took into account both the contributions of the stellar parameters ($\Delta [\text{X}/\text{Fe}]_{\textsl{param}}$) and the random error affecting the $EW$s or spectral fitting ($\Delta [\text{X}/\text{Fe}]_{\textsl{ran}}$). 
            The latter uncertainties were automatically generated by \textit{ASTRA} in the same way that is reported in \cite{lucchesi_extremely_2024}.

            To estimate the contribution from the stellar parameters, we ran new iterations of the abundance derivation, varying only one parameter at a time.
            Specifically, six runs were executed, with each run setting one of the parameters to one of the two extremes defined by its uncertainty \mbox{$T_{\textsl{eff}} = (4578 \, ; \, 4766) \,\text{K}$}, \mbox{$\log{g} = (1.00 \, ; \, 1.54) \, \text{dex}$}, and \mbox{$v_{\textsl{turb}} = (1.57 \, ; \, 1.93) \, \text{km s}^{-1}$}.
            Once all the abundances corresponding to the new stellar parameters were obtained, the uncertainty on mean abundances due to each individual parameter, e.g. $T_{\textsl{eff}}$, was calculated as 
            \begin{multline}
            \label{eq:param_uncert}
                \Delta [\text{X}/\text{Fe}]_{T_{\textsl{eff}}} = \\ \text{max}\left(\left|[\text{X}/\text{Fe}] - [\text{X}/\text{Fe}]_{T_{4578}}\right| \, ; \, \left|[\text{X}/\text{Fe}] - [\text{X}/\text{Fe}]_{T_{4766}}\right|\right)
            \end{multline}
            and the same for the other stellar parameters, where $[\text{X}/\text{Fe}]$ is the mean value reported in \autoref{tab:LTE_abundances} and $[\text{X}/\text{Fe}]_{\textsl{param}}$ indicates the abundance value obtained varying that specific parameter.
            The final uncertainties were thus determined by summing quadratically all the contributions, see \autoref{tab:LTE_abundances}.

            There were a few exceptions that also took into account other terms to evaluate the final uncertainty.
            One of these special cases was carbon, whose chemical abundance was calculated after fixing the oxygen abundance value, due to its effect on CO and CH molecules. 
            Thus, we ran the carbon abundance derivation varying the oxygen abundance to its extremes and then used an analogue of \autoref{eq:param_uncert} to calculate $\Delta [\text{C}/\text{Fe}]_{\text{O}}$.
            This term was then quadratically added to the overall uncertainty of carbon, but turned out to be negligible compared to the other contributions.

            If calculated as described above, the silicon abundance should have an uncertainty of $\Delta[\text{Si}/\text{Fe}] = \pm 0.21$. However, since both observed lines of this element exhibited significant and evident blends, it was deemed more conservative to assume the uncertainty of $\Delta[\text{Si}/\text{Fe}] = \pm 0.30$.

            Another exception was oxygen, for which the normal uncertainty derivation would yield a value of $\Delta[\text{O}/\text{Fe}] = \pm 0.20$. 
            Instead, a value of $\Delta[\text{O}/\text{Fe}] = \pm 0.25$ was assumed to maintain a more cautious approach, given that we only have a single line, which is also very weak and therefore sensitive to small fluctuation of the noise.

            It's worth noting that the uncertainties on solar abundances \citep{asplund_chemical_2009} proved to be negligible with respect to the other contributions considered.
            Moreover, considering that the majority of NLTE corrections in the literature are provided without uncertainties, we assumed the $\Delta [\text{X}/\text{Fe}]_{\textsl{LTE}}$ values as the final errors also in the 1D NLTE case, as reported in the last column of \autoref{tab:NLTE_abundances}.
        
%-----------------------------------------------------------------
    \section{DR20080 and the Local Group}
    \label{sec:DR20080_Local_Group}    

        \begin{figure}[t]
            \centering
            \includegraphics[width=0.95\hsize]{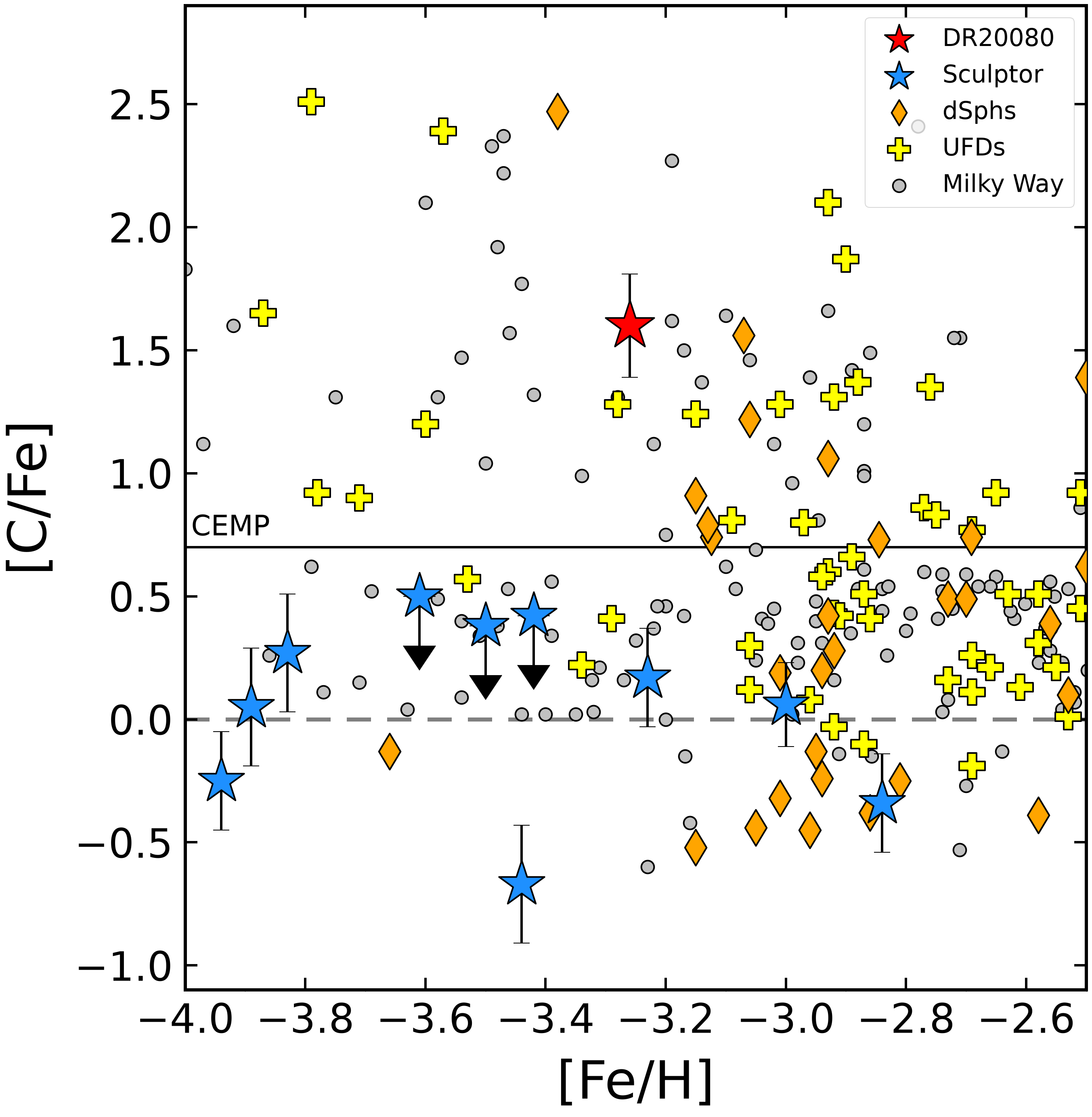}
            \caption{Comparison between $[\text{C}/\text{Fe}]$ abundance values, as a function of $[\text{Fe}/\text{H}]$, of DR20080 (displayed as a \textit{red star}) and of metal-poor stars in four different environments: Sculptor dSph (\textit{blue stars}), UFDs (\textit{yellow crosses}), dSphs (\textit{orange diamonds}), and MW halo (\textit{grey dots}); see references in \Autoref{subsec:Comparison_Data}.
            \textit{Horizontal black solid line} is the CEMP limit ($[\text{C}/\text{Fe}] \geq 0.7 $), \textit{gray dashed line} is the solar value.
            }
            \label{fig:C}
        \end{figure}   
        
        \begin{figure*}
                \centering
                \includegraphics[width=0.33\hsize]{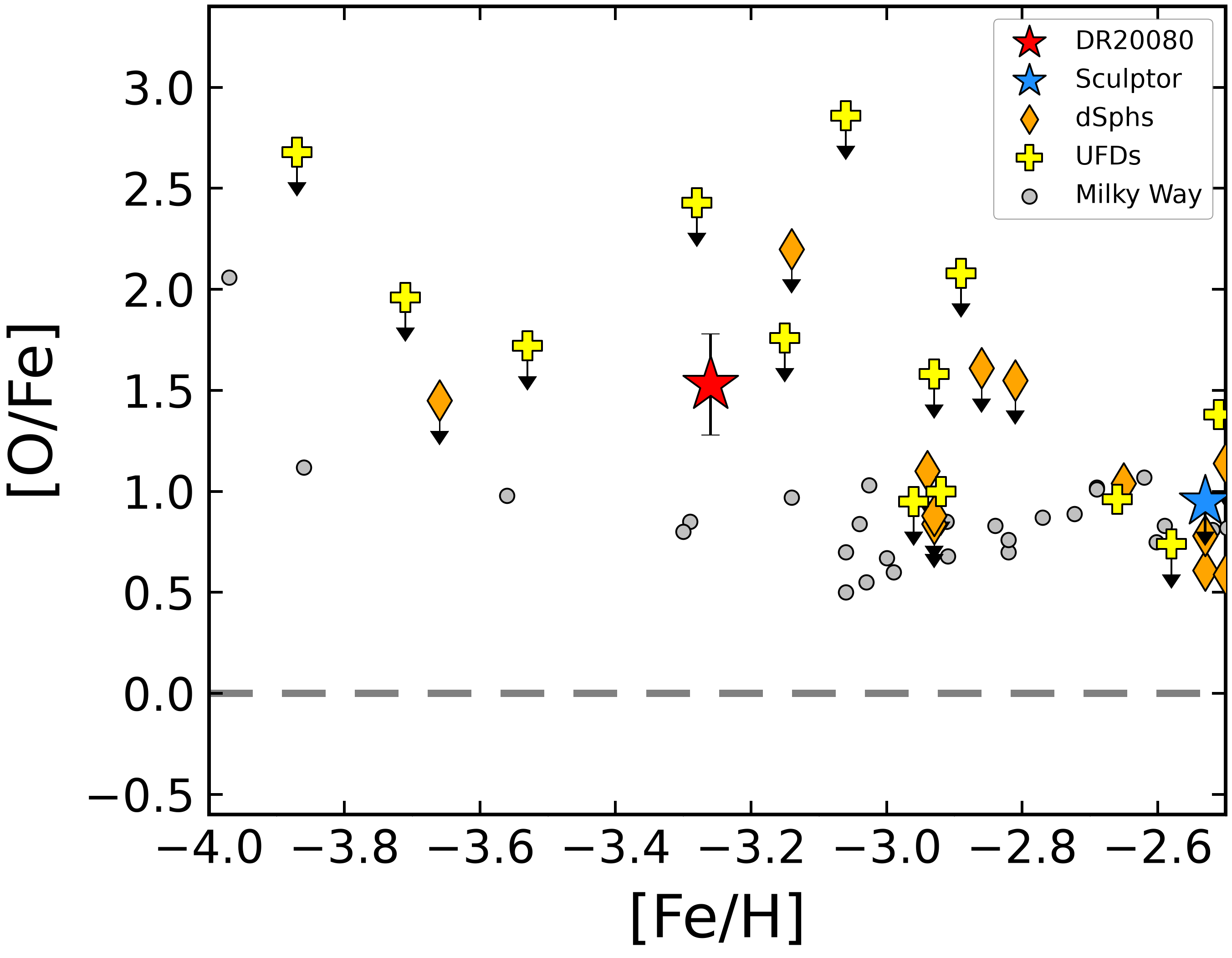}
                \includegraphics[width=0.33\hsize]{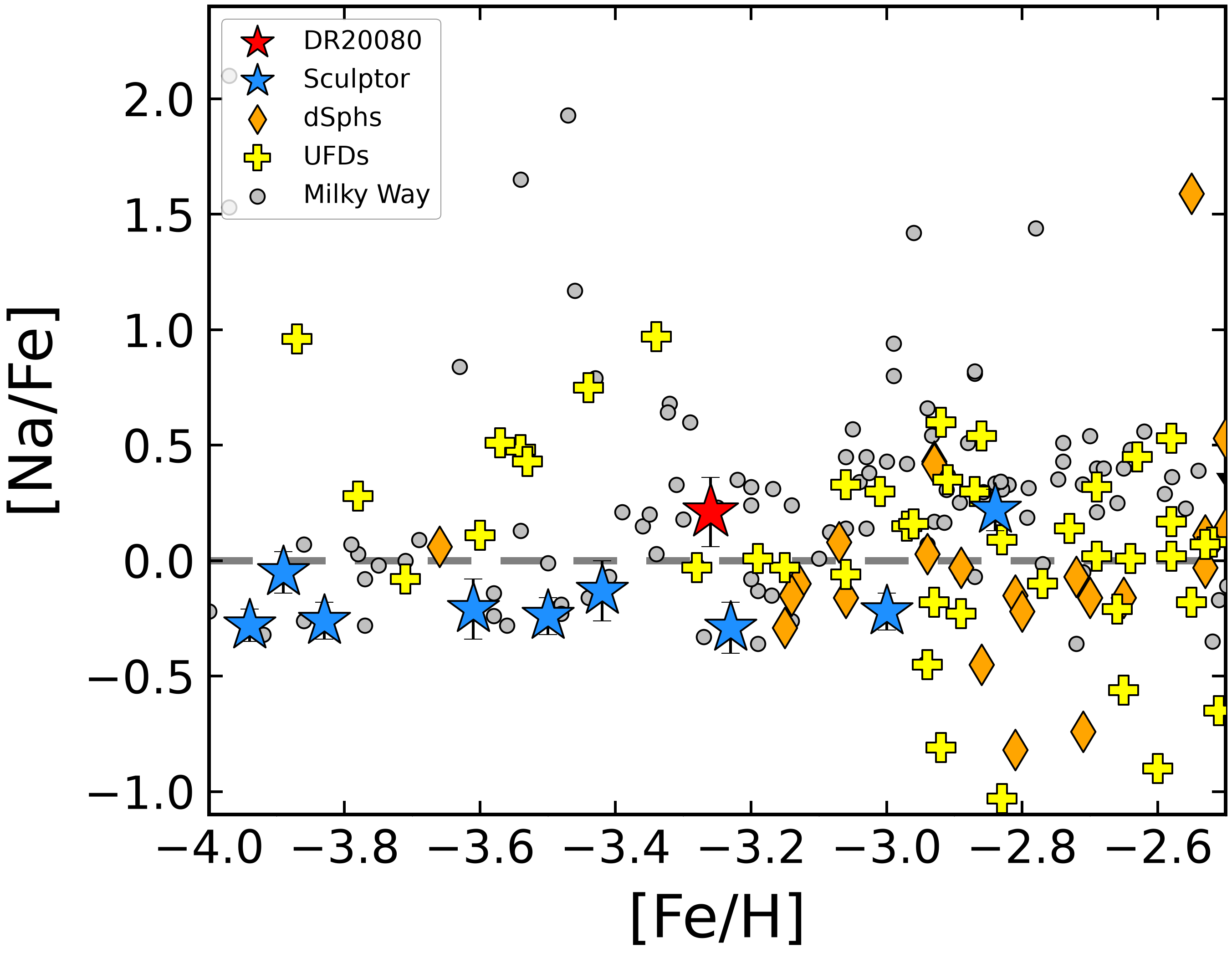}
                \includegraphics[width=0.33\hsize]{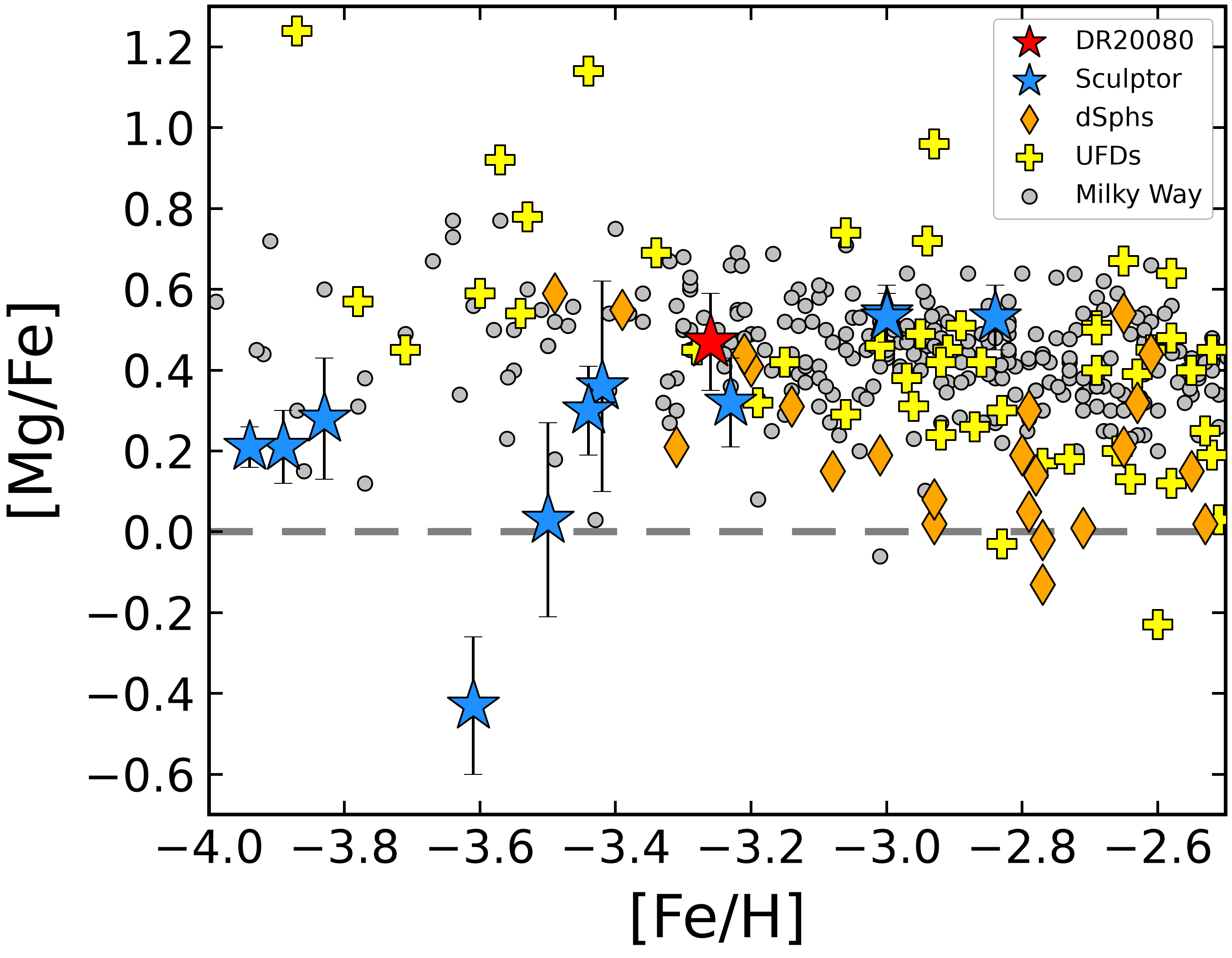}
                \\ \vspace{0.2cm}
                \includegraphics[width=0.33\hsize]{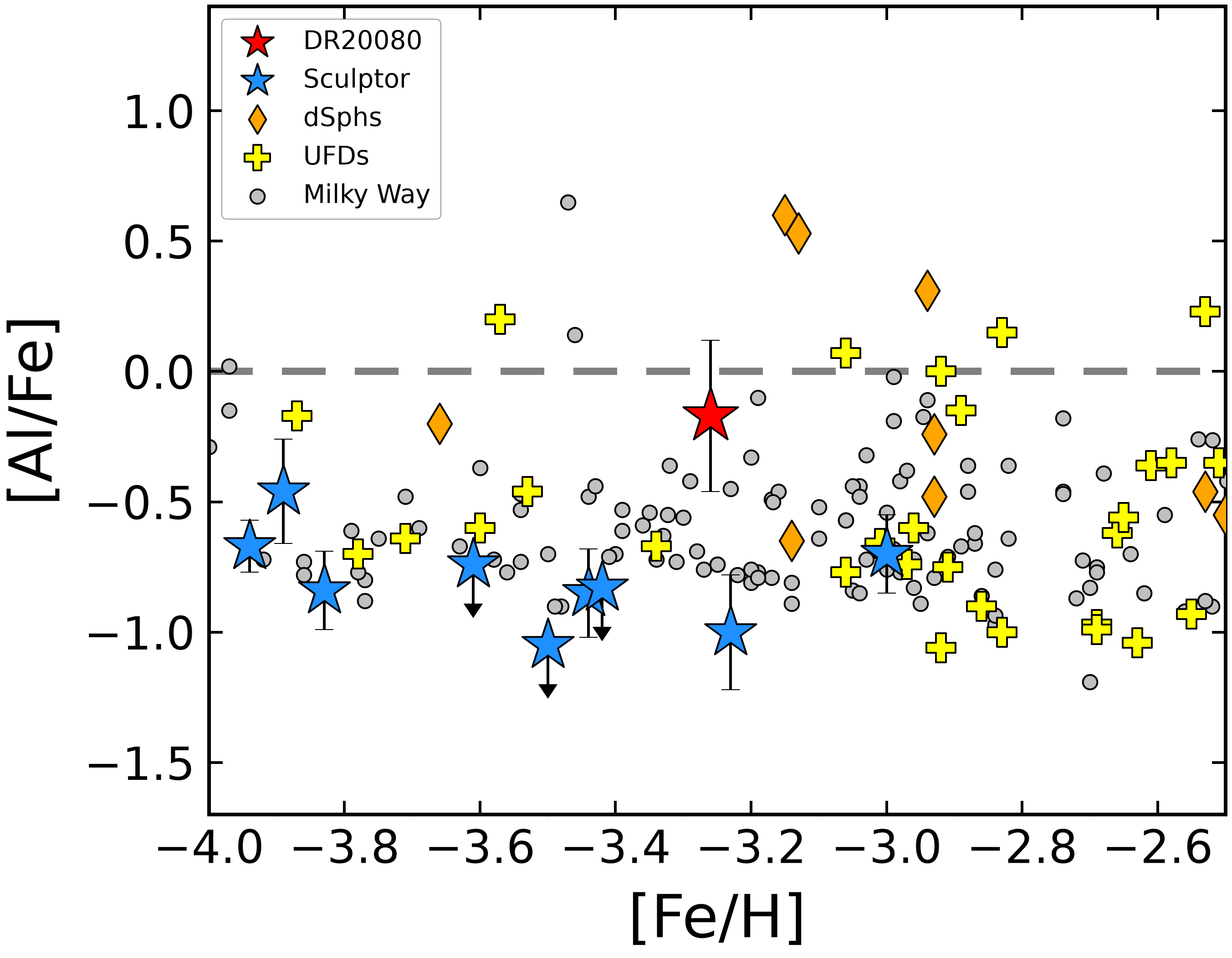}
                \includegraphics[width=0.33\hsize]{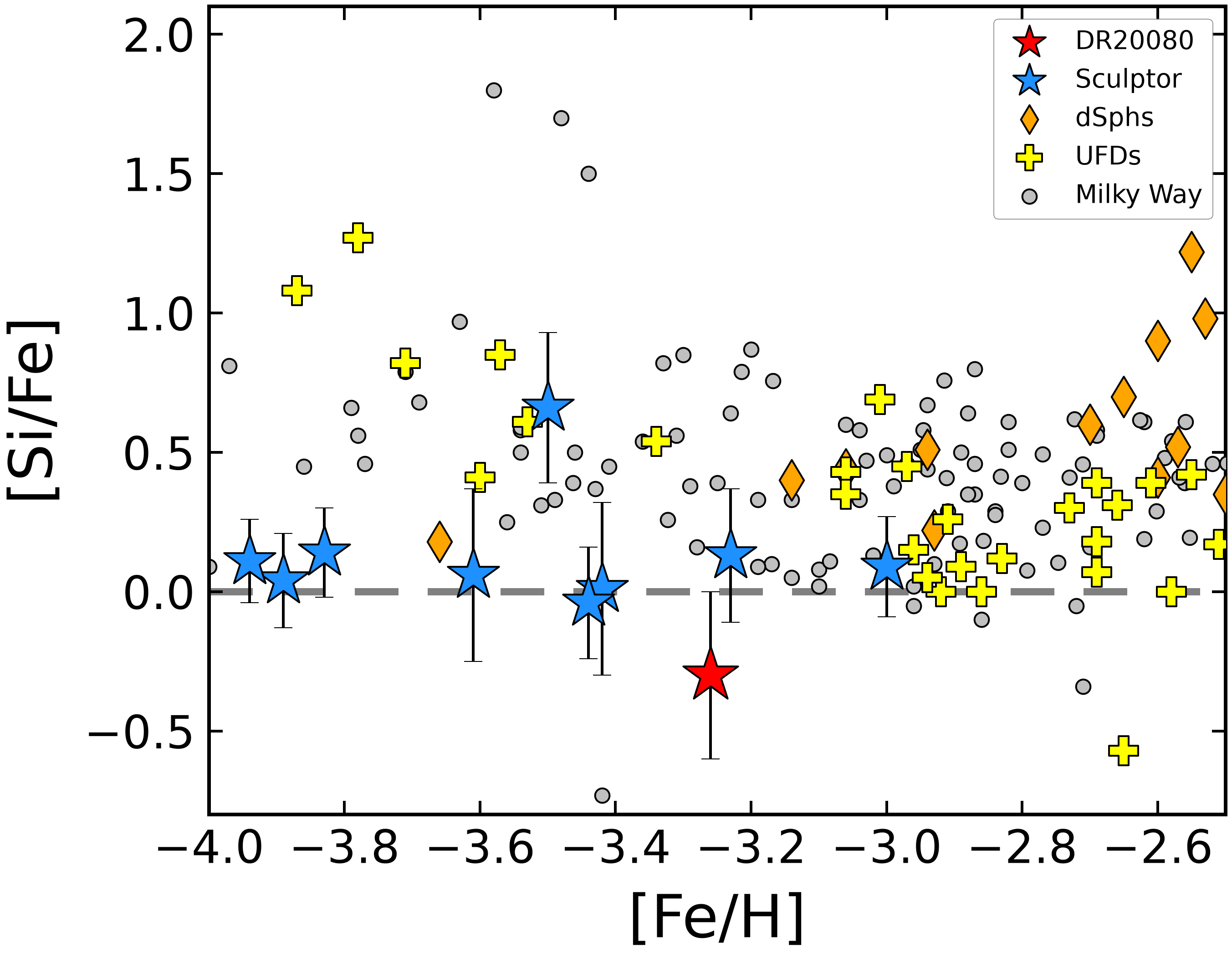}
                \includegraphics[width=0.33\hsize]{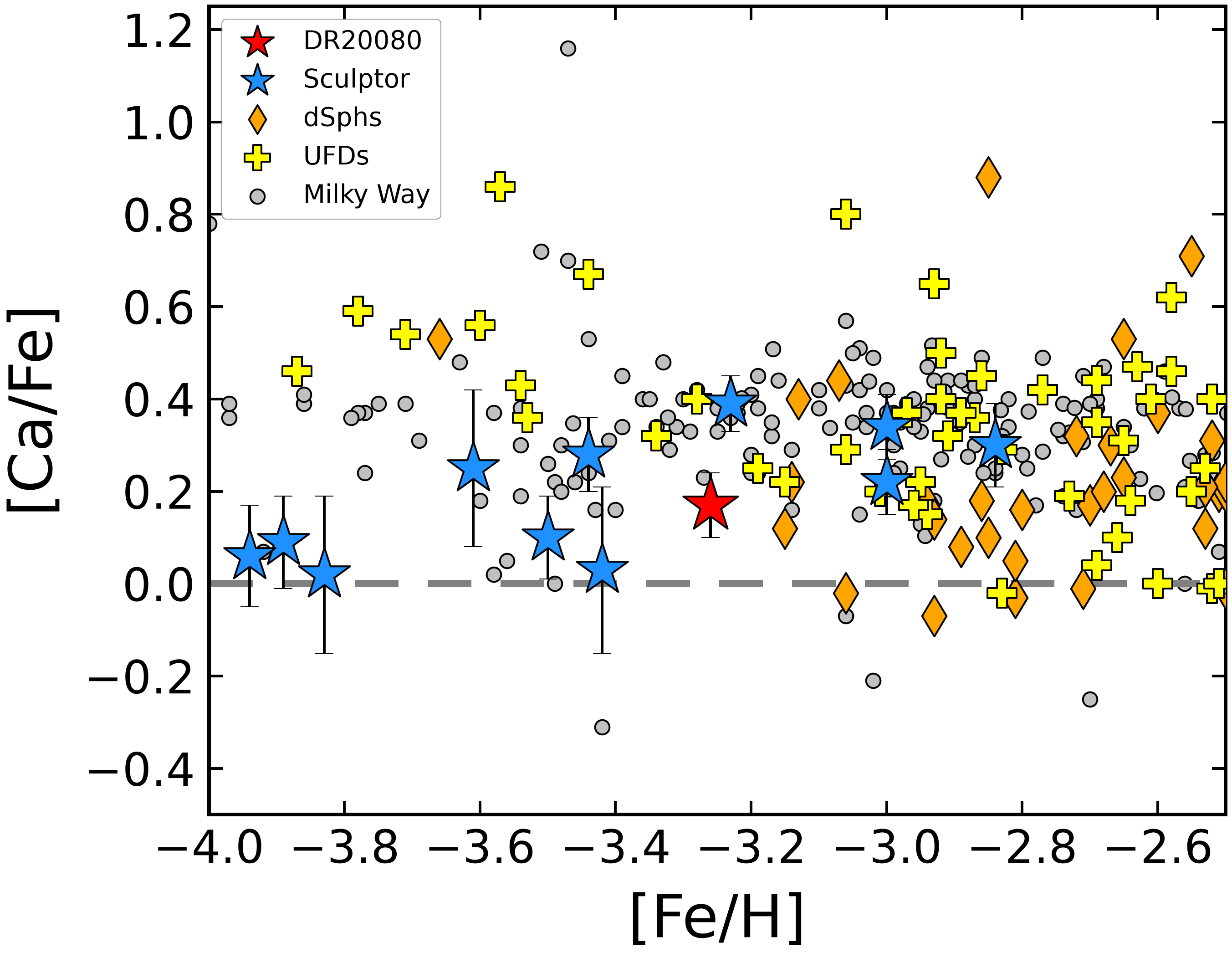}
                \\ \vspace{0.2cm}
                \includegraphics[width=0.33\hsize]{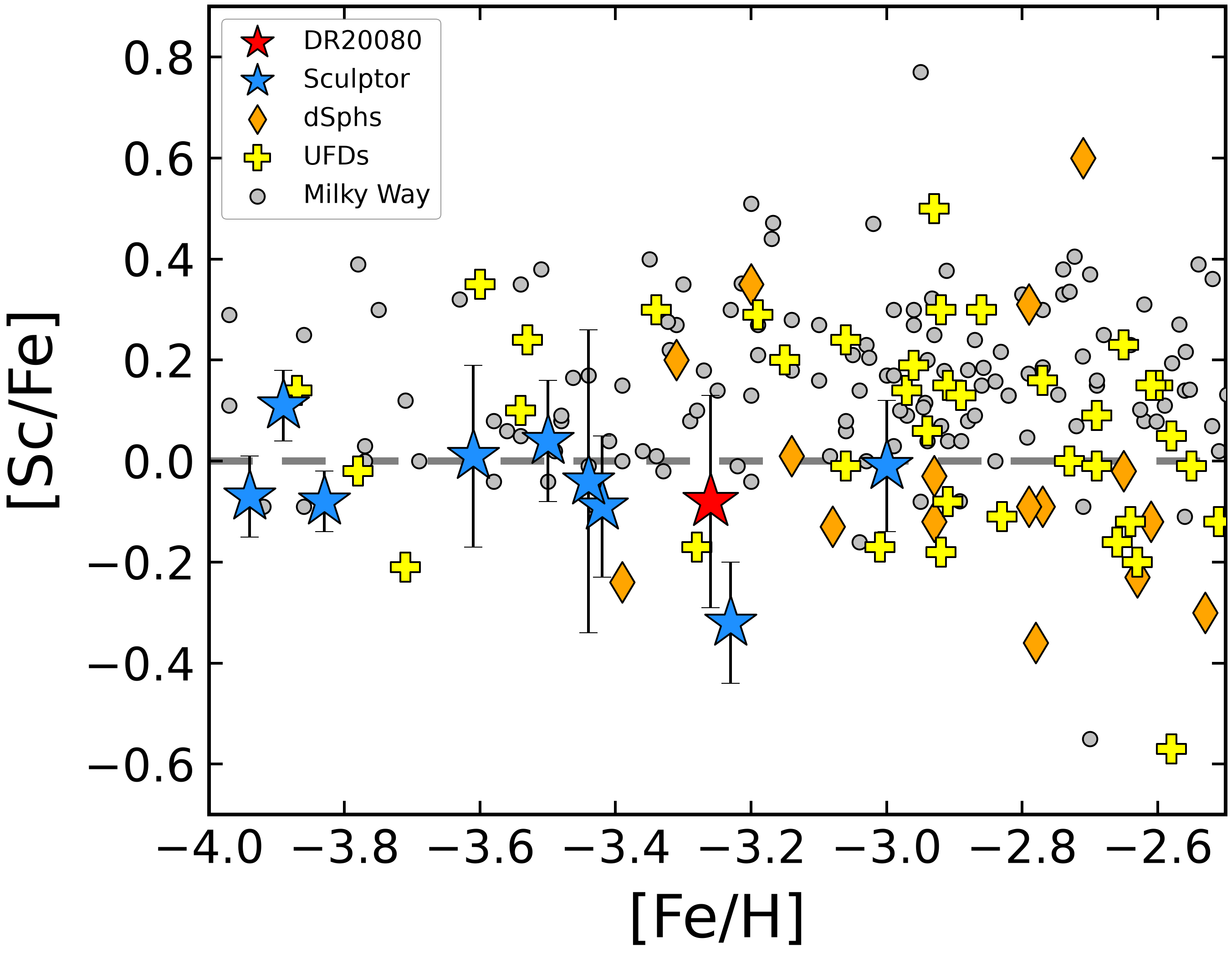}
                \includegraphics[width=0.33\hsize]{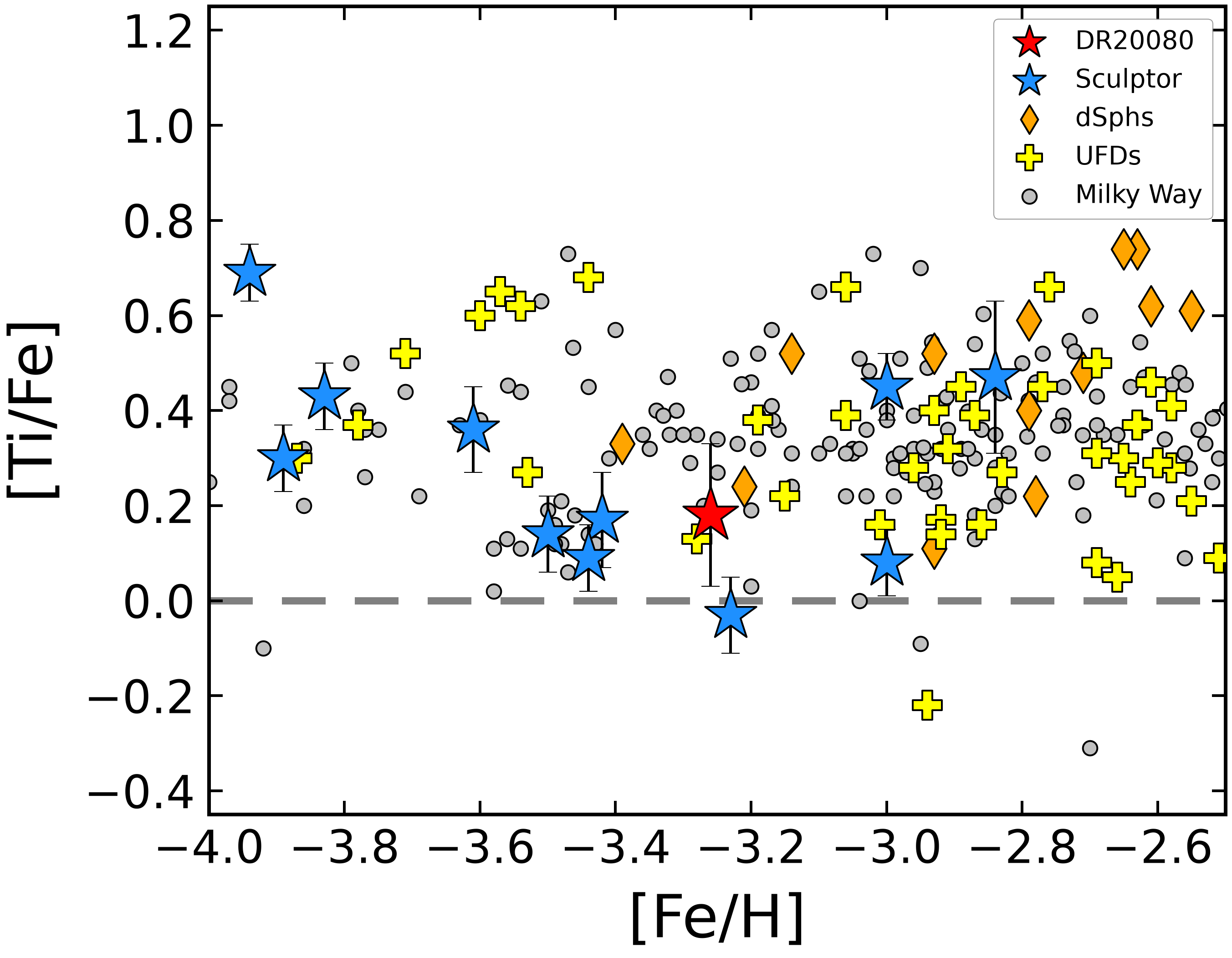}
                \includegraphics[width=0.33\hsize]{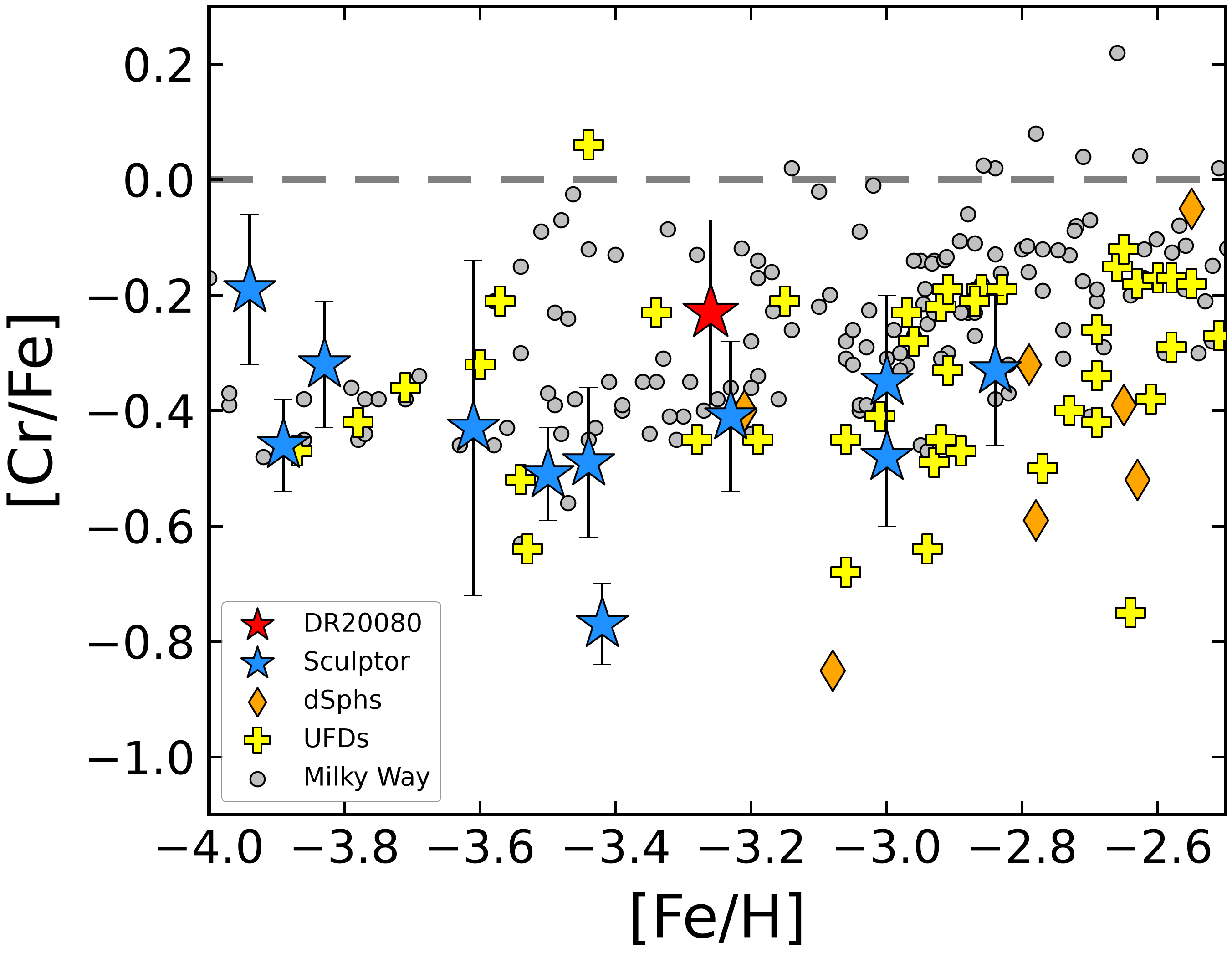}
                \\ \vspace{0.2cm}
                \includegraphics[width=0.33\hsize]{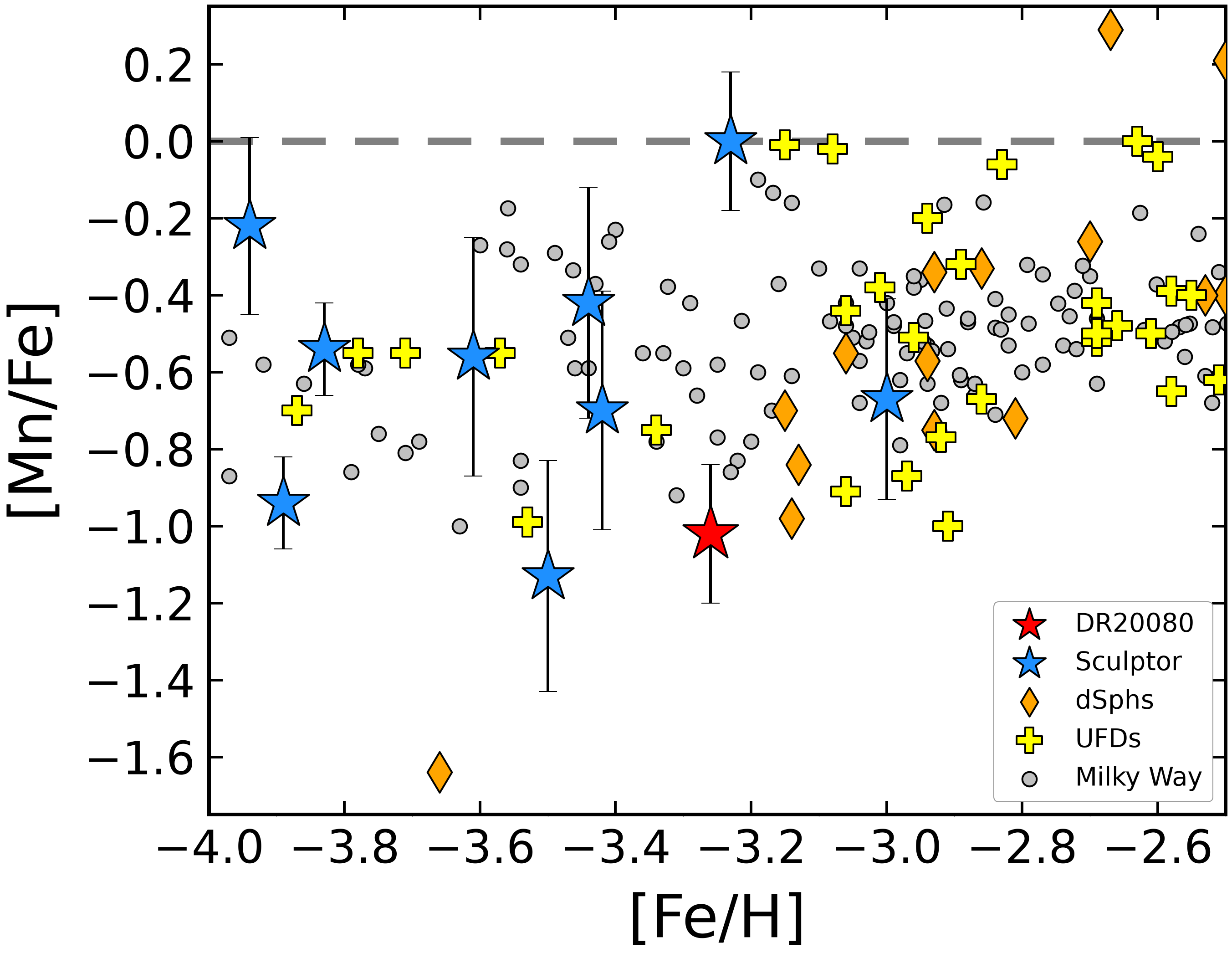}
                \includegraphics[width=0.33\hsize]{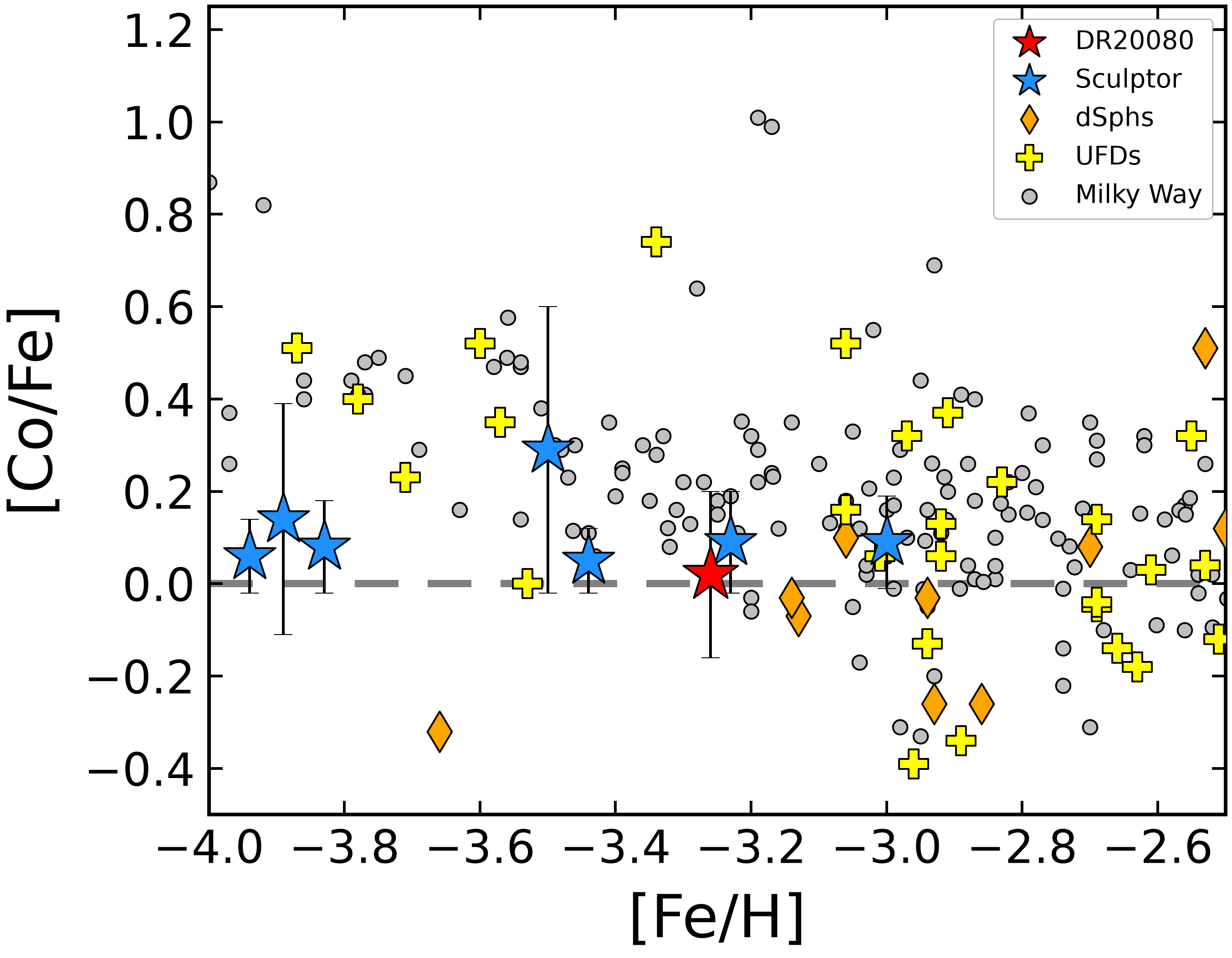}
                \includegraphics[width=0.33\hsize]{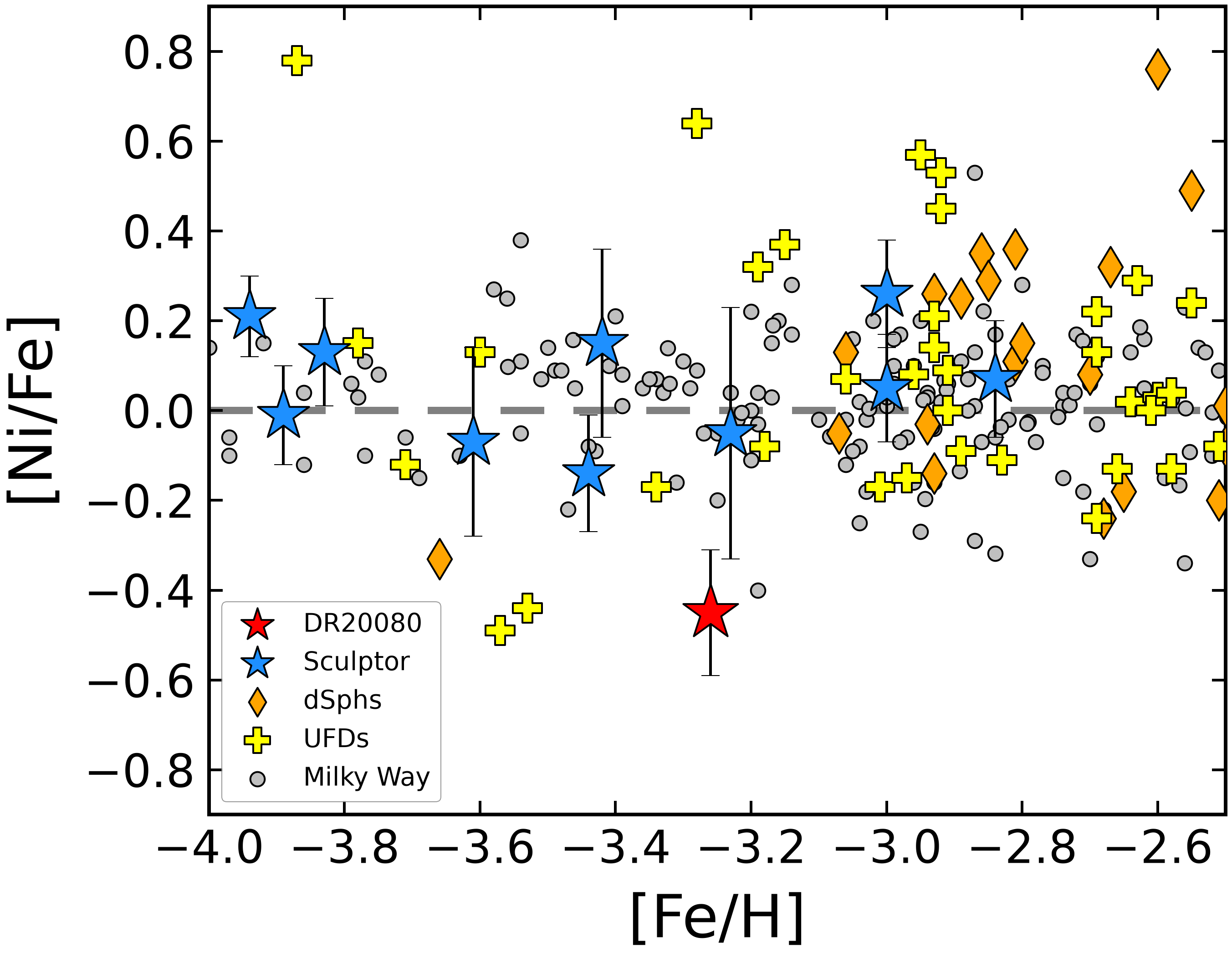}
                \caption{Same as \autoref{fig:C}, but for $[\text{O}/\text{Fe}]$, $[\text{Na}/\text{Fe}]$, $[\text{Mg}/\text{Fe}]$, $[\text{Al}/\text{Fe}]$, $[\text{Si}/\text{Fe}]$, $[\text{Ca}/\text{Fe}]$, $[\text{Sc}/\text{Fe}]$, $[\text{Ti}/\text{Fe}]$,  $[\text{Cr}/\text{Fe}]$, $[\text{Mn}/\text{Fe}]$, $[\text{Co}/\text{Fe}]$, and $[\text{Ni}/\text{Fe}]$. 
                See \Autoref{subsec:Comparison_Data} for references.
                }
                \label{fig:XFe}
        \end{figure*} 

        \begin{figure*}
            \centering
            \includegraphics[width=0.42\hsize]{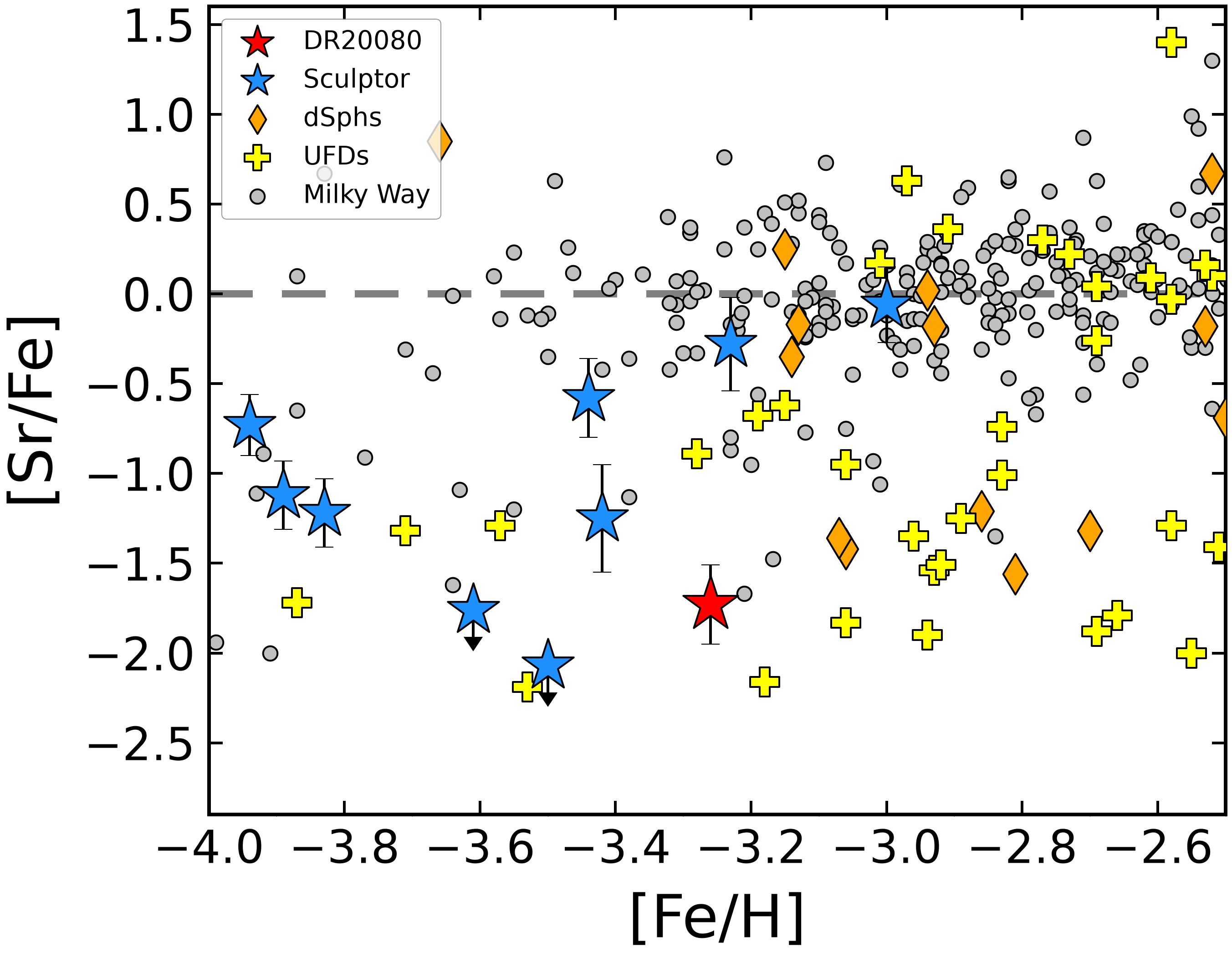}
            \includegraphics[width=0.42\hsize]{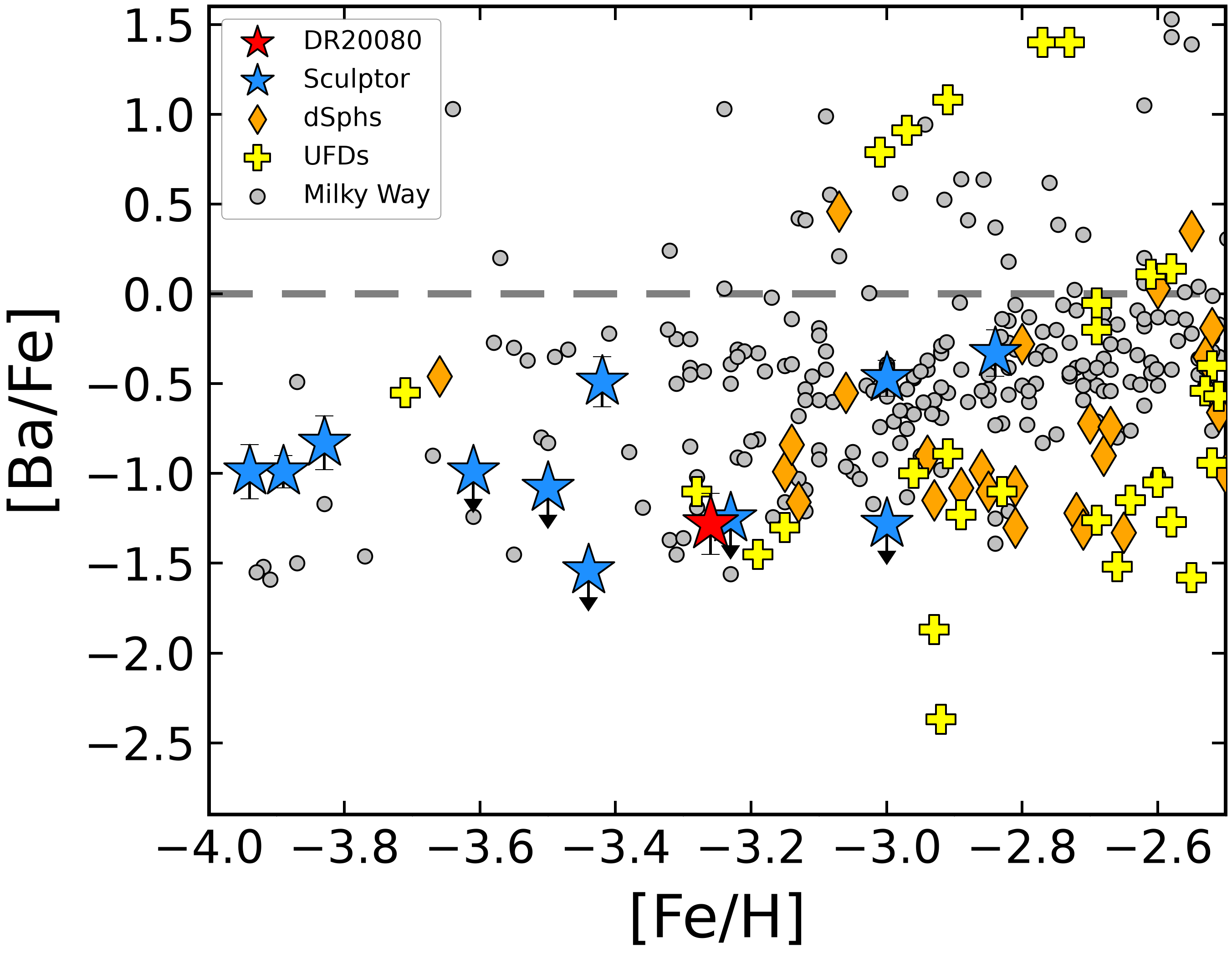} 
            \\ \vspace{0.2cm}
            \includegraphics[width=0.47\hsize]{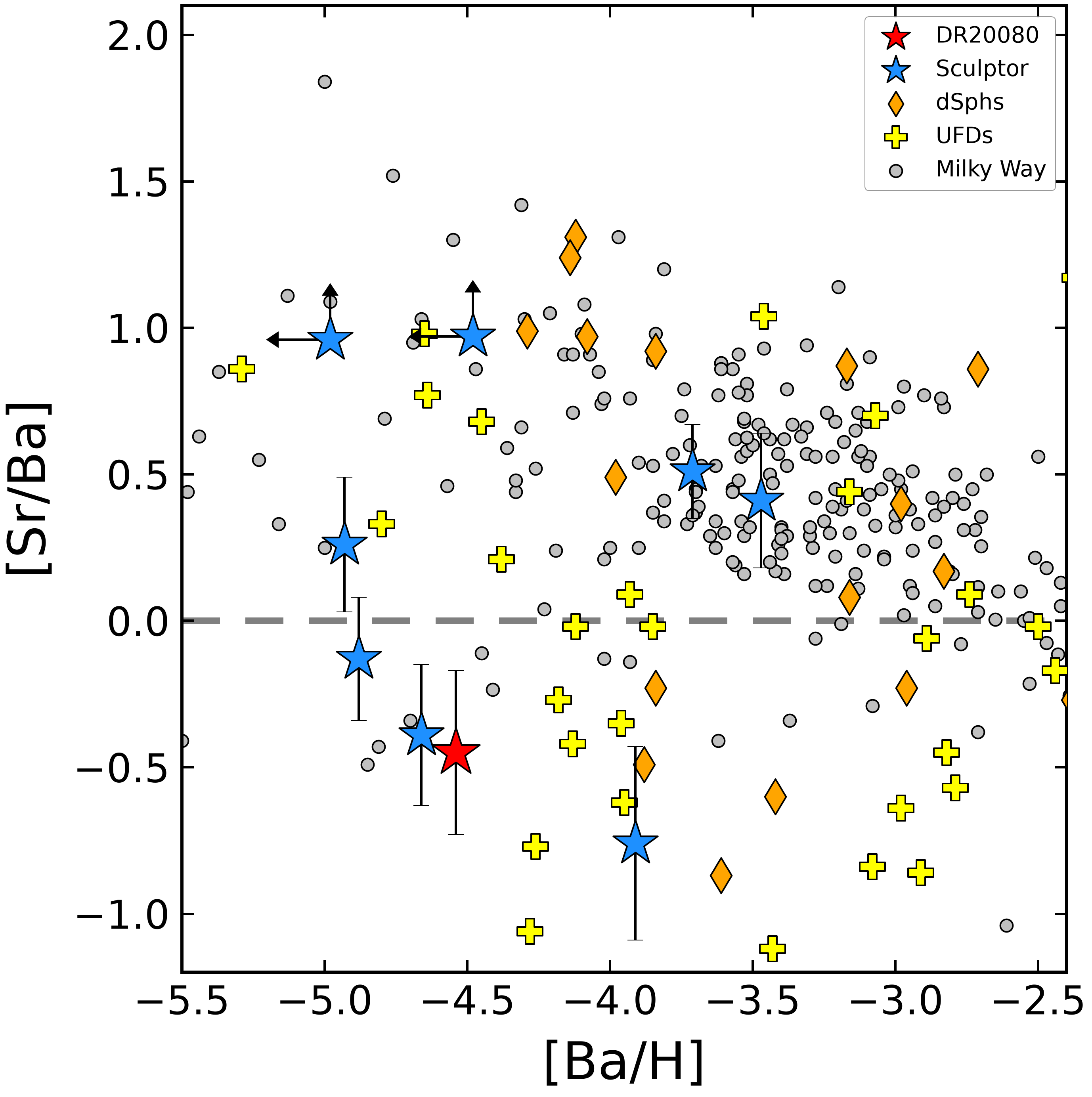}
            \caption{Same as \autoref{fig:C}, but for n-capture elements, $[\text{Sr}/\text{Fe}]$ and $[\text{Ba}/\text{Fe}]$ (\textit{top panels}), and $[\text{Sr}/\text{Ba}]$ as a function of $[\text{Ba}/\text{H}]$ (\textit{bottom panel}).
            References in \Autoref{subsec:Comparison_Data}.
            }
            \label{fig:SrBa}
        \end{figure*} 
    
        \begin{figure*}
            \centering
            \includegraphics[width=0.80\hsize]{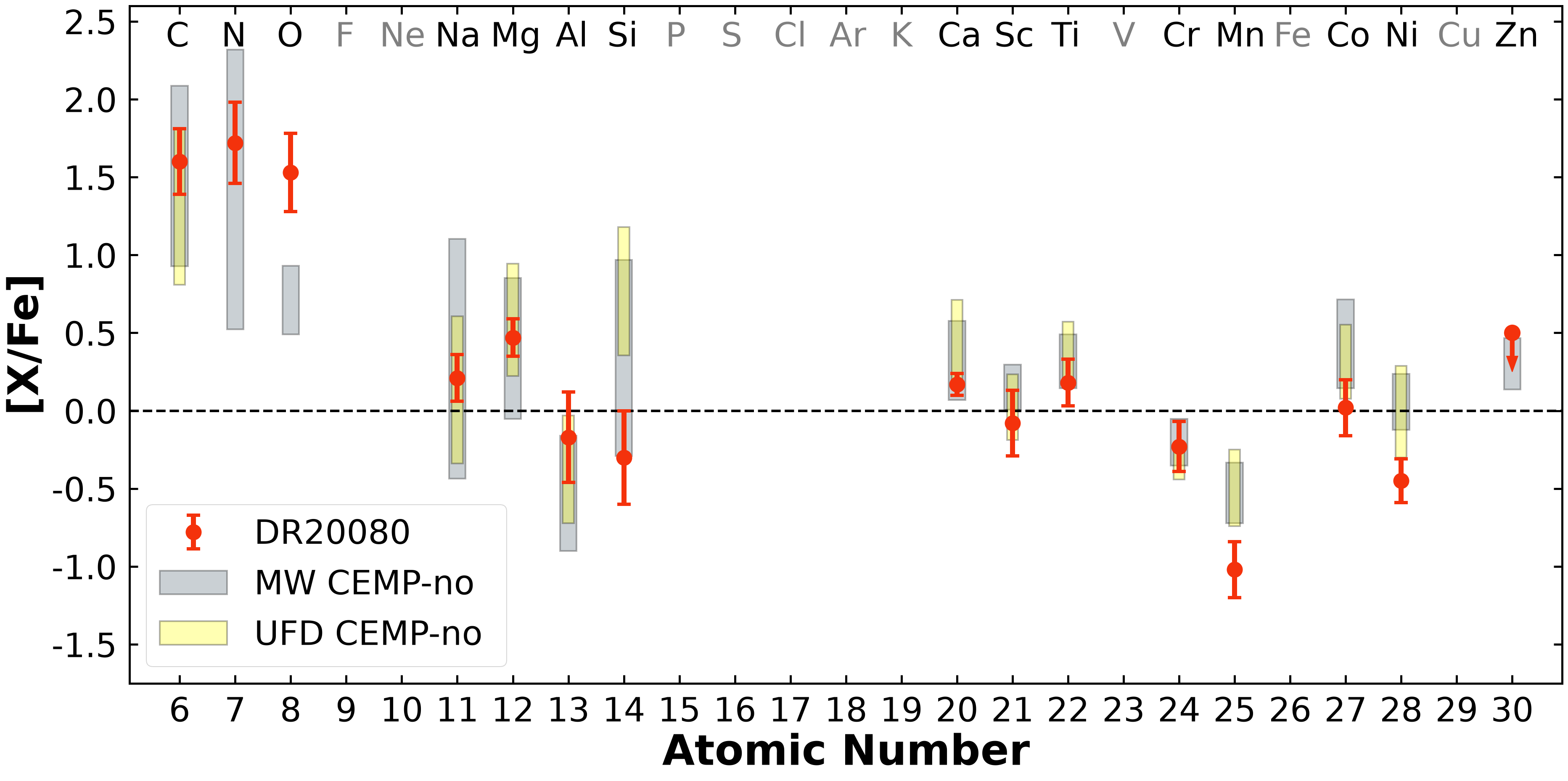}
            \caption{DR20080 LTE abundance ratios (\textit{red}) compared with mean abundances in MW CEMP-no stars $\langle[\text{X}/\text{Fe}]_{\text{MW CEMP-no}}\rangle \pm \sigma_{\text{X}}$ (\textit{grey}) and in UFD CEMP-no stars $\langle[\text{X}/\text{Fe}]_{\text{UFD CEMP-no}}\rangle \pm \sigma_{\text{X}}$ (\textit{yellow}).  
            Only elements measured both in DR20080 and in the MW sample are displayed. 
            Metallicity range of the MW and UFDs samples is $-4 \leq [\text{Fe}/\text{H}] \leq -2.5$.
            The mean metallicity of MW CEMP-no stars is $\langle[\text{Fe}/\text{H}]_{\text{MW CEMP-no}}\rangle = -3.23 \pm 0.36$, for the UFDs sample is$\langle[\text{Fe}/\text{H}]_{\text{UFD CEMP-no}}\rangle = -3.43 \pm 0.45$, while DR20080 has $[\text{Fe}/\text{H}] = -3.26 \pm 0.12$. 
            Compilation of \cite{vanni_characterizing_2023} for the MW sample, and of \cite{norris_boo-1137extremely_2010,frebel_high-resolution_2010,frebel_segue_2014,ji_detailed_2020} for the UFDs sample, see text for references.
            \autoref{fig:DR20080_vs_other_CEMP-no_[X/H]} shows the same for $[\text{X}/\text{H}]$.
            }
            \label{fig:DR20080_vs_other_CEMP-no_[X/Fe]}
        \end{figure*} 

        When we compare the derived chemical abundances of DR20080 with other very metal-poor ($[\text{Fe}/\text{H}] < -2$) stars belonging to the Sculptor dSph, for which data are available in the literature, or to other environments in the MW satellite system, such as the MW Stellar halo, other dSphs or UFDs, we begin to reveal how unique this CEMP-no star really is.

        \subsection{Comparison Data}
        \label{subsec:Comparison_Data}        
            In this Subsection, we provide all the references for the comparison samples used in \Cref{subsec:Early_chemical_enrichment,subsec:Comparison_other_CEMP-no}.
            \cref{fig:C,fig:SrBa,fig:XFe} show the elemental abundances of DR20080 (red star symbol) compared with other metal-poor stars with similar metallicities ($-4.0 < [\text{Fe}/\text{H}] < -2.5$) for which high-resolution analysis is available in the literature.
            
            The Sculptor stars at these low metallicities (blue stars) are from: \cite{frebel_linking_2010,tafelmeyer_extremely_2010,kirby_detailed_2012,jablonka_early_2015,simon_chemical_2015}; and \cite{skuladottir_zero-metallicity_2021,skuladottir_tracing_2024}.
            
            For the stars belonging to other dSphs (orange diamonds), in the case of carbon (\autoref{fig:C}) we used the literature compilation from \cite{lucchesi_extremely_2024}, which includes data from: \cite{tafelmeyer_extremely_2010,kirby_detailed_2012,venn_nucleosynthesis_2012,susmitha_abundance_2017,hansen_ages_2018,yoon_identification_2020,lucchesi_homogeneity_2020,lucchesi_extremely_2024}; and \cite{sestito_pristine_2024}, integrated with the abundances retrived from the SAGA database\footnote{\url{http://sagadatabase.jp/}} \citep[e.g.,][]{suda_stellar_2008,suda_stellar_2017}, including data from: \cite{fulbright_draco_2004, cohen_chemical_2009, cohen_chemical_2010}; and \cite{honda_enrichment_2011}.
            In \cref{fig:XFe,fig:SrBa}, the data for dSphs stars were taken from \cite{reichert_neutron-capture_2020} for Mg, Sc, Ti, Cr, Zn.
            For the other elements (Na, Si, Ca, Mn, Co, Ni, Sr and Ba), the abundances were obtained through the SAGA database, which includes data from: \cite{shetrone_abundance_2001,fulbright_draco_2004,sadakane_subaruhds_2004,koch_complexity_2008,aoki_chemical_2009,cohen_chemical_2009,cohen_chemical_2010,letarte_high-resolution_2010,tafelmeyer_extremely_2010,honda_enrichment_2011,kirby_detailed_2012,venn_nucleosynthesis_2012,lemasle_vltflames_2012,lemasle_vltflames_2014,hendricks_insights_2014,fabrizio_carina_2015,tsujimoto_chemical_2015,ural_inefficient_2015,francois_abundance_2016,norris_populations_2017}; and \cite{hansen_ages_2018}.
            These data include stars belonging to the following galaxies: Canes Venatici I, Carina, Draco, Fornax, Sagittarius, Sextans and Ursa Minor.
            
            The MW compilation (grey dots) is the same used in \cite{vanni_characterizing_2023}, which includes data from \cite{cayrel_first_2004,christlieb_he_2004,norris_he_2007,caffau_extremely_2011,yong_most_2013,hansen_exploring_2014,keller_single_2014,roederer_search_2014,frebel_sd_2015,li_high-resolution_2015,bonifacio_topos_2015,bonifacio_topos_2018, francois_chemical_2018,starkenburg_pristine_2018,aguado_back_2019}; and \cite{ezzeddine_evidence_2019}, and has been extended with the recent studies from \cite{lombardo_chemical_2022,lombardo_chemical_2025} and \cite{fernandes_de_melo_chemical_2024}. 
            
            The data for stars belonging to UFDs (yellow crosses) are retrieved from \cite{lai_feh_2011}; \cite{frebel_segue_2014}; \cite{roederer_detailed_2014}; \cite{kirby_triangulum_2017}; \cite{nagasawa_chemical_2018}; \cite{ji_chemical_2019}; \cite{marshall_chemical_2019}; \cite{ji_detailed_2020}; \cite{hansen_chemical_2020}; \cite{verdiani_carbon_2026}, and from the JINA database\footnote{\url{https://jinabase.pythonanywhere.com/}} \citep{abohalima_jinabasedatabase_2018}, covering data from: \cite{feltzing_evidence_2009,norris_boo-1137extremely_2010,simon_high-resolution_2010,frebel_high-resolution_2010,frebel_chemical_2016,gilmore_elemental_2013,ishigaki_chemical_2014,roederer_detailed_2016,francois_abundance_2016}; and \cite{ji_r-process_2016,ji_high-resolution_2016}.   
            The sample includes stars from Bo\"{o}tes I, Bo\"{o}tes II, Canes Venatici II, Carina II, Carina III, Coma Berenices, Grus I, Grus II, Hercules, Horologium I, Leo IV, Reticulum II, Segue 1, Segue 2, Triangulum II, Tucana III, Tucana IV and Ursa Major II.
            
            Given that NLTE corrections have been published for only a handful of these stars, and even then, typically for just a few elements, we focused on the LTE abundances for this comparison.
            All these abundances are referred to the solar abundances from \cite{asplund_chemical_2009}.
            Additionally, carbon abundances taken into account were corrected for internal mixing \citep{placco_carbon-enhanced_2014}.
            In the case of duplicate entries among the various references, the measurements from the original source were preferred.

        \subsection{Early chemical enrichment in different environments}
        \label{subsec:Early_chemical_enrichment}  
            Let us first examine the case of carbon.
            As shown in \autoref{fig:C}, the $[\text{C}/\text{Fe}]$ ratio for DR20080 lies well above the CEMP threshold ($[\text{C}/\text{Fe}] \geq 0.7 $), confirming its CEMP classification, as previously inferred from the medium-resolution measures presented in \cite{skuladottir_tracing_2024}.
            Moreover, it is the only star in Sculptor, at such low metallicity, exhibiting such behaviour, more closely resembling the trend observed among the members of the MW halo or of the UFDs reported in the figure.
            This supports the idea that dSphs experienced a very different early chemical enrichment history compared to UFDs and the MW halo.

            \autoref{fig:XFe} shows the derived $[\text{X}/\text{Fe}]$ abundances for elements from O to Ni (see also \autoref{fig:XFe_upplims} in \autoref{App:abun_ratios} for the plots of N and of the other analysed elements, for which we estimated an upper limit).
            The first panel of \autoref{fig:XFe} shows an enhancement in O relative to the few MW stars with available measurements in the literature.
            Indeed, this element is particularly challenging to measure at these $[\text{Fe}/\text{H}]$ unless it is significantly enhanced, as also indicated by the lack of measurements at $[\text{O}/\text{Fe}] < +0.5$.
            As illustrated in the second panel, Na exhibits an enhancement in the DR20080 abundance pattern compared to the Sculptor average.
            A similar behaviour is observed for Al.
            In contrast, DR20080 is among the stars with the lowest abundance of Si within the considered metallicity range.
            Indeed, $[\text{Si}/\text{Fe}]$ is underabundant even when compared to stars from other environments (see \autoref{fig:XFe}).
            Moreover, the iron-peak elements Mn and Co follow the lower edge of the distribution and Ni is found to be underabundant, contrasting with the general trend in Sculptor that aligns more closely with the populations of UFDs and the MW halo.
            On the other hand, the other elements shown in \autoref{fig:XFe}, such as the \textalpha-elements or the iron-peak element Cr, are fully consistent with the overall trend in Sculptor, in dSphs, in UFDs and in the MW.
    
            DR20080 shows no enhancement in n-capture elements, like strontium, barium and europium (see \Cref{tab:LTE_abundances,tab:NLTE_abundances} and the top panels of \autoref{fig:SrBa}).
            As highlighted in \cite{frebel_near-field_2015}, studies on UFDs have revealed that the abundances of Sr and Ba, along with their ratio $[\text{Sr}/\text{Ba}]$, may serve as criteria to distinguish, at similar $[\text{Fe}/\text{H}]$, which stars were formed from more pristine gas  \citep[e.g.,][]{ji_chemical_2019}.
            Specifically, lower abundances of n-capture elements seem to indicate a formation in an earlier gas cloud.
            Notably, the $[\text{Sr}/\text{Ba}]$ ratio, displayed in the bottom panel of \autoref{fig:SrBa} as a function of $[\text{Ba}/\text{H}]$, appears to be underabundant with respect to the typical stellar populations represented in the plot.        
            This seems to provide further evidence supporting the conclusion that DR20080 formed in a primitive environment.

        \subsection{Comparison to other CEMP-no stars}
        \label{subsec:Comparison_other_CEMP-no}          
            To further explore the unusual abundance pattern of DR20080, a comparison was made between this star and other CEMP-no stars from the MW halo and from UFDs with similar metallicities (\autoref{fig:DR20080_vs_other_CEMP-no_[X/Fe]}), i.e. $\pm 0.75 \,\text{dex}$ from the value of DR20080.
            For this purpose, we considered the 39 MW CEMP-no stars with $-4 \leq [\text{Fe}/\text{H}] \leq -2.5$ within the same literature compilation from \cite{vanni_characterizing_2023} used for \cref{fig:C,fig:XFe,fig:SrBa} (see \Autoref{subsec:Comparison_Data}), which contains both main-sequence and RGB stars.
            We also considered 8 CEMP-no stars belonging to UFDs, identified within the same metallicity range in the same literature compilation described in \Autoref{subsec:Comparison_Data}.
            As for the previous case, we only considered 1D LTE abundance values for this analysis.
    
            \begin{figure}[t]
                \centering
                \includegraphics[width=\hsize]{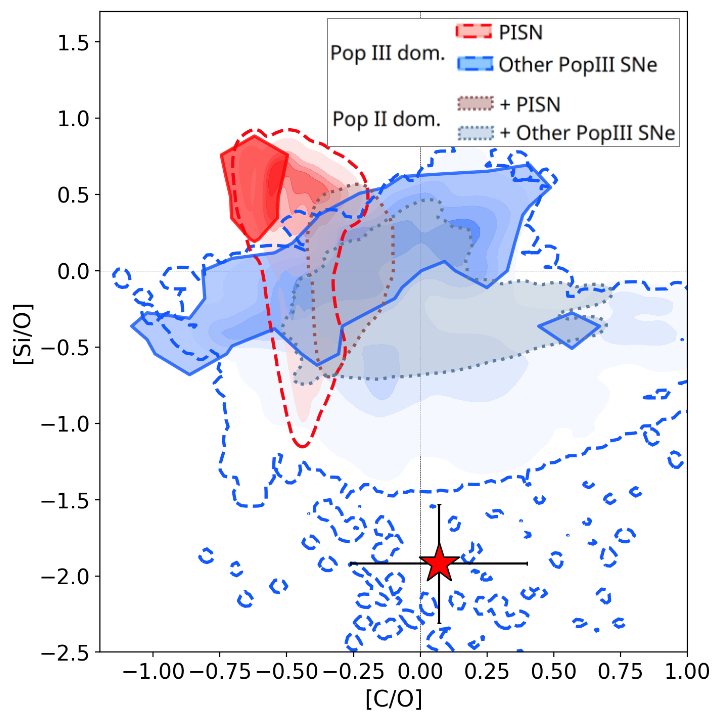}
                \caption{\textit{Red star} represents DR20080's NLTE abundance ratios, with their uncertainties, while coloured contours are SN yield models. 
                \textit{Solid red} and \textit{blue} lines indicate environments enriched exclusively by PISNe or other Pop~III SNe respectively, \textit{dashed red} and \textit{blue} primarily by PISNe or other Pop~III SNe, but with some contribution from normal Pop~II SNe, and \textit{dotted brown} and \textit{grey} mainly by normal Pop~II SNe with minor contributions from PISNe or other Pop~III SNe respectively.
                Adapted from \cite{vanni_chemical_2024}.
                Models from \cite{salvadori_probing_2019} and \cite{vanni_characterizing_2023}. 
                }
                \label{fig:enrichment_models}
            \end{figure}
            
            \begin{figure*}
                \centering
                \includegraphics[width=0.9\hsize]{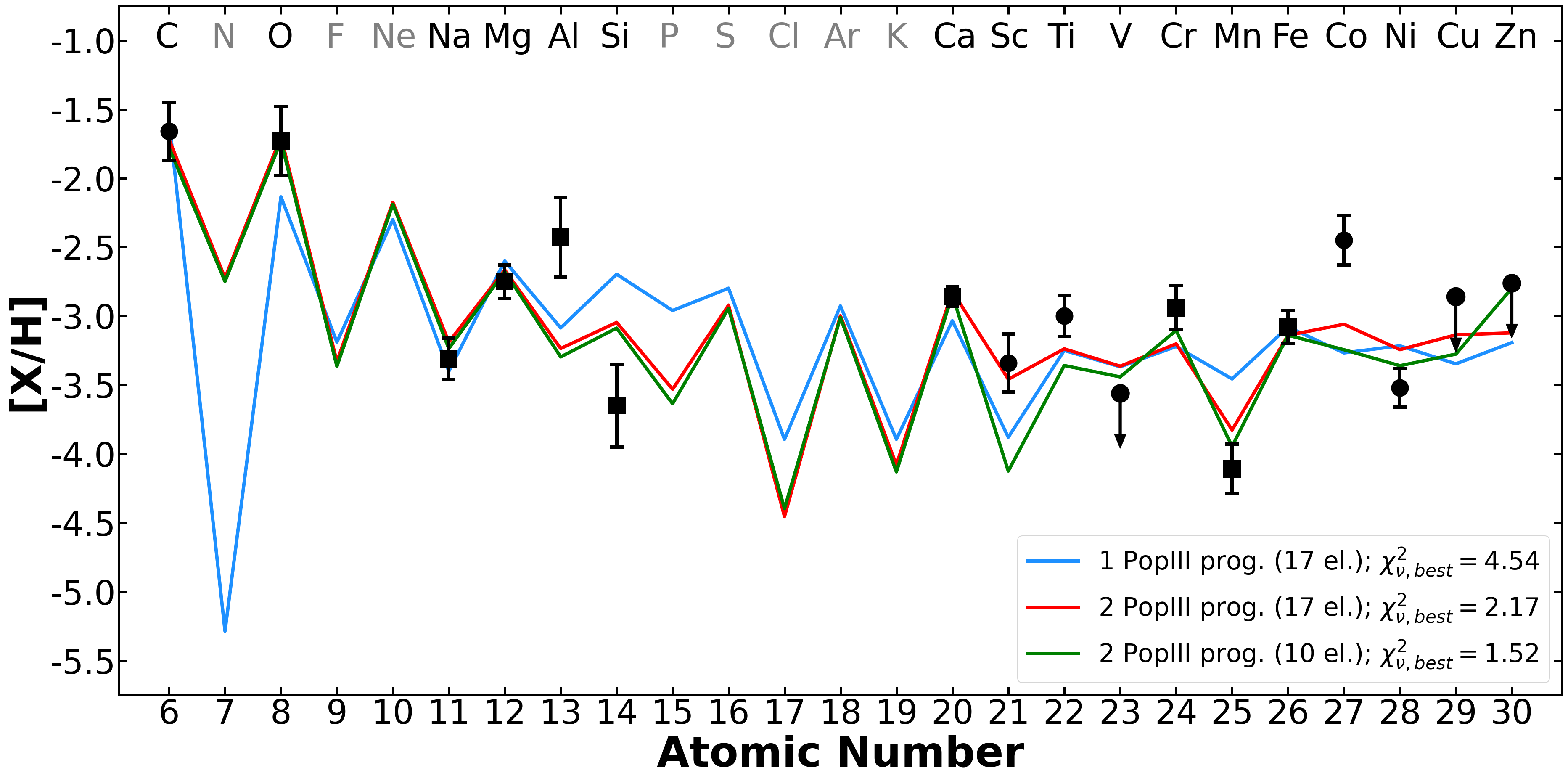}
                \caption{Best-fit models of DR20080's abundance pattern, all made considering both Pop~III and Pop~II SN yields models: 
                1)~\textit{blue line}, only one progenitor star (a Pop~III fSN with $m_{\text{PopIII}} = 11.4 \text{M}_{\odot}$ and $E_{\text{SN}} = 0.6 \cdot 10^{51} \,\text{erg}$); 
                2)~\textit{red line}, it represents the fit with two progenitors (95\% Pop~III fSN with $m_{\text{PopIII}} = 43 \text{M}_{\odot}$ and $E_{\text{SN}} = 0.6 \cdot 10^{51} \,\text{erg}$; 5\% Pop~III HN with $m_{\text{PopIII}} = 19.4 \, \text{M}_{\odot}$ and $E_{\text{SN}} = 10 \cdot 10^{51} \,\text{erg}$.) which considers all the measured elements (\textit{squares} and \textit{circles}); 
                3)~\textit{green line}, it also represents the fit with two progenitors (85\% Pop~III fSN with $m_{\text{PopIII}} = 43 \text{M}_{\odot}$ and $E_{\text{SN}} = 0.6 \cdot 10^{51} \,\text{erg}$; 15\% Pop~III HN with $m_{\text{PopIII}} = 50 \, \text{M}_{\odot}$ and $E_{\text{SN}} = 10 \cdot 10^{51} \,\text{erg}$.), but only accounting for those elements considered as reliable (\textit{squares}, see text for more details).
                }
                \label{fig:DR20080_progenitors}
            \end{figure*}
    
            To make this comparison, we calculated the mean of the abundance ratios of each element X, $\langle[\text{X}/\text{Fe}]_{\text{CEMP-no}}\rangle$, in the two samples, and we assumed the standard deviation, $\sigma_{\text{X}}$, as the indication of the scatter of these values.
            These mean abundances, alongside the derived abundances of DR20080, are displayed in \autoref{fig:DR20080_vs_other_CEMP-no_[X/Fe]}.
            It is worth noticing that not all stars in the sample had abundance derivations for every element displayed, so each mean value is based only on the available data within the sample.
    
            The first thing we notice is that carbon and nitrogen are well aligned with the typical MW CEMP-no stars, while the oxygen abundance in DR20080 is significantly enhanced.
            The carbon abundance is also consistent with that of CEMP-no stars in UFDs.
            In contrast, the iron-peak elements, namely manganese and nickel, are noticeably under-abundant relative to both populations (see also \autoref{fig:XFe}).
            It is also worth mentioning silicon and cobalt, which, although consistent with the mean values of CEMP-no stars in the MW, appear to be slightly under-abundant.
            Moreover, while Co is consistent with the UFD sample, Si is significantly under-abundant relative to it.
    
            Even though for single elements such as carbon, DR20080 appears to be more similar to what is observed in the MW halo or in UFDs than in Sculptor or other dSphs, as shown for instance in \autoref{fig:C}, the overall abundance pattern reveals that DR20080 is intrinsically different even from the CEMP-no stars in the MW halo or in UFDs.
            This, in turn, suggests that its chemical enrichment history also differs from that of such an environment.
            Such observation may therefore support the hypothesis of different populations of CEMP-no stars, each significantly shaped by the environment in which they formed.
    
            However, it is worth noting that, according to the recent study by \cite{arroyo-polonio_chemo-dynamics_2024}, DR20080 is part of a very metal-poor subpopulation of Sculptor's stars that may not have formed in situ, but may instead be the remnant of a past merger with a UFD.
            This possibility is also predicted by cosmological models for the Local Group formation \citep[e.g.,][]{salvadori_carbon-enhanced_2015,koutsouridou_nefertiti_2026}, as mergers are a natural consequence of hierarchical structure formation in the $\Lambda$CDM cosmological model.
            However, quantifying the contribution of such events to the stellar population of a Sculptor-like galaxy is challenging, as individual UFDs are expected to contribute only a small fraction of its stellar mass, while the number of CEMP-no stars accreted depends on the star formation history of the accreted UFD and its Pop~III IMF \citep[e.g.,][]{rossi_crimsons_2026}.
            According to this interpretation, if DR20080 indeed did not form in situ, it provides further indirect evidence of the rarity of CEMP-no stars in dSphs, still leaving open the question regarding the differences in their formation in this class of galaxies and the puzzle that distinguishes them from those in UFDs and in the MW halo.
    
            \begin{figure}
                \centering
                \begin{subfigure}[t]{\hsize}
                    \centering
                    \includegraphics[width=\textwidth]{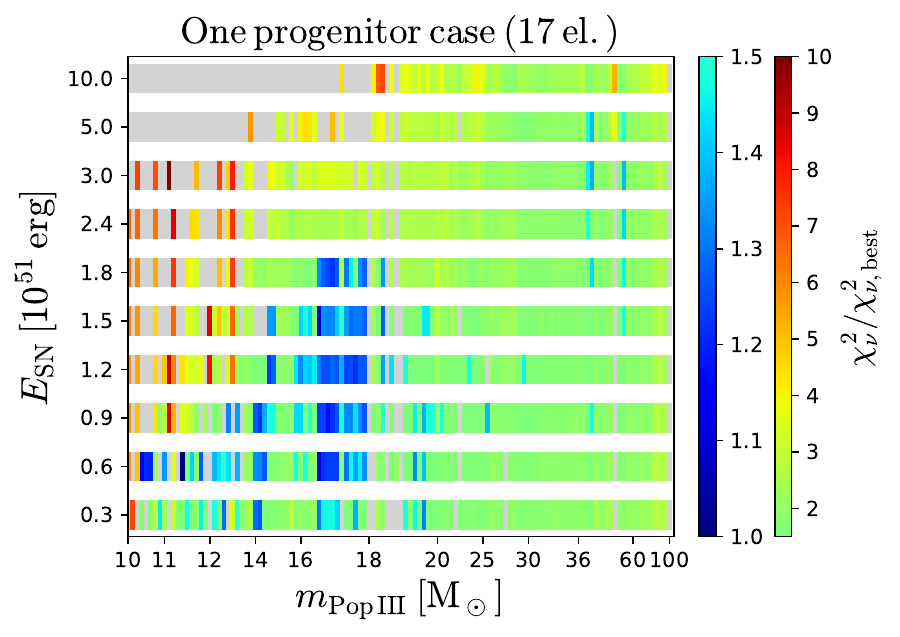}
                    \caption{Best fit: Pop~III fSN with $E_{\text{SN}} = 0.6 \cdot 10^{51} \,\text{erg}$ ; $m_{\text{PopIII}} = 11.4 \,\text{M}_{\odot}$
                    }
                    \label{subfig:1prog}
                \end{subfigure}
                \\
                \vspace{0.2cm}
                \begin{subfigure}[t]{\hsize}
                    \centering
                    \includegraphics[width=\textwidth]{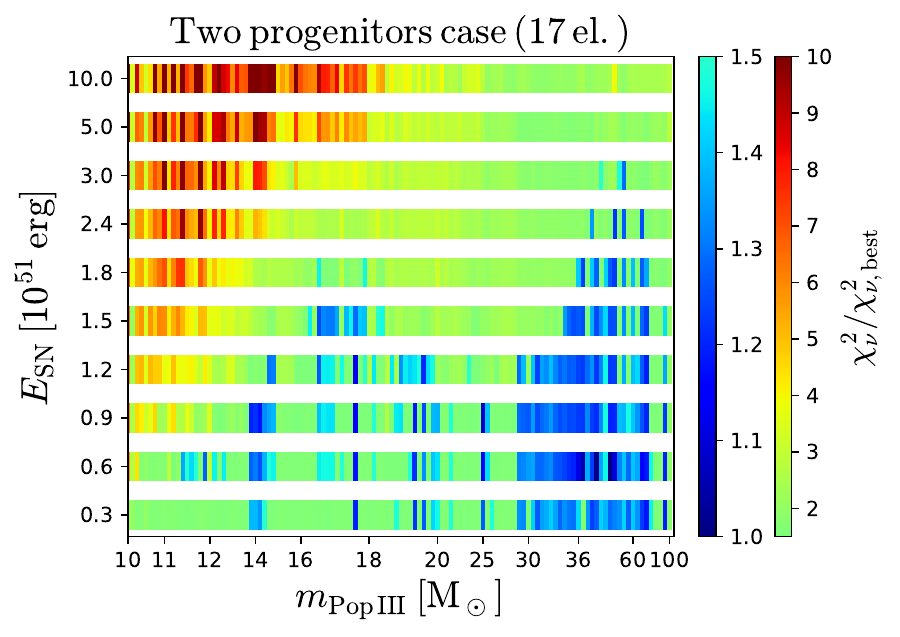}
                    \caption{Best fit: 95\% Pop~III fSN with $E_{\text{SN}} = 0.6 \cdot 10^{51} \,\text{erg}$ ; $m_{\text{PopIII}} = 43 \,\text{M}_{\odot}$
                    }
                    \label{subfig:2prog}
                \end{subfigure}
                \\
                \vspace{0.2cm}
                \begin{subfigure}[t]{\hsize}
                    \centering
                    \includegraphics[width=\textwidth]{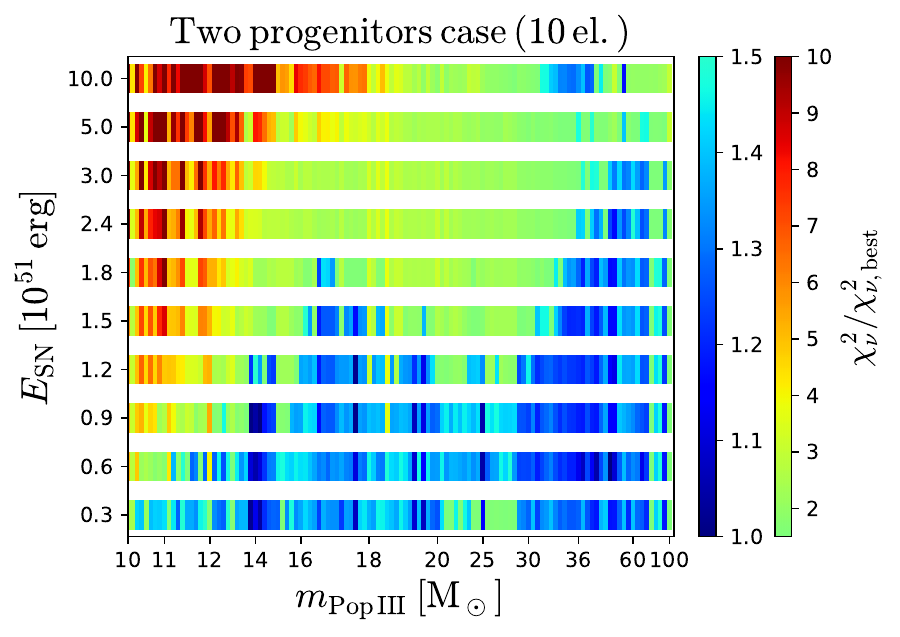}
                    \caption{Best fit: 85\% Pop~III fSN with $E_{\text{SN}} = 0.6 \cdot 10^{51} \,\text{erg}$ ; $m_{\text{PopIII}} = 43 \,\text{M}_{\odot}$
                    }
                    \label{subfig:2prog_trusted}
                \end{subfigure}
                \\
                \caption{Parameter space (progenitor mass, $m_{\text{PopIII}}$, on the x-axis; SN energy, $E_{\text{SN}}$, on the y-axis) for the three fits displayed in \autoref{fig:DR20080_progenitors} ((\subref{subfig:1prog}) \textit{blue fit}, (\subref{subfig:2prog}) \textit{red fit}, and (\subref{subfig:2prog_trusted}) \textit{green fit} respectively), with every bin colour-coded according to the goodness of the fit ($\chi^2_{\nu}/\chi^2_{\nu,\textsl{best}}$).
                \textit{Darkest blue bins} correspond to the best fits.
                The two progenitors plots only show the parameter space corresponding to the major contributor to the enrichment.
                }
                \label{fig:prog}
            \end{figure}

%-----------------------------------------------------------------
    \section{Pop~III progenitor of DR20080}
    \label{sec:PopIII_progenitor_DR20080}

        Since Pop~III stars cannot be observed directly, our understanding of them relies on indirect evidence, particularly from the chemical abundances found in the stellar photospheres of ancient, metal-poor stars that may be their direct descendants.
        Their chemical yields depend primarily on the stellar mass ($m_{\text{PopIII}}$) and the explosion energy of its SN ($E_{\text{SN}}$), making the analysis of stellar abundances a valuable tool for constraining these factors and gaining insight into the elusive Pop~III stars.

        As discussed in \Autoref{sec:DR20080_Local_Group}, DR20080 shows signs of having formed in a gas cloud mainly enriched by Pop~III stars, as we can infer from its CEMP-no nature and from the low $[\text{Sr}/\text{Ba}]$ and $[\text{Ba}/\text{Fe}]$ ratios.
        Moreover, if we examine the chemical enrichment models from various types of SNe (including Pair-Instability SNe, PISNe), originating both from Pop~III and Pop~II stars, presented in \cite{vanni_chemical_2024} and displayed in \autoref{fig:enrichment_models}, we can find more clues pointing towards the Pop~III enrichment.
        In this plot, DR20080 (red star) occupies a region, $[\text{Si}/\text{O}]_{\textsl{NLTE}} = -1.92 \pm 0.39$ and $[\text{C}/\text{O}]_{\textsl{NLTE}} = 0.07 \pm 0.32$, that is sparsely populated by stellar yield models. 
        The only models that are consistent with the abundances derived for this CEMP-no star are those prevalently enriched by Pop~III SNe.
        It is important to notice that the conclusion remains the same in the LTE case, with $[\text{Si}/\text{O}] = -1.83 \pm 0.39$ and $[\text{C}/\text{O}] = 0.07 \pm 0.32$.

        \subsection{Possible spinstar progenitor}
        \label{subsec:possible_spistar_prog}
            This value of $[\text{C}/\text{O}]$ is in good agreement with the predicted yields of spinstar progenitors, as shown, for example, in Fig. 6 of \cite{choplin_probing_2026}. 
            The abundances of n-capture elements $[\text{Sr}/\text{Fe}]$, $[\text{Ba}/\text{Fe}]$ and $[\text{Sr}/\text{Ba}]$ also fall within the ranges predicted by the spinstar models presented in \cite{rizzuti_explaining_2025}.
            However, $\log{\text{C}/\text{O}} = -0.19$ falls outside the range covered by the models of \cite{rizzuti_explaining_2025}, while the Na and Mg abundances are more consistent with the non-rotating or slowly rotating models of \cite{choplin_probing_2026}. 
            Moreover, \cite{rizzuti_explaining_2025} point out that their spinstar models struggle to reproduce the abundances of C, N and O of CEMP-no stars in the metallicity range $-4 < [\text{Fe}/\text{H}] < -2$, tending to underestimate their abundances. 
            The isotopic ratio $^{12}\text{C}/^{13}\text{C}$ is generally considered one of the most powerful diagnostics for investigating a possible contribution from spinstar progenitors \citep[e.g.,][]{rizzuti_explaining_2025}.
            However, since DR20080 is an evolved giant, its present-day $^{12}\text{C}/^{13}\text{C}$ ratio has been altered by internal mixing and therefore does not provide a direct indication of the isotopic ratio in the star at birth (see also \autoref{subsec:CNO}). 
            Overall, although a spinstar origin for the progenitor of DR20080 cannot be ruled out.

        \subsection{Possible SN progenitors}
        \label{subsec:possible_SN_prog}
            To derive the properties of DR20080's progenitor stars, we fitted the observed 1D NLTE abundance pattern using a $\chi^2$ minimisation procedure applied to a grid of SN yield models, as also done in \cite{koutsouridou_true_2024}, using the general analytical calculations derived by \cite{salvadori_probing_2019} and \cite{vanni_characterizing_2023}. 
            For Pop~III stars, the type II and pair-instability SN yields were taken from \cite{heger_nucleosynthetic_2002,heger_nucleosynthesis_2010}  and from \cite{woosley_evolution_1995} and \cite{limongi_presupernova_2018} for Pop~II progenitors.
            For Type Ia SN yields (for both Pop~III and Pop~II stars), we adopted the models of \cite{iwamoto_nucleosynthesis_1999} and \cite{leung_explosive_2018}.
    
            In total, three fits were performed under different assumptions (see \autoref{fig:DR20080_progenitors}): the first (blue line) assumes that DR20080 formed from gas enriched by a single progenitor, while the second (red line) and the third (green line) assume enrichment by two distinct progenitors. 
            The latter two cases differ in the number of considered elements. 
            In the first and second fits, all the measured elements from carbon to zinc were included.
            The only exception was nitrogen, which was always excluded due to the internal mixing that makes N abundance unreliable for this purpose. 
            The third fit only included those elements whose abundance theoretical values are in best agreement with the observed ones for the overall stellar population \citep[e.g.,][]{kobayashi_origin_2020,koutsouridou_large_2025}, namely O, Na, Mg, Al, Si, Ca, Cr, Mn, Fe, and Cu (indicated by the \textit{squares} in \autoref{fig:DR20080_progenitors}), which we will refer to as reliable elements. 
    
            In all three cases, the best-fit model corresponds to a pure Pop~III enrichment.
            Indeed, in the first case the minimum $\chi^2_{\nu,\textsl{best}} = 4.54$ is obtained for a Pop~III progenitor with a mass of $m_{\text{PopIII}} = 11.4 \text{M}_{\odot}$, which exploded as a fSN with an energy of $E_{\text{SN}} = 0.6 \cdot 10^{51} \,\text{erg}$.
            In the second fit ($\chi^2_{\nu,\textsl{best}} = 2.17$), assuming two progenitors, the major contributor to the enrichment accounts for 95\% of the abundance pattern, and corresponds to a Pop~III fSN with $m_{\text{PopIII}} = 43 \text{M}_{\odot}$ and $E_{\text{SN}} = 0.6 \cdot 10^{51} \,\text{erg}$.
            The remaining 5\% originates from a Pop~III hypernova (HN) of mass $m_{\text{PopIII}} = 19.4 \, \text{M}_{\odot}$ and energy $E_{\text{SN}} = 10 \cdot 10^{51} \,\text{erg}$.
            Finally, the third fit ($\chi^2_{\nu,\textsl{best}} = 1.52$), which assumes two progenitors but only includes the 10 reliable elements, yields the same main progenitor as in the previous case, which now contributes 85\% of the total abundance pattern.
            The remaining 15\% comes from a Pop~III HN with $m_{\text{PopIII}} = 50 \, \text{M}_{\odot}$ and $E_{\text{SN}} = 10 \cdot 10^{51} \,\text{erg}$.
            These results indicate that the two-progenitor solutions generally provide a better match to the observed abundances than the single-progenitor case.
            However, we should notice that some elements, such as Ti and Co, are still not well reproduced by the models.
            Still, these elements are not usually well reproduced by the theoretical yields, also when considering Pop~II stars yields \citep[e.g.,][]{romano_quantifying_2010,kobayashi_origin_2020,koutsouridou_large_2025}.
            In addition, there could also be some effects of underlying physics of Pop~III stars that are not yet accounted for in current models.
            
            \autoref{fig:DR20080_progenitors} presents the best-fit models, but it is also important to explore the parameter space in order to assess the possible degeneracies among the models.
            The fit results for the three studied cases are shown in \cref{subfig:1prog,subfig:2prog,subfig:2prog_trusted}, where different combinations of $m_{\text{PopIII}}$ and $E_{\text{SN}}$ are colour-coded according to their $\chi^2_{\nu}/\chi^2_{\nu,\textsl{best}}$ values.
            For the two-progenitor cases, in \autoref{subfig:2prog} and \autoref{subfig:2prog_trusted}, only the parameter space for the main progenitor is plotted. 
            For each primary contributor, the fitting procedure identifies the properties of the secondary progenitor that minimise $\chi^2_\nu$.
            The best-fit solution ($\chi^2_{\nu,\textsl{best}}$) corresponds to the combination of primary and secondary progenitors that minimises the overall $\chi^2_\nu$. 
            From these two plots, it appears that, although restricting the fit to the most reliable elements yields a lower $\chi^2_{\nu,\textsl{best}}$ value, the reduced number of constraints significantly increases the degeneracy of the final solution.
            However, it is worth noting that, despite the increased degeneracy, the result for the best fit main progenitor is the same in both cases, further reinforcing the robustness of this outcome.
            Regarding the secondary progenitor, being only a minor contributor the degeneracy of the possible solutions is higher, but is remarkable that in both cases the large majority of solutions (i.e. those fits with $\chi^2_\nu < 1.5 \,\chi^2_{\nu,\textsl{best}}\,$; 97.5\% accounting for the complete abundance pattern and 78.9\% when only considering reliable elements) point towards a Pop~III HN as the secondary progenitor of DR20080. 
    
            Notably, the secondary progenitor of DR20080 in our best-fit model is consistent with the main progenitor inferred for the most metal-poor star in Sculptor, AS0039: a Pop~III HN with $m_{\text{PopIII}} \approx 20 \, \text{M}_{\odot}$ and $E_{\text{SN}} = 10 \cdot 10^{51} \,\text{erg}$ \citep{skuladottir_zero-metallicity_2021,skuladottir_tracing_2024}.
            This may indicate a connection between the formation environments.
            Particularly, it is possible that the same Pop~III HN enriched both AS0039 and DR20080.
            Moreover, it should be noticed that both stars belong to the subpopulation identified by \cite{arroyo-polonio_chemo-dynamics_2024} and interpreted as the remnant of a past UFD merger.
            However, given the uncertainties on the secondary progenitor of DR20080, this interpretation should be treated with caution.
    
            The results of this study strongly suggest that DR20080 is a strong candidate for being a direct descendant of Pop~III stars.
            The evidence points to a Pop~III star with a mass of $m_{\text{PopIII}} = 43 \text{M}_{\odot}$ that exploded as a fSN with an energy of $E_{\text{SN}} = 0.6 \cdot 10^{51} \,\text{erg}$ as the main contributor to its chemical enrichment, with a minor contribution from a HN.
            This study confirms the crucial role of CEMP-no stars in trying to unveil the mystery behind the first stars ever formed in the Universe.

%-----------------------------------------------------------------

    \section{Discussion and Conclusions}
    \label{sec:Conclusions}

        DR20080 is a CEMP-no star within the Sculptor dSph, the first confirmed so far at extremely low metallicities ($[\text{Fe}/\text{H}] \leq -3 $) and one of the very few identified in a dSph to date. 
        Such stars are indeed very rare in dSphs, which can be the manifestation of an underlying difference in the star formation conditions in this class of galaxy.
        Moreover, according to the recent work of \cite{arroyo-polonio_chemo-dynamics_2024}, DR20080 belongs to a subpopulation of stars that may have formed in a UFD and were then accreted into Sculptor.
        
        In this work, we conducted the analysis of the high-resolution spectra of this star, taken with VLT/UVES in three spectral bands, all within the visible range.
        These high-quality spectra allowed for the determination of chemical abundances for 17 elements and the estimation of upper limits for 4 others (see \Autoref{sec:Chemical_abundances}).
        This makes the analysis presented here one of the most detailed studies of chemical abundances ever conducted in a dwarf galaxy at such low metallicities.
        In particular, the high abundance of carbon and the underabundance of barium, along with the non-detection of europium ($[\text{Eu}/\text{Fe}] < +0.50$), confirm that this star belongs to the CEMP-no class.
        Subsequently, state-of-the-art corrections for NLTE effects were applied to all the abundances of elements for which such corrections were available.

        When comparing the chemical abundances of DR20080 with other stars of similar metallicity observed in Sculptor, dSphs, UFDs, and the MW halo, it becomes evident that the derived pattern is unique, at least in terms of what has been observed in the literature so far, both among stars in Sculptor and among extremely metal-poor stars in general.
        This conclusion is particularly true in relation to other CEMP-no stars in the MW halo and in UFDs, and points towards the existence of a dependency between CEMP-no properties and the type of galaxy in which they formed (see \Autoref{sec:DR20080_Local_Group}).

        The high-resolution spectral analysis of DR20080 also gave us various hints about the extremely metal-poor gas from which this star formed.
        First of all, CEMP-no stars are generally thought to have formed in an environment enriched by Pop~III stars exploded as fSNe. 
        Furthermore, the low $[\text{Si}/\text{O}]$ ratio indicates that Pop~III stars are the most probable main contributors to the chemical composition of DR20080 atmosphere.
        Finally, by comparing the derived 1D NLTE abundances with theoretical SN yield models (considering both single- and two-progenitor scenarios, for Pop~III and Pop~II stars), the case of DR20080 is best reproduced by a two–Pop~III-progenitor model.
        In this scenario, the dominant contributor is a Pop~III fSN of mass $m_{\text{PopIII}} = 43 \text{M}_{\odot}$ and explosion energy $E_{\text{SN}} = 0.6 \cdot 10^{51} \,\text{erg}$, while the secondary progenitor is a Pop~III HN with $m_{\text{PopIII}} = 19.4 \, \text{M}_{\odot}$ and  $E_{\text{SN}} = 10 \cdot 10^{51} \,\text{erg}$ (see \Autoref{sec:PopIII_progenitor_DR20080}). 
        Despite the degeneracies affecting the determination of the secondary progenitor, our best fit is consistent with the proposed progenitor of the star AS0039, the most metal-poor star known in Sculptor \citep{skuladottir_zero-metallicity_2021,skuladottir_tracing_2024}, which may suggest a connection between the two formation environments.
        These results lead us to conclude that DR20080 is an excellent candidate for being a direct descendant of Pop~III stars, expanding our view of the first stellar population.
        
        Our results further highlight the importance of increasing the sample of stars observed in dwarf galaxies, particularly through high-resolution spectroscopy.
        Indeed, this would improve the statistics on the rarity of CEMP-no stars within dSphs and help to clarify the role of host galaxy characteristics on CEMP-no properties, thereby expanding our understanding of the possible environment dependence of this class of stars.
        In this context, the 4DWARFS survey \citep{skuladottir_4most_2023}, which saw its first light in October 2025, represents one of the most promising opportunities to shed light on these aspects by increasing the sample of observed stars in dwarf galaxies by orders of magnitude.
        Once a potential CEMP-no candidate is identified, this work also highlights how crucial it is to perform a high-quality follow-up, which can lead to significantly different results compared to low-resolution studies, and help break the degeneracy among progenitor models thanks to the larger number of detected elements.

%-----------------------------------------------------------------
    \begin{acknowledgements}
        This project has received funding from the European Research Council (ERC) under the European Union’s Horizon 2020 research and innovation programme (grant agreement No. 101117455).
        Based on observations carried out with ESO/VLT/UVES at the La Silla Paranal Observatory (program ID 110.2407, PI: Á. Skúladóttir).
    \end{acknowledgements}

    \bibliography{bibliography_bibtex}

%-------------------------------------------------------------------

    \begin{appendix}

        \section{Individual los velocities}
        \label{App:Rad_Vel}

            \vspace{-6mm}

            \begin{strip}
                \centering
                \captionof{table}{True los velocity value for each band and exposure.}
                \label{tab:true_vrad}
            
                \begin{tabular}{cccccccc}
                    \toprule
                    & & \multicolumn{2}{c}{B-band} & \multicolumn{2}{c}{Rl-band} & \multicolumn{2}{c}{Ru-band} \\
                    \midrule
                    Exposure & Date & $v_{\textsl{los}}$ & $\Delta v_{\textsl{los}}$ & $v_{\textsl{los}}$ & $\Delta v_{\textsl{los}}$ & $v_{\textsl{los}}$ & $\Delta v_{\textsl{los}}$  \\
                    & $[\text{dd}/\text{mm}/\text{yyyy}]$ & $[\text{km}/\text{s}]$ & $[\text{km}/\text{s}]$ & $[\text{km}/\text{s}]$ & $[\text{km}/\text{s}]$ & $[\text{km}/\text{s}]$ & $[\text{km}/\text{s}]$ \\
                    \midrule
                    $1$ & 25/10/2022 & $119.20$ & $\pm 0.30$ & $119.25$ & $\pm 0.24$ & $121.36$ & $\pm 0.70$ \\
                    $2$ & 29/10/2022 & $121.00$ & $\pm 0.31$ & $120.33$ & $\pm 0.26$ & $119.65$ & $\pm 0.71$ \\
                    $3$ & 18/11/2022 & $121.42$ & $\pm 0.41$ & $120.66$ & $\pm 0.31$ & $120.02$ & $\pm 0.69$ \\
                    $4$ & 18/11/2022 & $123.57$ & $\pm 0.50$ & $121.68$ & $\pm 0.27$ & $119.94$ & $\pm 0.71$ \\
                    $5$ & 20/11/2022 & $119.36$ & $\pm 0.28$ & $120.07$ & $\pm 0.24$ & $120.96$ & $\pm 0.62$ \\
                    $6$ & 21/11/2022 & $121.02$ & $\pm 0.34$ & $120.83$ & $\pm 0.25$ & $118.70$ & $\pm 0.72$ \\
                    $7$ & 21/11/2022 & $120.12$ & $\pm 0.26$ & $120.24$ & $\pm 0.22$ & $117.82$ & $\pm 0.66$ \\
                    $8$ & 22/11/2022 & $119.53$ & $\pm 0.32$ & $119.47$ & $\pm 0.26$ & $117.27$ & $\pm 0.58$ \\
                    \bottomrule
                \end{tabular}
            \end{strip}
            
            \noindent The observed los velocity for each exposure, $v_{\textsl{los}_i}^{\textsl{obs}}$, where $i$ indicates $i$-th exposure, was measured exploiting one of the features of the program \textit{DAOSPEC}.
            This tool (see also \Autoref{subsec:Tools}) takes a literature list of lines of interest and an observed spectrum as input and, through a cross-check, outputs the observed los velocity associated with that specific spectrum, as well as its variance $\sigma_{v_{\textsl{los,i}}^{\textsl{obs}}}$ and the number of lines $n_{\textsl{lines}}$ that were used for the measurement.

\begin{figure}[t]
                \centering
                \includegraphics[width=0.90\hsize]{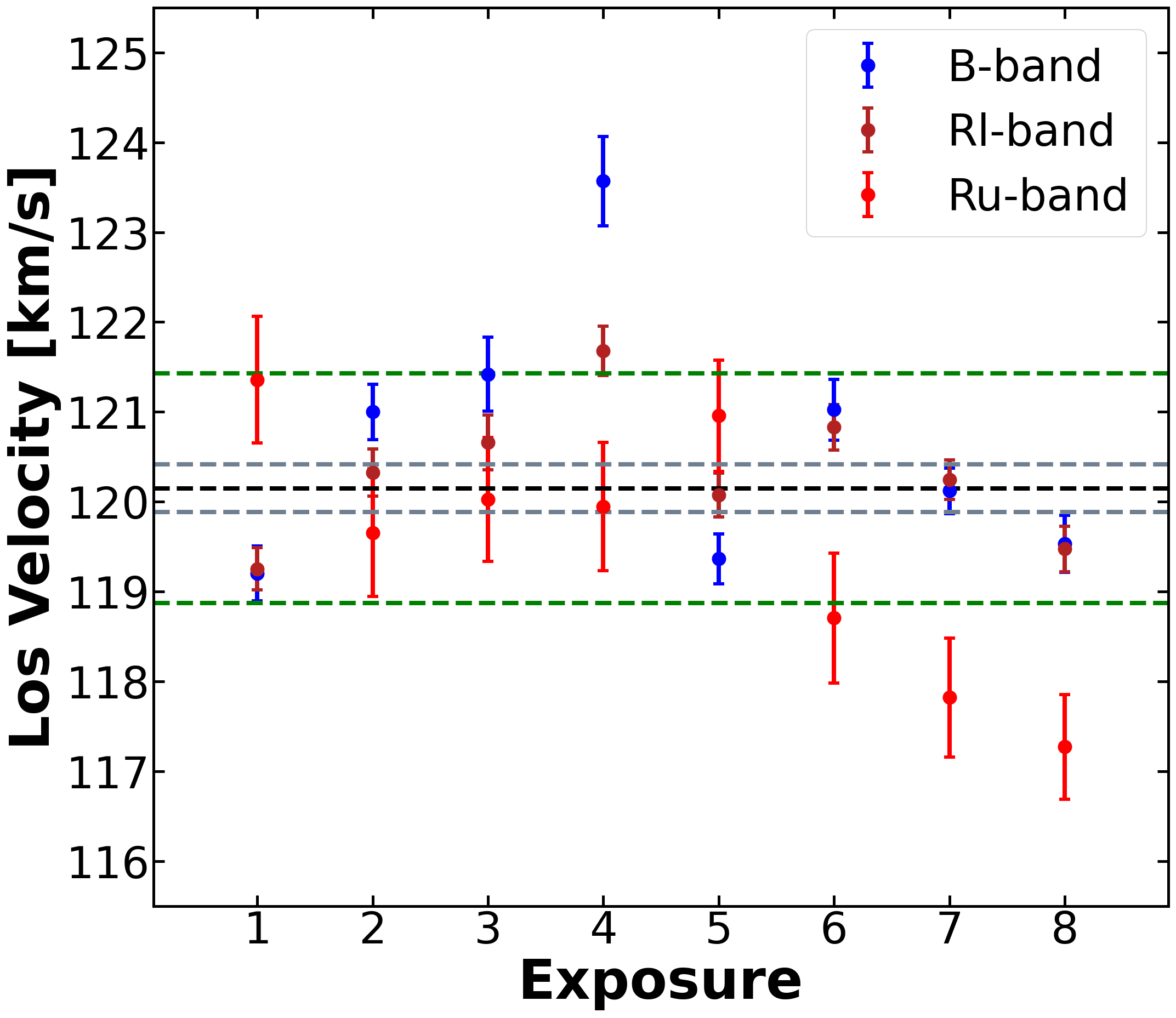}
                \\
                \vspace{2mm}
                \includegraphics[width=0.90\hsize]{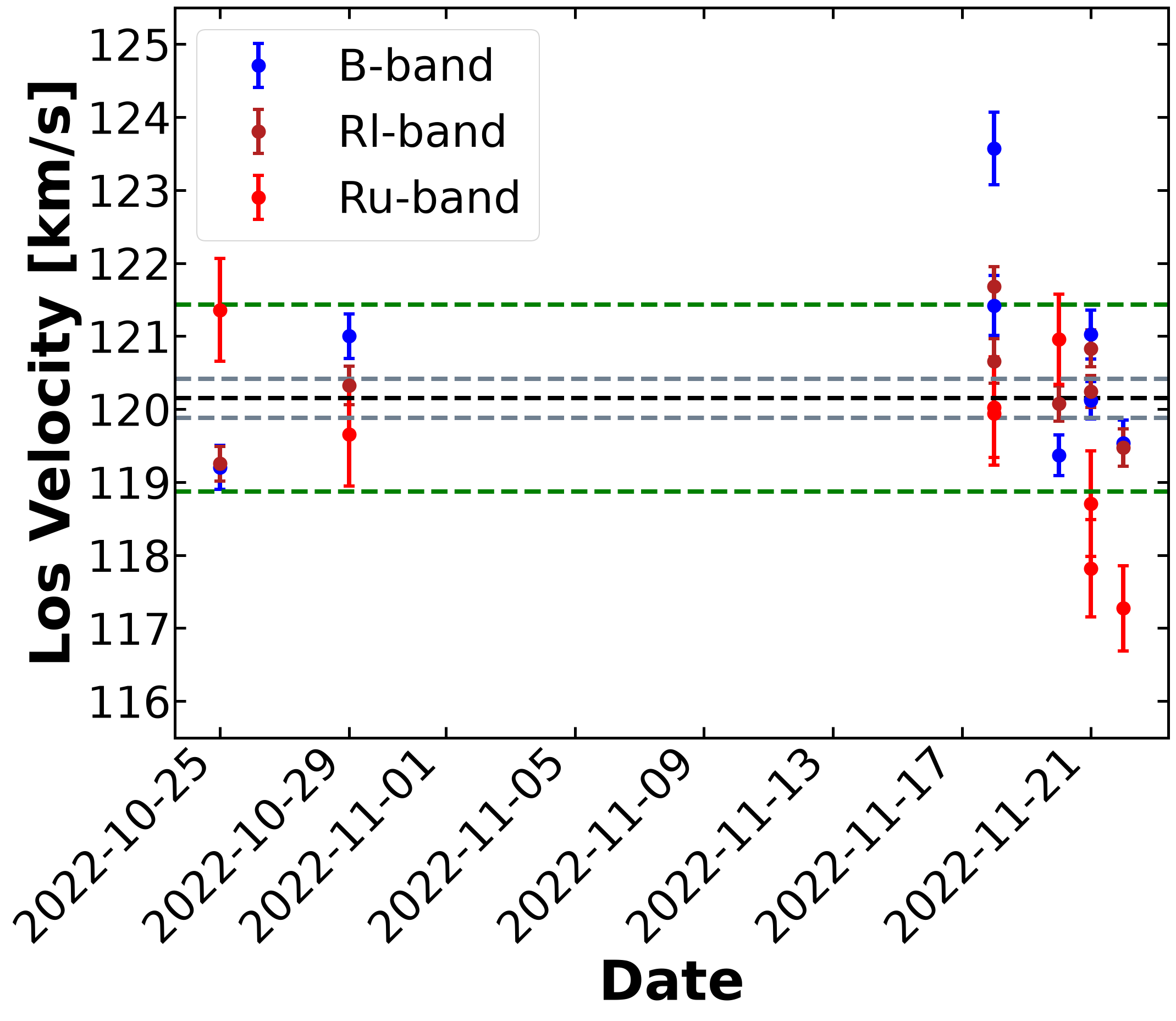}
                \caption{Los velocities of all 24 exposures after the heliocentric velocity correction, plotted as a function of the exposure number (\textit{top panel}) and the observation date (\textit{bottom panel}).
                Each band is represented by a different colour. 
                The \textit{black dashed line} is the arithmetic mean value of $v_{\textsl{los}}$, the \textit{grey} ones are the two extremes defined by $\Delta v_{\textsl{los}}$ and the \textit{green} ones by the standard deviation, $\sigma$, from the mean.}
                \label{fig:radial_vel}
            \end{figure}

            The true los velocity of DR20080, $v_{\textsl{los}_i}$, is the observed value corrected to the heliocentric reference frame, while the uncertainty was evaluated as
                \begin{equation*}
                \label{eq:delta_los}
                    \Delta v_{\textsl{los}_i} = \dfrac{\sigma_{v_{\textsl{los,i}}^{\textsl{obs}}}}{\sqrt{n_{\textsl{lines}_i} - 1}} \,\text{.}
                \end{equation*}
            All the results are presented in \autoref{tab:true_vrad} and plotted in \autoref{fig:radial_vel}.

            The mean value was estimated as an arithmetic mean of all the velocity measurements, while the uncertainty was estimated using the following equation   
            \begin{equation*}
                \Delta v_{\textsl{los}} = \dfrac{\sigma}{\sqrt{n_{\textsl{obs}} - 1}} \,\text{,}
            \end{equation*}
            where $\sigma$ is the standard deviation between all the measurements and $n_{\textsl{obs}} = 24$ is the total number of exposures. 
            This method gave the result $v_{\textsl{los}} = 120.15 \pm 0.27 \,\text{km s}^{-1}$.

            \clearpage

        \section{Analysed lines}              
        \label{App:Lines}
        
            \subsection{Line list}
            \label{subapp:line_list}                
                \autoref{tab:Linelist} presents the line list, together with the corresponding $EW$ value derived using \textit{ASTRA}.
                The line list is a compilation of the VALD database \citep{piskunov_vald_1995,ryabchikova_vienna_1997,kupka_vald-2_1999,kupka_vald-2_2000} and the works from \cite{tafelmeyer_extremely_2010,van_der_swaelmen_chemical_2013,jablonka_early_2015} (see also \Autoref{subsec:Tools}).
                When the $EW$ value is reported in parentheses, the corresponding abundance ratio $[\text{X}/\text{Fe}]$, listed in the fourth column, was derived through spectral synthesis rather than from the $EW$ measurement.
                In these cases, the reported $EW$ value should be regarded only as indicative and was not used for a reliable abundance determination(e.g., due to line blending).            
                Moreover, for all individual lines for which a NLTE correction is available in the literature for our adopted stellar parameters, the corresponding correction is reported ($\Delta [\text{X}/\text{H}]_{\textsl{NLTE}}$).
                In the same table are also reported the two molecular bands analysed in this work, namely CH and CN, together with their reference wavelength.
                Finally, we also report the upper limits derived for individual lines that were too weak to be detected, but correspond to elements whose abundance ratios are of particular interest.

            \subsection{Additional spectral fits}
            \label{subapp:additional_spectral_fits}
\begin{figure}[h]
                \centering
                \includegraphics[width=0.40\textwidth]{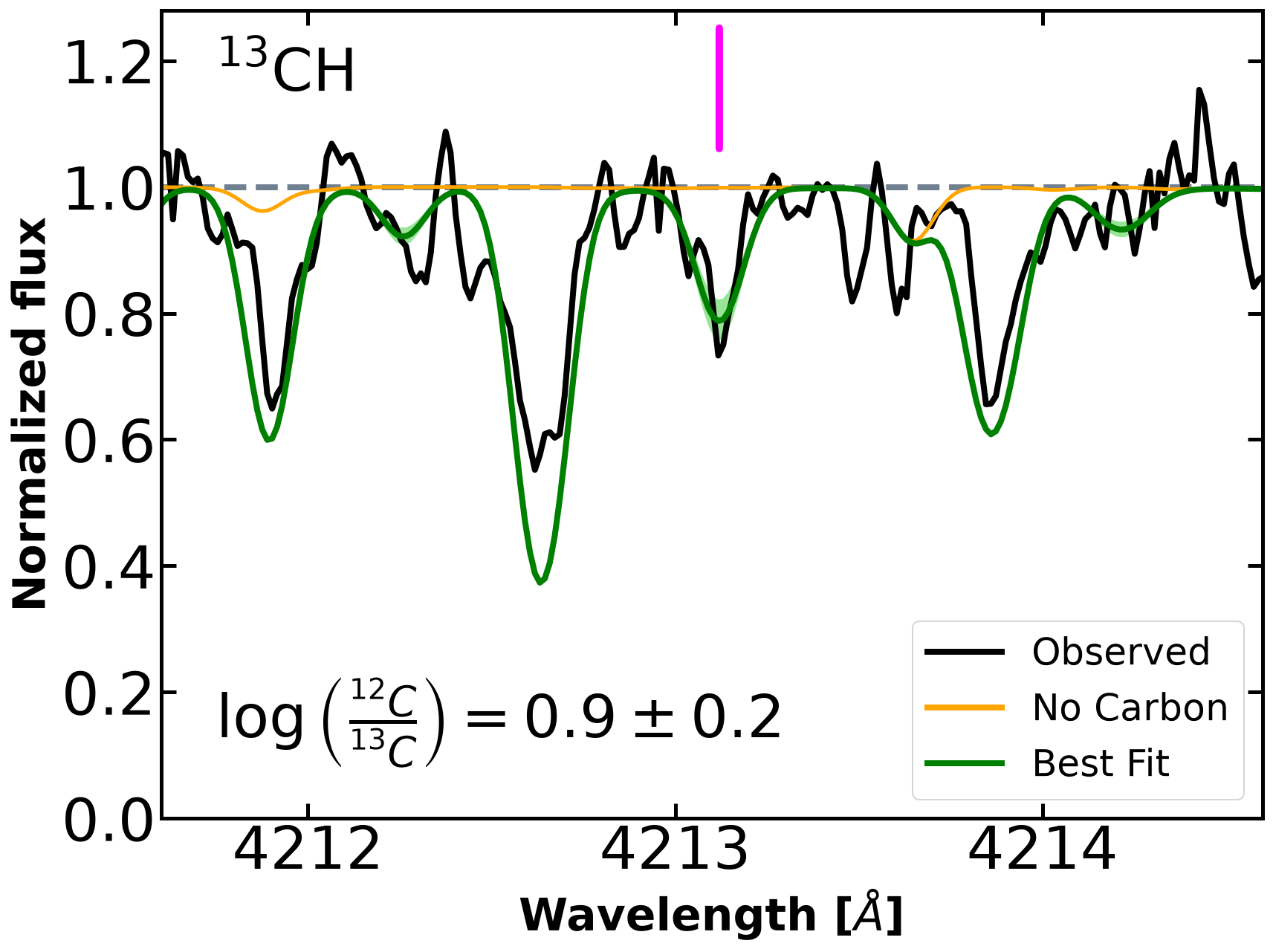}
                \caption{As \autoref{fig:spectral_lines}, but for the $^{13}$\ion{CH}{} absorption feature at $\sim 4213\,\text{\AA}$ $-$ B-band.
                }
                \label{fig:13CH_4213}
            \end{figure}

\begin{figure}[h]
                \centering
                \includegraphics[width=0.40\textwidth]{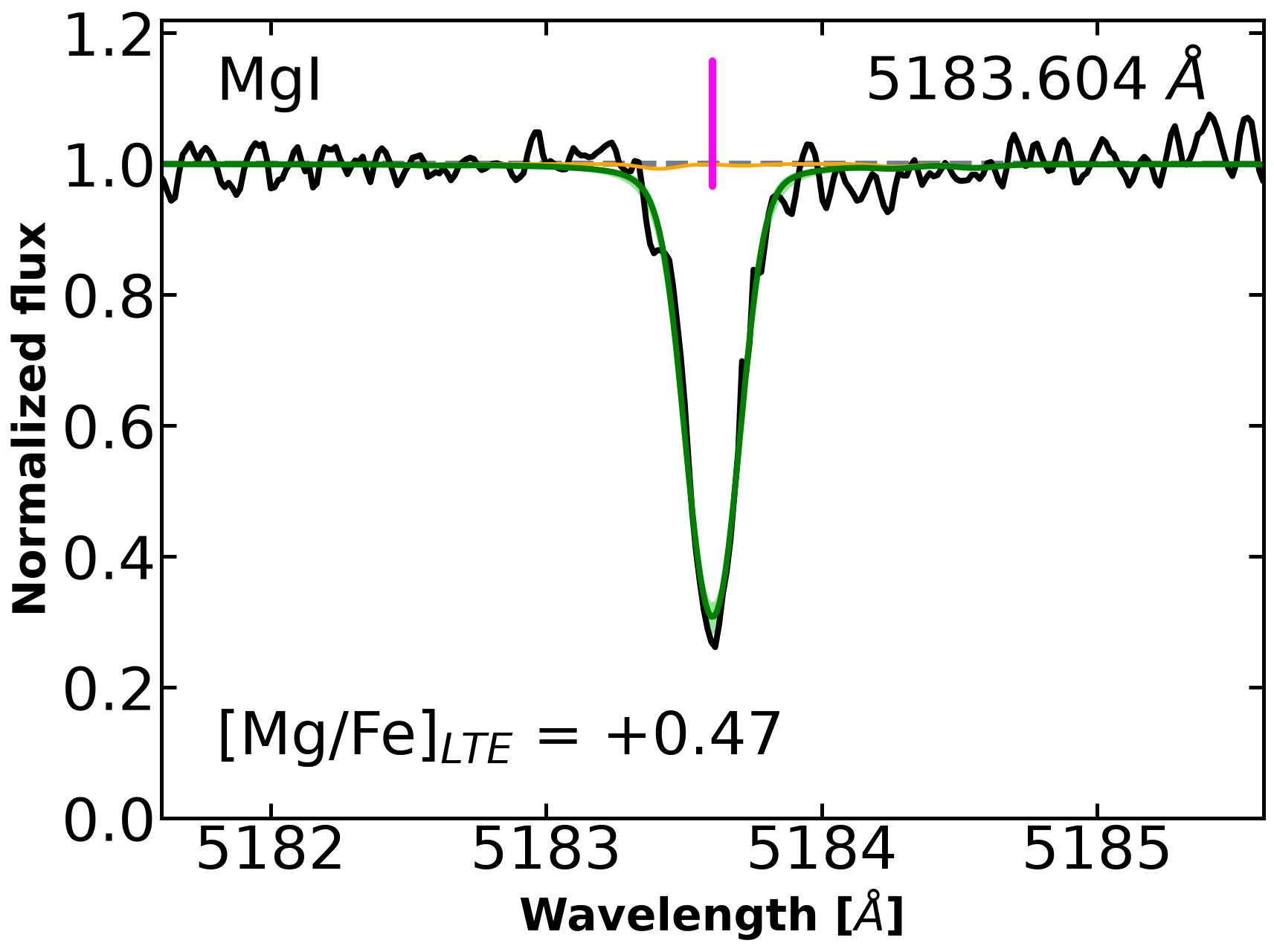}
                \caption{As \autoref{fig:spectral_lines}, but for the \ion{Mg}{i} absorption line at $5183.604\,\text{\AA}$ $-$ Rl-band.
                }
                \label{fig:Mg_5183}
            \end{figure}

                \autoref{fig:CN_3883} reports the CN molecular band around $3883 \,\text{\AA}$ used to derive the abundance of nitrogen.
                In \autoref{fig:13CH_4213} is reported one of the $^{13}$CH absorption features \citep{masseron_ch_2014} used for the estimation of $\log{\left(\frac{^{12}\text{C}}{^{13}\text{C}}\right)}$.
                \autoref{fig:Mg_5183} shows one of the three \ion{Mg}{i} lines used to derive the abundance value of magnesium.

            \begin{table}[h]
                \caption{Line list (see text for references), derived LTE abundances and NLTE corrections for the star DR20080. 
                Abundances of lines with the $EW$ reported in parentheses were derived with spectral synthesis. 
                The abundances are all referred to $[\text{Fe}/\text{H}]_{\textsl{mean}} = -3.26$.
                }
                \label{tab:Linelist}
                \centering
                {\fontsize{9}{12}\selectfont
                    \tabcolsep=0.09cm   
                    \begin{tabular}{ccccccc}
                        \toprule
                        Ion & \textlambda & $\chi_{\textsl{ex}}$ & $\log{(gf)}$ & $EW$ & $[\text{X}/\text{Fe}]$ & $\Delta [\text{X}/\text{H}]_{\textsl{NLTE}}$  \\
                        & $[\AA]$ & $[\text{eV}]$ & & $[\text{m}\AA]$ & $[\text{dex}]$ & $[\text{dex}]$ \\
                        \midrule
                        C$_{\textsl{obs}}$ (CH) & $4300$ & $-$ & $-$ & $(-)$ & $+1.07$ & $-$      \\
                        \midrule
                        \ion{N}{} (CN) & $3883$ & $-$ & $-$ & $-$ & $+1.72$ & $-$      \\
                        \midrule
                        \ion{O}{i} & $6300.304$ & $0.000$ & $-9.750$ & $(12.5)$ & $+1.53$ & $+0.000$ \\
                        \midrule
                        \ion{Na}{i} & $5889.951$ & $0.000$ & $0.108$ & $(138.4)$ & $+0.33$ & $-0.313$ \\
                        \ion{Na}{i} & $5895.924$ & $0.000$ & $-0.194$ & $(112.7)$ & $+0.09$ & $-0.198$ \\
                        \midrule
                        \ion{Mg}{i} & $5172.684$ & $2.712$ & $-0.450$ & $(155.2)$ & $+0.61$ & $+0.001$ \\
                        \ion{Mg}{i} & $5183.604$ & $2.717$ & $-0.239$ & $(158.4)$ & $+0.48$ & $+0.002$ \\
                        \ion{Mg}{i} & $5528.405$ & $4.346$ & $-0.498$ & $(35.0)$ & $+0.33$ & $+0.118$ \\
                        \midrule
                        \ion{Al}{i} & $3961.520$ & $0.014$ & $-0.323$ & $(125.0)$ & $-0.17$ & $+1.00$  \\
                        \midrule
                        \ion{Si}{i} & $3905.523$ & $1.909$ & $-1.041$ & $(-)$ & $-0.30$ & $-0.091$ \\
                        \midrule
                        \ion{Ca}{i} & $4226.728$ & $0.000$ & $0.244$ & $(146.7)$ & $+0.03$ & $+0.004$ \\
                        \ion{Ca}{i} & $4425.437$ & $1.879$ & $-0.358$ & $(27.7)$ & $+0.31$ & $+0.297$ \\
                        \ion{Ca}{i} & $4454.779$ & $1.899$ & $0.258$ & $(59.5)$ & $+0.22$ & $+0.308$ \\
                        \ion{Ca}{i} & $6122.217$ & $1.886$ & $-0.316$ & $(27.7)$ & $+0.14$ & $+0.270$ \\
                        \ion{Ca}{i} & $6162.173$ & $1.899$ & $-0.090$ & $45.8$ & $+0.28$ & $+0.264$ \\
                        \ion{Ca}{i} & $6439.075$ & $2.526$ & $0.390$ & $(21.7)$ & $+0.04$ & $+0.248$ \\
                        \midrule
                        \ion{Sc}{ii} & $4246.822$ & $0.315$ & $0.24$ & $(100.5)$ & $-0.11$ & $-$      \\
                        \ion{Sc}{ii} & $4226.728$ & $0.000$ & $0.244$ & $(146.7)$ & $+0.03$ & $-$      \\
                        \midrule
                        \ion{Ti}{i} & $4981.730$ & $0.848$ & $0.570$ & $(17.2)$ & $-0.25$ & $+0.495$ \\
                        \ion{Ti}{i} & $4991.066$ & $0.836$ & $0.450$ & $27.4$ & $+0.11$ & $+0.528$ \\
                        \ion{Ti}{i} & $5014.276$ & $0.813$ & $0.040$ & $(14.3)$ & $-0.06$ & $+0.569$ \\
                        \ion{Ti}{i} & $5210.384$ & $0.048$ & $-0.820$ & $(17.3)$ & $+0.14$ & $+0.653$ \\
                        \midrule
                        \ion{Ti}{ii} & $3759.293$ & $0.607$ & $0.280$ & $(138.6)$ & $-0.28$ & $-0.112$ \\
                        \ion{Ti}{ii} & $3761.322$ & $0.574$ & $0.180$ & $(120.0)$ & $-0.23$ & $-0.064$ \\
                        \ion{Ti}{ii} & $3913.461$ & $1.116$ & $-0.360$ & $(88.9)$ & $-0.25$ & $+0.124$ \\
                        \ion{Ti}{ii} & $4028.338$ & $1.892$ & $-0.920$ & $35.2$ & $+0.34$ & $+0.139$ \\
                        \ion{Ti}{ii} & $4290.215$ & $1.165$ & $-0.870$ & $(84.6)$ & $+0.20$ & $+0.155$ \\
                        \ion{Ti}{ii} & $4395.031$ & $1.084$ & $-0.540$ & $(93.5)$ & $+0.24$ & $+0.104$ \\
                        \ion{Ti}{ii} & $4399.765$ & $1.237$ & $-1.200$ & $80.7$ & $+0.66$ & $+0.154$ \\
                        \ion{Ti}{ii} & $4417.713$ & $1.165$ & $-1.190$ & $(61.6)$ & $+0.38$ & $+0.132$ \\
                        \ion{Ti}{ii} & $4418.331$ & $1.237$ & $-1.990$ & $49.0$ & $+0.82$ & $+0.132$ \\
                        \ion{Ti}{ii} & $4443.801$ & $1.080$ & $-0.710$ & $89.3$ & $+0.17$ & $+0.009$ \\
                        \ion{Ti}{ii} & $4468.493$ & $1.131$ & $-0.630$ & $(100.8)$ & $+0.09$ & $+0.011$ \\
                        \ion{Ti}{ii} & $5129.156$ & $1.892$ & $-1.340$ & $(16.0)$ & $+0.20$ & $+0.099$ \\
                        \ion{Ti}{ii} & $5188.687$ & $1.582$ & $-1.050$ & $35.9$ & $-0.02$ & $+0.101$ \\
                        \ion{Ti}{ii} & $5226.539$ & $1.566$ & $-1.260$ & $33.3$ & $+0.12$ & $+0.101$ \\
                        \ion{Ti}{ii} & $5336.786$ & $1.582$ & $-1.600$ & $20.9$ & $+0.20$ & $+0.118$ \\
                        \midrule
                        \ion{V}{i} & $4379.230$ & $0.301$ & $0.580$ & $-$ & $<-0.10$ & $-$      \\
                        \ion{V}{i} & $4384.713$ & $0.287$ & $0.410$ & $-$ & $<+0.70$ & $-$      \\
                        \midrule
                        \ion{Cr}{i} & $4254.352$ & $0.000$ & $-0.090$ & $(104.4)$ & $-0.25$ & $+0.644$ \\
                        \ion{Cr}{i} & $5206.023$ & $0.941$ & $0.020$ & $(64.0)$ & $-0.20$ & $+0.533$ \\
                        \ion{Cr}{i} & $5208.409$ & $0.941$ & $0.170$ & $(77.1)$ & $-0.19$ & $+0.526$ \\
                        \ion{Cr}{i} & $5409.784$ & $1.030$ & $-0.670$ & $(18.2)$ & $-0.30$ & $+0.478$ \\
                        \midrule
                        \ion{Mn}{i} & $4030.750$ & $0.000$ & $-0.49$ & $(98.6)$ & $-0.94$ & $+0.153$ \\
                        \ion{Mn}{i} & $4033.060$ & $0.000$ & $-1.817$ & $(76.6)$ & $-1.09$ & $+0.184$ \\
                        \bottomrule
                    \end{tabular}
                    }
            \end{table}

            \begin{table}[htp]
                \setcounter{table}{0}
                \centering
                \caption{continued.}
                \centering
                {\fontsize{9}{12}\selectfont
                    \tabcolsep=0.09cm
                    \begin{tabular}{ccccccc}
                        \toprule
                        Ion & \textlambda & $\chi_{\textsl{ex}}$ & $\log{(gf)}$ & $EW$ & $[\text{X}/\text{Fe}]$ & $\Delta [\text{X}/\text{H}]_{\textsl{NLTE}}$  \\
                        & $[\AA]$ & $[\text{eV}]$ & & $[\text{m}\AA]$ & $[\text{dex}]$ & $[\text{dex}]$ \\
                        \midrule
                        \ion{Fe}{i} & $4021.866$ & $2.759$ & $-0.729$ & $36.5$ & $-0.04$ & $-$      \\
                        \ion{Fe}{i} & $4032.627$ & $1.485$ & $-2.377$ & $25.6$ & $-0.17$ & $-$      \\
                        \ion{Fe}{i} & $4062.441$ & $2.845$ & $-0.862$ & $22.6$ & $-0.12$ & $-$      \\
                        \ion{Fe}{i} & $4067.978$ & $3.211$ & $-0.472$ & $24.9$ & $-0.02$ & $-$      \\
                        \ion{Fe}{i} & $4076.629$ & $3.211$ & $-0.529$ & $21.8$ & $-0.05$ & $-$      \\
                        \ion{Fe}{i} & $4107.488$ & $2.832$ & $-0.879$ & $36.7$ & $+0.19$ & $-$      \\
                        \ion{Fe}{i} & $4132.899$ & $2.845$ & $-1.006$ & $34.9$ & $+0.29$ & $-$      \\
                        \ion{Fe}{i} & $4134.677$ & $2.832$ & $-0.649$ & $31.8$ & $-0.15$ & $-$      \\
                        \ion{Fe}{i} & $4147.668$ & $1.485$ & $-2.104$ & $43.1$ & $-0.09$ & $-$      \\
                        \ion{Fe}{i} & $4153.899$ & $3.397$ & $-0.321$ & $30.1$ & $+0.16$ & $-$      \\
                        \ion{Fe}{i} & $4154.499$ & $2.832$ & $-0.688$ & $33.8$ & $-0.07$ & $-$      \\
                        \ion{Fe}{i} & $4175.636$ & $2.845$ & $-0.827$ & $40.2$ & $+0.22$ & $-$      \\
                        \ion{Fe}{i} & $4181.754$ & $2.832$ & $-0.371$ & $47.4$ & $-0.11$ & $-$      \\
                        \ion{Fe}{i} & $4187.039$ & $2.450$ & $-0.548$ & $56.9$ & $-0.21$ & $+0.170$ \\
                        \ion{Fe}{i} & $4187.795$ & $2.425$ & $-0.554$ & $79.6$ & $+0.28$ & $-$      \\
                        \ion{Fe}{i} & $4191.430$ & $2.469$ & $-0.666$ & $56.2$ & $-0.09$ & $-$      \\
                        \ion{Fe}{i} & $4222.213$ & $2.450$ & $-0.967$ & $50.2$ & $+0.07$ & $+0.178$ \\
                        \ion{Fe}{i} & $4227.426$ & $3.332$ & $0.266$ & $48.5$ & $-0.14$ & $-$      \\
                        \ion{Fe}{i} & $4233.603$ & $2.482$ & $-0.604$ & $64.0$ & $+0.02$ & $-$      \\
                        \ion{Fe}{i} & $4238.810$ & $3.397$ & $-0.233$ & $26.3$ & $-0.02$ & $-$      \\
                        \ion{Fe}{i} & $4250.119$ & $2.469$ & $-0.405$ & $69.5$ & $-0.07$ & $-$      \\
                        \ion{Fe}{i} & $4282.403$ & $2.176$ & $-0.779$ & $67.7$ & $-0.08$ & $+0.190$ \\
                        \ion{Fe}{i} & $4442.339$ & $2.198$ & $-1.255$ & $59.8$ & $+0.20$ & $-$      \\
                        \ion{Fe}{i} & $4447.717$ & $2.223$ & $-1.342$ & $38.2$ & $-0.09$ & $+0.197$ \\
                        \ion{Fe}{i} & $4494.563$ & $2.198$ & $-1.136$ & $56.9$ & $+0.02$ & $+0.202$ \\
                        \ion{Fe}{i} & $4859.741$ & $2.876$ & $-0.764$ & $36.8$ & $+0.05$ & $-$      \\
                        \ion{Fe}{i} & $4871.318$ & $2.865$ & $-0.363$ & $53.4$ & $-0.06$ & $-$      \\
                        \ion{Fe}{i} & $4872.138$ & $2.882$ & $-0.567$ & $42.7$ & $-0.03$ & $-$      \\
                        \ion{Fe}{i} & $4890.755$ & $2.876$ & $-0.394$ & $43.2$ & $-0.20$ & $+0.179$ \\
                        \ion{Fe}{i} & $4891.492$ & $2.851$ & $-0.112$ & $62.0$ & $-0.17$ & $+0.179$ \\
                        \ion{Fe}{i} & $4903.310$ & $2.882$ & $-0.926$ & $22.4$ & $-0.09$ & $+0.185$ \\
                        \ion{Fe}{i} & $4918.994$ & $2.865$ & $-0.342$ & $56.9$ & $-0.02$ & $-$      \\
                        \ion{Fe}{i} & $4920.502$ & $2.833$ & $0.068$ & $79.4$ & $-0.02$ & $-$      \\
                        \ion{Fe}{i} & $4938.814$ & $2.876$ & $-1.077$ & $22.0$ & $+0.04$ & $-$      \\
                        \ion{Fe}{i} & $5006.119$ & $2.833$ & $-0.638$ & $47.6$ & $+0.06$ & $+0.197$ \\
                        \ion{Fe}{i} & $5041.756$ & $1.485$ & $-2.203$ & $51.2$ & $+0.04$ & $-$      \\
                        \ion{Fe}{i} & $5049.820$ & $2.279$ & $-1.355$ & $36.2$ & $-0.11$ & $+0.201$ \\
                        \ion{Fe}{i} & $5068.766$ & $2.940$ & $-1.042$ & $23.6$ & $+0.11$ & $-$      \\
                        \ion{Fe}{i} & $5171.596$ & $1.485$ & $-1.793$ & $81.7$ & $+0.21$ & $-$      \\
                        \ion{Fe}{i} & $5191.455$ & $3.039$ & $-0.551$ & $32.4$ & $-0.08$ & $+0.191$ \\
                        \ion{Fe}{i} & $5192.344$ & $2.998$ & $-0.421$ & $43.6$ & $-0.05$ & $-$      \\
                        \ion{Fe}{i} & $5194.941$ & $1.557$ & $-2.090$ & $58.1$ & $+0.13$ & $+0.198$ \\
                        \ion{Fe}{i} & $5202.336$ & $2.176$ & $-1.838$ & $25.5$ & $+0.01$ & $+0.197$ \\
                        \ion{Fe}{i} & $5216.274$ & $1.608$ & $-2.150$ & $47.3$ & $+0.05$ & $+0.193$ \\
                        \ion{Fe}{i} & $5232.940$ & $2.940$ & $-0.058$ & $57.7$ & $-0.23$ & $+0.182$ \\
                        \ion{Fe}{i} & $5266.555$ & $2.998$ & $-0.386$ & $47.2$ & $-0.02$ & $-$      \\
                        \ion{Fe}{i} & $5281.790$ & $3.039$ & $-0.834$ & $23.7$ & $+0.01$ & $+0.194$ \\
                        \ion{Fe}{i} & $5283.621$ & $3.241$ & $-0.432$ & $27.7$ & $-0.06$ & $-$      \\
                        \ion{Fe}{i} & $5324.179$ & $3.211$ & $-0.103$ & $45.7$ & $-0.08$ & $+0.217$ \\
                        \ion{Fe}{i} & $5328.531$ & $1.557$ & $-1.850$ & $80.1$ & $+0.30$ & $-$      \\
                        \ion{Fe}{i} & $5332.899$ & $1.557$ & $-2.777$ & $21.1$ & $+0.08$ & $-$      \\
                        \ion{Fe}{i} & $5424.068$ & $4.320$ & $0.520$ & $23.0$ & $+0.16$ & $-$      \\
                        \ion{Fe}{i} & $5572.842$ & $3.397$ & $-0.275$ & $29.8$ & $-0.00$ & $+0.210$ \\
                        \ion{Fe}{i} & $6136.615$ & $2.453$ & $-1.400$ & $24.3$ & $-0.18$ & $+0.179$ \\
                    \end{tabular}
                    }
            \end{table} 

            \begin{table}[htp]
                \setcounter{table}{0}
                \centering
                {\fontsize{9}{12}\selectfont
                    \tabcolsep=0.09cm
                    \begin{tabular}{ccccccc}
                        \toprule
                        Ion & \textlambda & $\chi_{\textsl{ex}}$ & $\log{(gf)}$ & $EW$ & $[\text{X}/\text{Fe}]$ & $\Delta [\text{X}/\text{H}]_{\textsl{NLTE}}$  \\
                        & $[\AA]$ & $[\text{eV}]$ & & $[\text{m}\AA]$ & $[\text{dex}]$ & $[\text{dex}]$ \\
                        \midrule
                        \ion{Fe}{i} & $6137.691$ & $2.588$ & $-1.403$ & $24.6$ & $-0.00$ & $+0.175$ \\
                        \ion{Fe}{i} & $6191.558$ & $2.433$ & $-1.417$ & $28.3$ & $-0.10$ & $+0.201$ \\
                        \ion{Fe}{i} & $6230.722$ & $2.559$ & $-1.281$ & $32.1$ & $-0.01$ & $+0.203$ \\
                        \ion{Fe}{i} & $6252.555$ & $2.404$ & $-1.687$ & $32.1$ & $+0.21$ & $+0.201$ \\
                        \ion{Fe}{i} & $6393.601$ & $2.433$ & $-1.432$ & $24.9$ & $-0.17$ & $+0.209$ \\
                        \ion{Fe}{i} & $6430.845$ & $2.176$ & $-2.006$ & $27.9$ & $+0.15$ & $+0.000$ \\
                        \ion{Fe}{i} & $6494.980$ & $2.404$ & $-1.273$ & $47.5$ & $+0.05$ & $-$      \\
                        \midrule
                        \ion{Fe}{ii} & $4233.162$ & $2.583$ & $-1.900$ & $(59.3)$ & $-0.04$ & $-0.006$ \\
                        \ion{Fe}{ii} & $4923.921$ & $2.891$ & $-1.320$ & $76.7$ & $+0.00$ & $+0.047$ \\
                        \ion{Fe}{ii} & $5018.436$ & $2.891$ & $-1.220$ & $86.6$ & $+0.10$ & $-$      \\
                        \ion{Fe}{ii} & $5197.567$ & $3.230$ & $-2.100$ & $22.0$ & $+0.09$ & $+0.000$ \\
                        \ion{Fe}{ii} & $5275.997$ & $3.199$ & $-1.940$ & $(24.5)$ & $-0.08$ & $-$      \\
                        \midrule
                        \ion{Co}{i} & $3894.077$ & $1.049$ & $0.100$ & $(64.5)$ & $+0.37$ & $+0.811$ \\
                        \ion{Co}{i} & $3995.307$ & $0.923$ & $-0.22$ & $(59.7)$ & $-0.18$ & $+0.649$ \\
                        \ion{Co}{i} & $4118.773$ & $1.049$ & $-2.011$ & $(37.8)$ & $-0.02$ & $+0.891$ \\
                        \ion{Co}{i} & $4121.318$ & $0.923$ & $-1.489$ & $(55.5)$ & $-0.10$ & $+0.816$ \\
                        \midrule
                        \ion{Ni}{i} & $5476.904$ & $1.826$ & $-0.780$ & $(27.3)$ & $-0.45$ & $+0.185$ \\
                        \midrule
                        \ion{Cu}{i} & $5105.537$ & $1.389$ & $-1.54$ & $-$ & $<+0.40$ & $-$      \\
                        \ion{Cu}{i} & $5700.237$ & $1.642$ & $-2.58$ & $-$ & $<+1.70$ & $-$      \\
                        \ion{Cu}{i} & $5782.127$ & $1.642$ & $-1.91$ & $-$ & $<+1.00$ & $-$      \\
                        \midrule
                        \ion{Zn}{i} & $4722.153$ & $4.030$ & $-0.338$ & $-$ & $<+0.70$ & $-$      \\
                        \ion{Zn}{i} & $4810.528$ & $4.078$ & $-0.137$ & $-$ & $<+0.50$ & $-$      \\
                        \midrule
                        \ion{Sr}{ii} & $4077.709$ & $0.000$ & $0.167$ & $(70.9)$ & $-1.73$ & $+0.288$ \\
                        \ion{Sr}{ii} & $4215.519$ & $0.000$ & $-0.145$ & $(65.5)$ & $-1.73$ & $+0.329$ \\
                        \midrule
                        \ion{Ba}{ii} & $4934.076$ & $0.000$ & $-0.15$ & $(36.1)$ & $-1.28$ & $+0.206$ \\
                        \midrule
                        \ion{Eu}{ii} & $3907.106$ & $0.207$ & $0.171$ & $-$ & $<+1.00$ & $-$      \\
                        \ion{Eu}{ii} & $4129.708$ & $0.000$ & $0.22$ & $-$ & $<+0.50$ & $-$      \\
                        \ion{Eu}{ii} & $4205.042$ & $0.000$ & $0.21$ & $-$ & $<+0.50$ & $-$      \\
                        \bottomrule
                    \end{tabular}
                }
            \end{table} 
            \clearpage

            \begin{strip}
                \centering
                \includegraphics[width=0.77\textwidth]{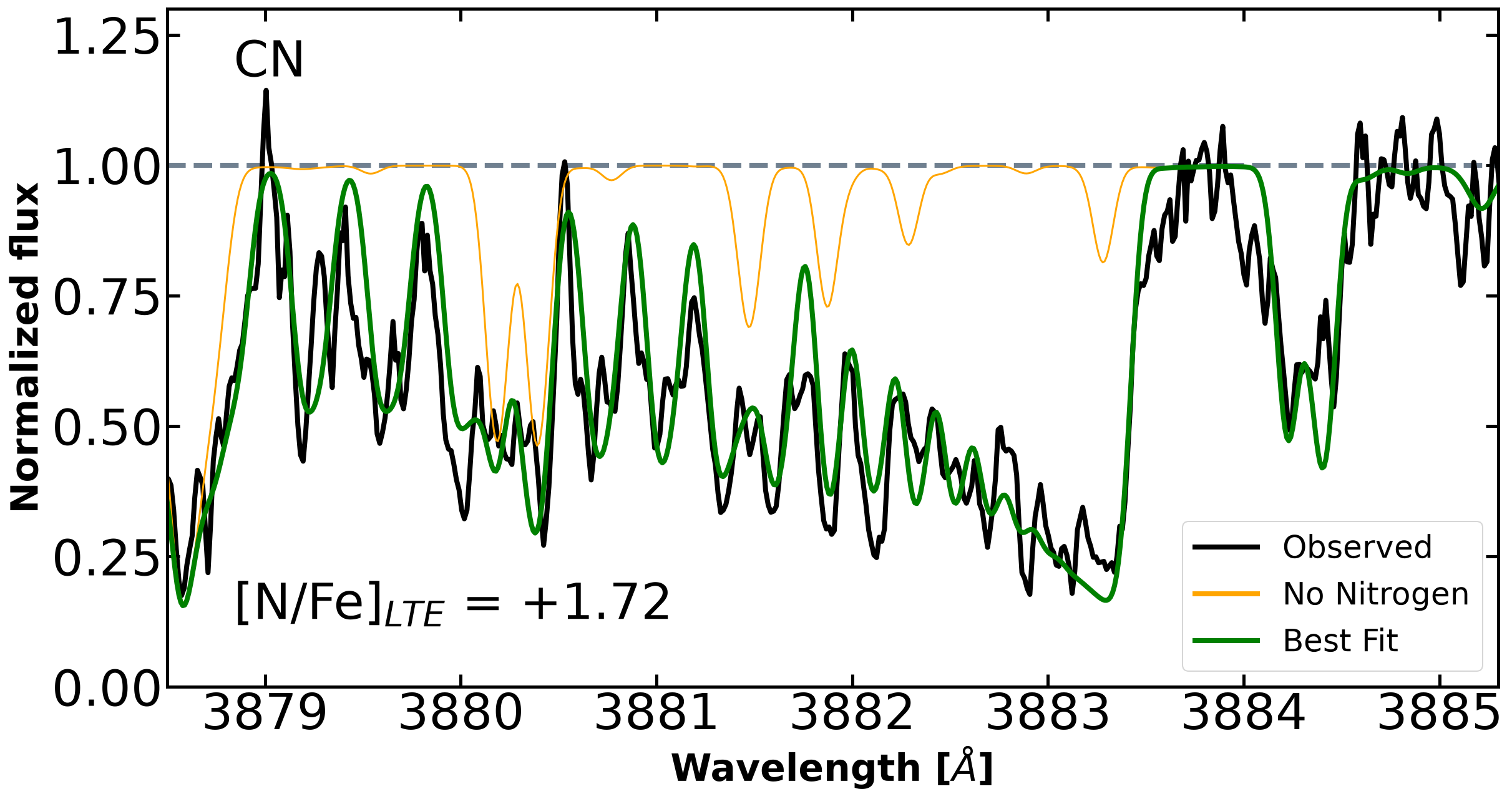}
                \captionof{figure}{As \autoref{fig:spectral_lines}, but for the CN molecular band ($\sim 3883\,\text{\AA}$) $-$ B-band.}
                \label{fig:CN_3883}
            \end{strip}

        \section{Abundance ratios}
        \label{App:abun_ratios}

            \begin{figure*}
                \centering
                \includegraphics[width=0.33\hsize]{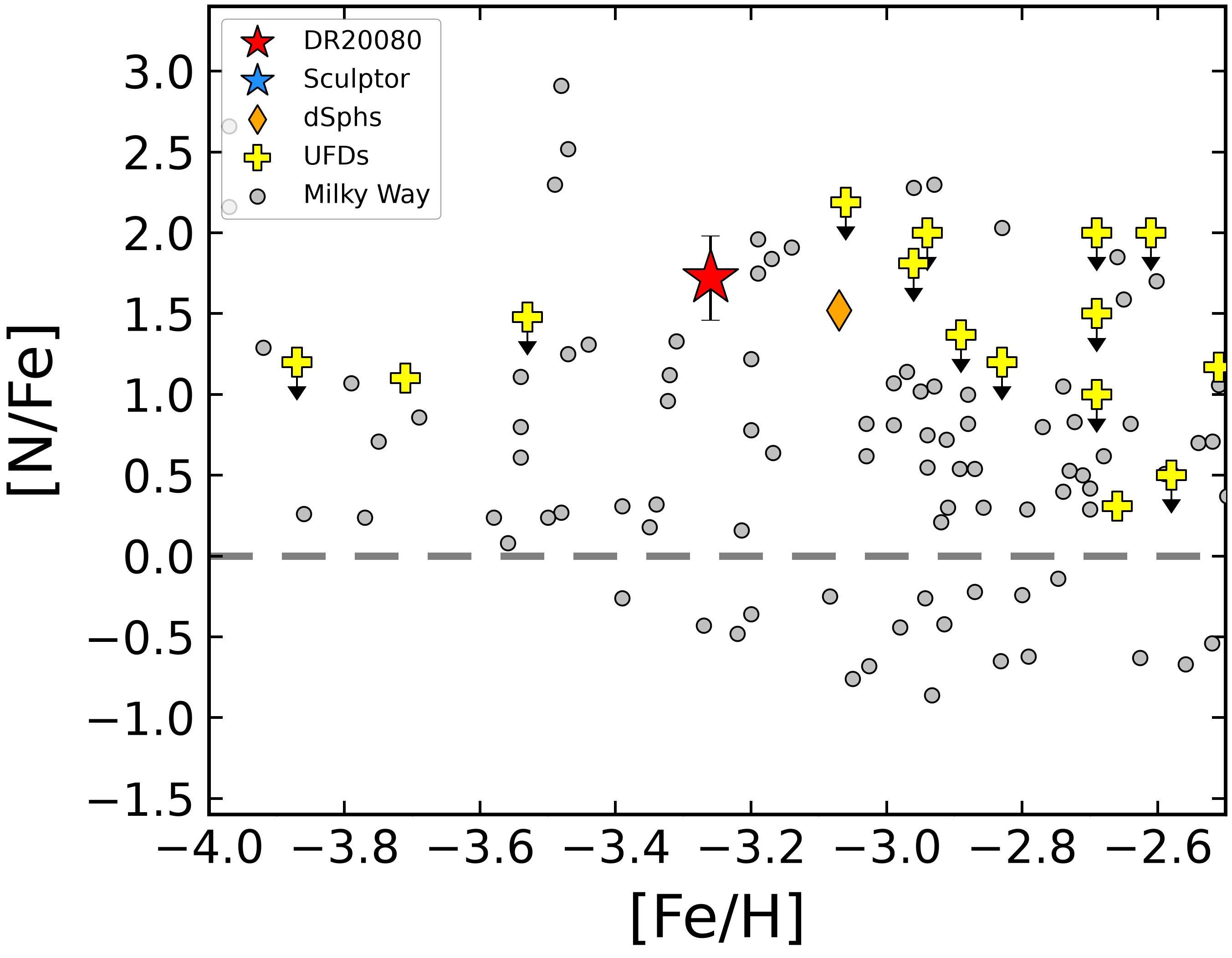}
                \includegraphics[width=0.33\hsize]{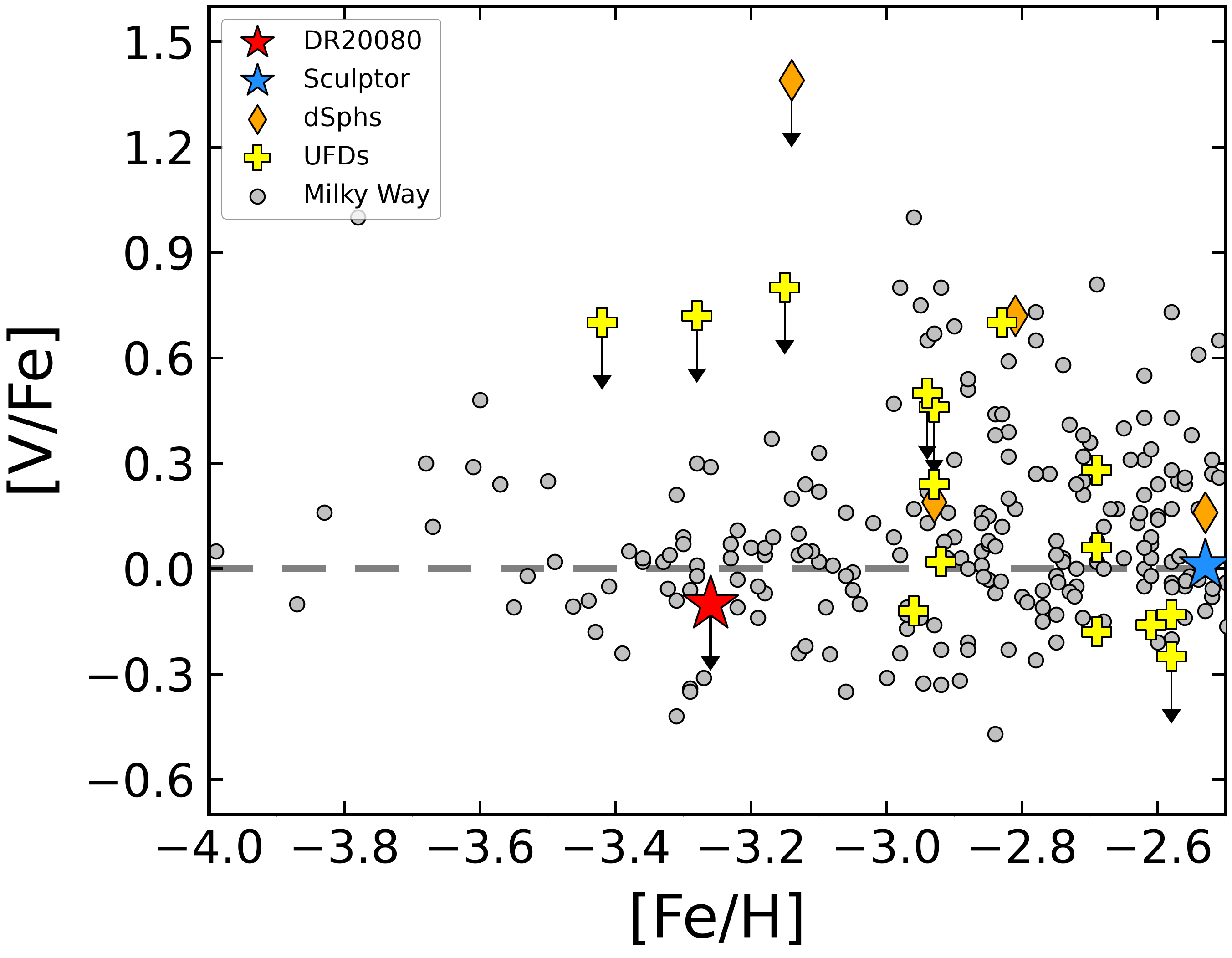}
                \includegraphics[width=0.33\hsize]{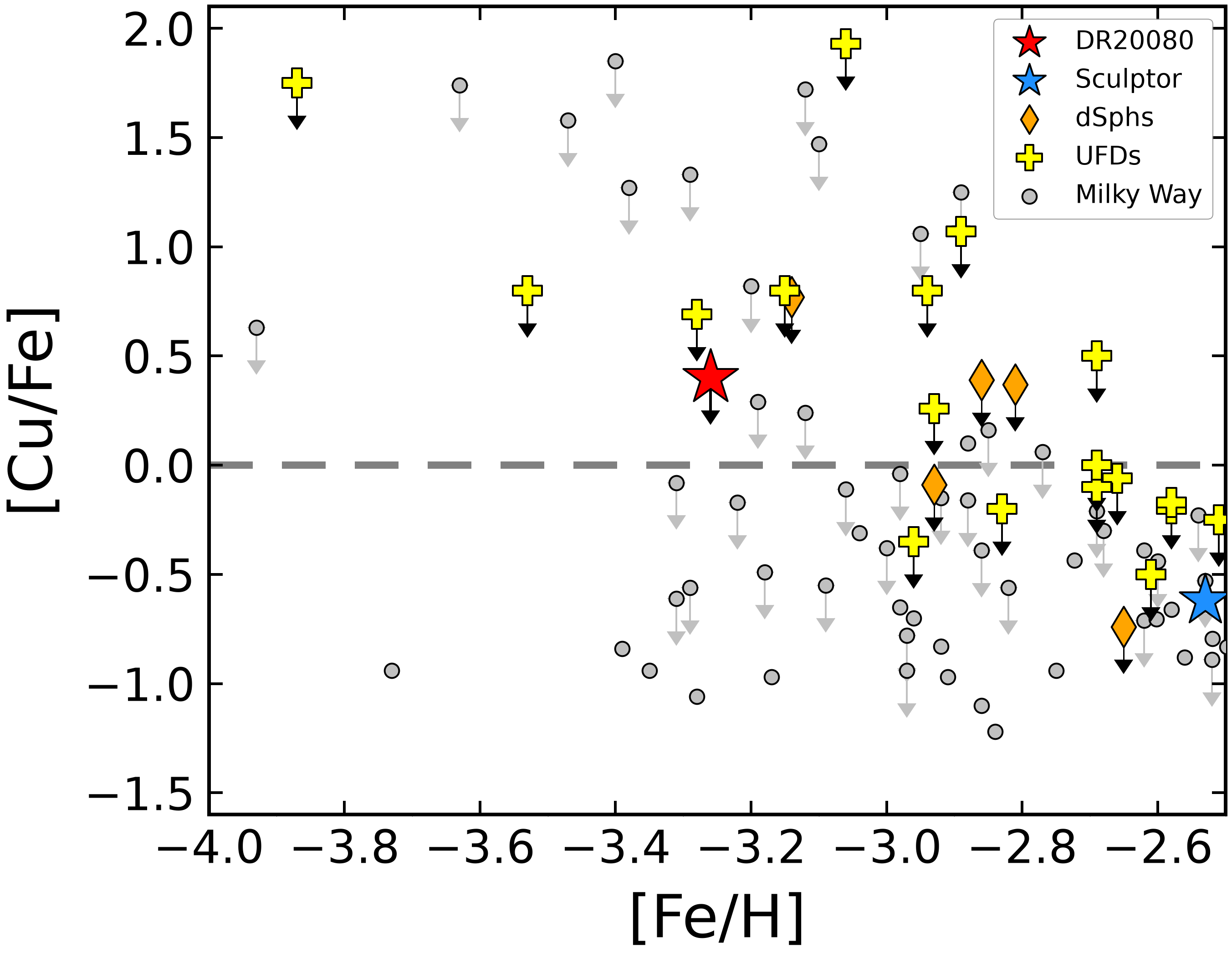}
                \\ \vspace{0.2cm}
                \includegraphics[width=0.33\hsize]{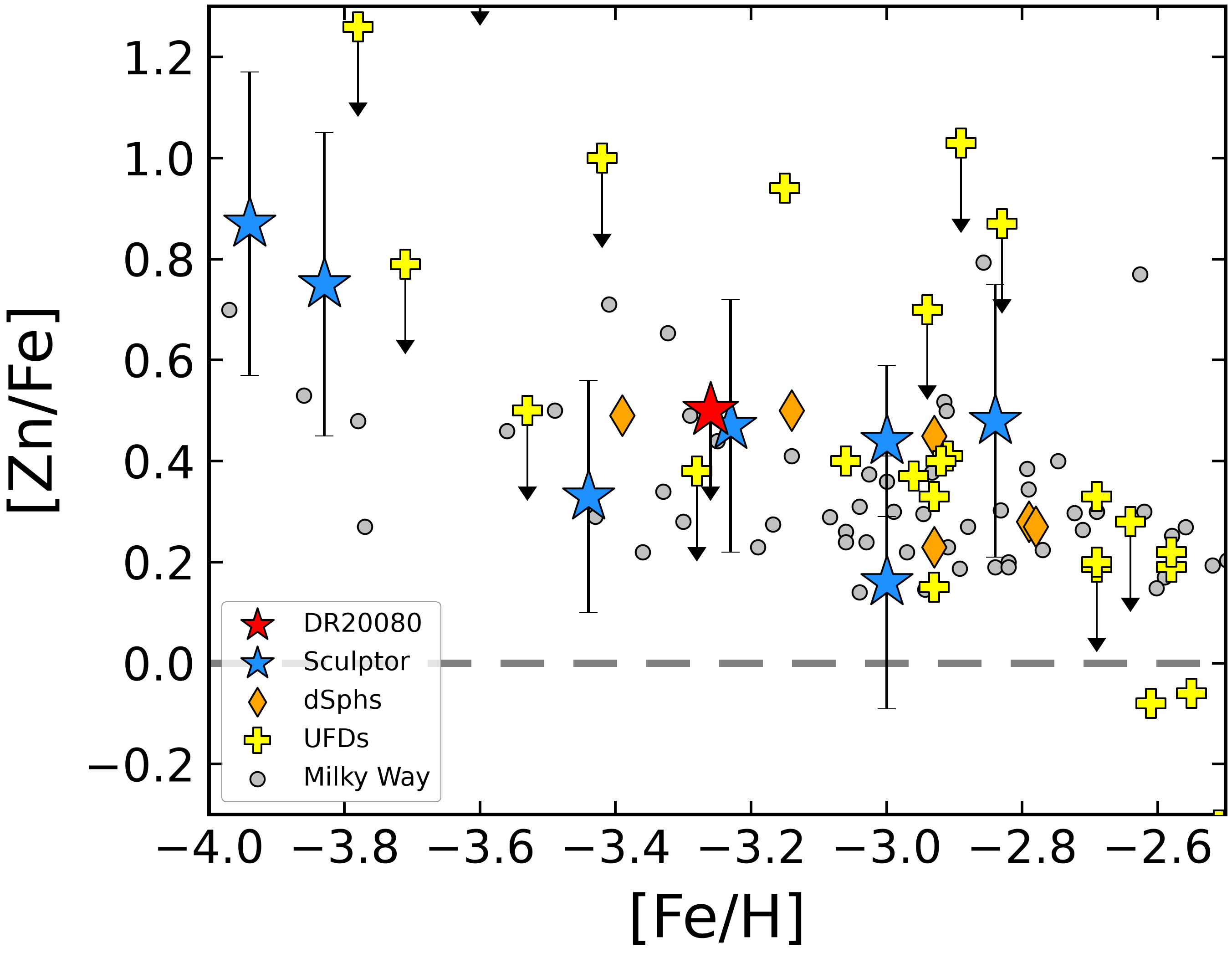}
                \includegraphics[width=0.33\hsize]{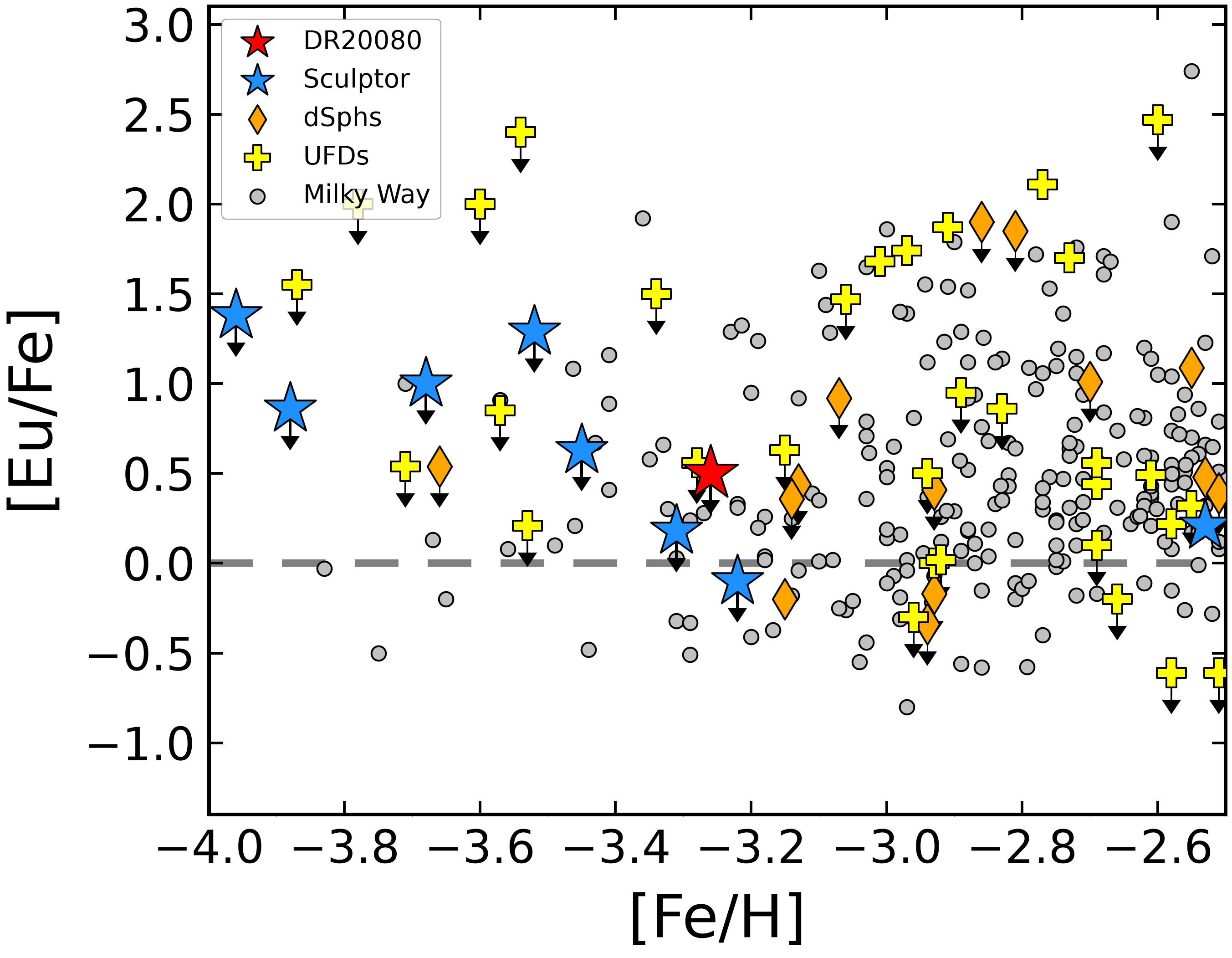}
                
                \caption{Same as \autoref{fig:C}, but for $[\text{N}/\text{Fe}]$, $[\text{V}/\text{Fe}]$, $[\text{Cu}/\text{Fe}]$, $[\text{Zn}/\text{Fe}]$, and $[\text{Eu}/\text{Fe}]$. 
                See \Autoref{subsec:Comparison_Data} and \autoref{App:abun_ratios} for references.
                }
                \label{fig:XFe_upplims}
            \end{figure*} 
            
            \begin{figure*}[t]
                \centering
                \includegraphics[width=0.8\textwidth]{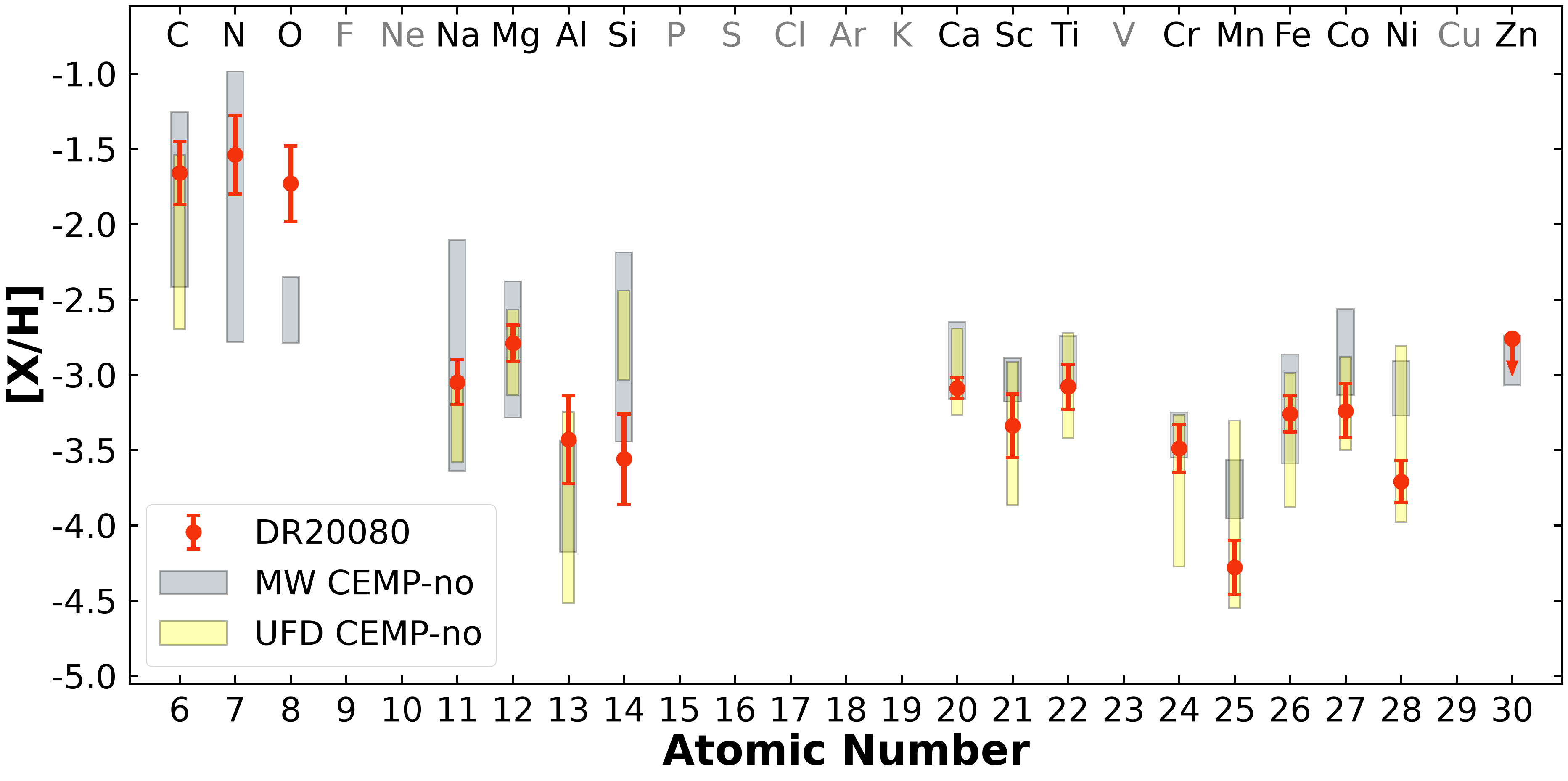}
                \caption{As \autoref{fig:DR20080_vs_other_CEMP-no_[X/Fe]}, but the abundances are reported as $[\text{X}/\text{H}]$. 
                        }
                \label{fig:DR20080_vs_other_CEMP-no_[X/H]}
            \end{figure*}

            \noindent \autoref{fig:XFe_upplims} shows the N abundance together with the upper limits derived for V, Cu, Zn and Eu in DR20080, compared with stars in the MW and its satellite system. 
            The corresponding references are reported in \Autoref{subsec:Comparison_Data} and are the same as for \Cref{fig:C,fig:XFe,fig:SrBa}.
            For V, Cu and Eu, the MW comparison sample consists of a compilation of data from \cite{lombardo_chemical_2022,lombardo_chemical_2025} together with measurements retrieved from the JINA database, including data from:  \cite{mcwilliam_spectroscopic_1995,norris_extremely_1997,westin_r-process-enriched_2000,aoki_chemical_2002,aoki_subaruhds_2002,aoki_spectroscopic_2005,aoki_high-resolution_2013,johnson_abundances_2002,sneden_extremely_2003,honda_spectroscopic_2004,honda_spectroscopic_2011,barbuy_new_2005,barklem_hamburgeso_2005,masseron_cs_2006,masseron_holistic_2010,masseron_lithium_2012,preston_ty_2006,sivarani_first_2006,frebel_discovery_2007,frebel_chemical_2007,cohen_new_2008,cohen_normal_2013,lai_detailed_2008,hayek_hamburgeso_2009,behara_three_2010,roederer_characterizing_2010,hollek_chemical_2011,allen_elemental_2012,ishigaki_chemical_2013,placco_metal-poor_2013,placco_metal-poor_2014,caliskan_chemical_2014,mashonkina_hamburgeso_2014,siqueira_mello_high-resolution_2014,spite_low_2014,hansen_elemental_2015,jacobson_high-resolution_2015}; and \cite{li_spectroscopic_2015}.
            
            In \autoref{fig:DR20080_vs_other_CEMP-no_[X/H]} are reported the same data as in \autoref{fig:DR20080_vs_other_CEMP-no_[X/Fe]}, but with the abundance ratios referred to hydrogen, $[\text{X}/\text{H}]$, instead of iron.
        
    \end{appendix}

\end{document}